\documentclass[acmlarge]{acmart}

\usepackage{amsmath,amsfonts}
\usepackage{balance,stfloats,hyperref,soul}
\usepackage[italic]{mathastext}
\usepackage{graphicx}
\usepackage{textcomp}
\usepackage{xcolor}
\usepackage[shortlabels]{enumitem}
\usepackage{tcolorbox}
\usepackage{xspace}
\usepackage{url}
\usepackage{booktabs}
\usepackage[nameinlink,noabbrev,capitalise]{cleveref}
\usepackage{algorithm}
\usepackage{caption}
\usepackage{subcaption}
\usepackage{gensymb}
\usepackage[noend]{algpseudocode}
\usepackage{wrapfig}
\usepackage{hyperref}
\usepackage{hanging}
\usepackage{colortbl}
\usepackage{tabularx}
\usepackage{pifont}
\definecolor{mygray}{HTML}{d6d0d6}

\setcopyright{acmlicensed}
\copyrightyear{2025}
\acmYear{2025}
\acmDOI{XXXXXXX.XXXXXXX}

\acmConference[IMWUT'25]{Proc. ACM Interact. Mob. Wearable Ubiquitous Technol.}{October 12 -- October 16,
  2025}{Espoo, Finland}
\acmISBN{978-1-4503-XXXX-X/18/06}

\newtheorem{thm}{Theorem}[section] 

\newcommand{\thistheoremname}{}
\newtheorem{genericthm}[thm]{\thistheoremname}

\begin{document}

\title{Ads that Talk Back: Implications and Perceptions of Injecting Personalized Advertising into LLM Chatbots}






\author{Brian Jay Tang}
\email{bjaytang@umich.edu}
\affiliation{
    \institution{University of Michigan}
    \city{Ann Arbor}
    \country{MI, USA}
}

\author{Kaiwen Sun}
\email{kaiwsun@iu.edu}
\affiliation{
    \institution{Indiana University Bloomington$^\ast$}
    \department{(Work done while at University of Michigan)}
    \city{Bloomington}
    \country{IN, USA}
}

\author{Noah T. Curran}
\email{ntcurran@umich.edu}
\affiliation{
    \institution{University of Michigan}
    \city{Ann Arbor}
    \country{MI, USA}
}

\author{Florian Schaub}
\email{fschaub@umich.edu}
\affiliation{
    \institution{University of Michigan}
    \city{Ann Arbor}
    \country{MI, USA}
}

\author{Kang G. Shin}
\email{kgshin@umich.edu}
\affiliation{
    \institution{University of Michigan}
    \city{Ann Arbor}
    \country{MI, USA}
}


\begin{abstract}
Recent advances in large language models (LLMs) have enabled the creation of
highly effective chatbots. However, the compute costs of widely deploying LLMs have raised questions about profitability. Companies have proposed exploring ad-based revenue streams for monetizing LLMs, which could serve as the new de facto platform for advertising. 
This paper investigates the implications of personalizing LLM advertisements to individual users via a between-subjects experiment with 179 participants.
We developed a chatbot that embeds personalized 
product advertisements within LLM responses, inspired by similar forays by 
AI companies. The evaluation of our benchmarks showed that ad injection only 
slightly impacted LLM performance, particularly response desirability.
Results revealed that participants struggled to detect ads, and even preferred LLM responses with hidden advertisements. Rather than clicking on our advertising disclosure, participants tried changing their advertising 
settings using natural language queries. We created an advertising dataset and an open-source LLM, Phi-4-Ads, fine-tuned to serve ads and flexibly adapt to user preferences.

\end{abstract}

\maketitle

\section{Introduction}\label{sec:introduction}

Recent advances in natural language processing (NLP) and large language 
models (LLMs) have enabled the creation of increasingly effective 
conversational AI agents. These LLMs are useful for a variety of tasks, 
such as information retrieval, writing/coding assistance, task automation, 
Q\&A, and embodiment in robots/systems. 
OpenAI's ChatGPT~\cite{introducingchatgpt}, one of the earliest LLM-powered chatbots, has prompted 
other companies to train and integrate LLMs into their own search
engines and products. 
Microsoft developed Bing Chatbot and Copilot, while Google and Amazon 
introduced Gemini and Rufus AI~\cite{RajivMehta2024rufus}. 
Many other online services, such as Snapchat, Quizlet, Instacart, 
and Shopify, now integrate LLMs into their own 
services and existing chat/search platforms~\cite{Snapchat2023}. 

\begin{wrapfigure}{r}{0.38\textwidth}
  \centering
  \vspace{-6mm}
  \includegraphics[width=0.38\columnwidth]{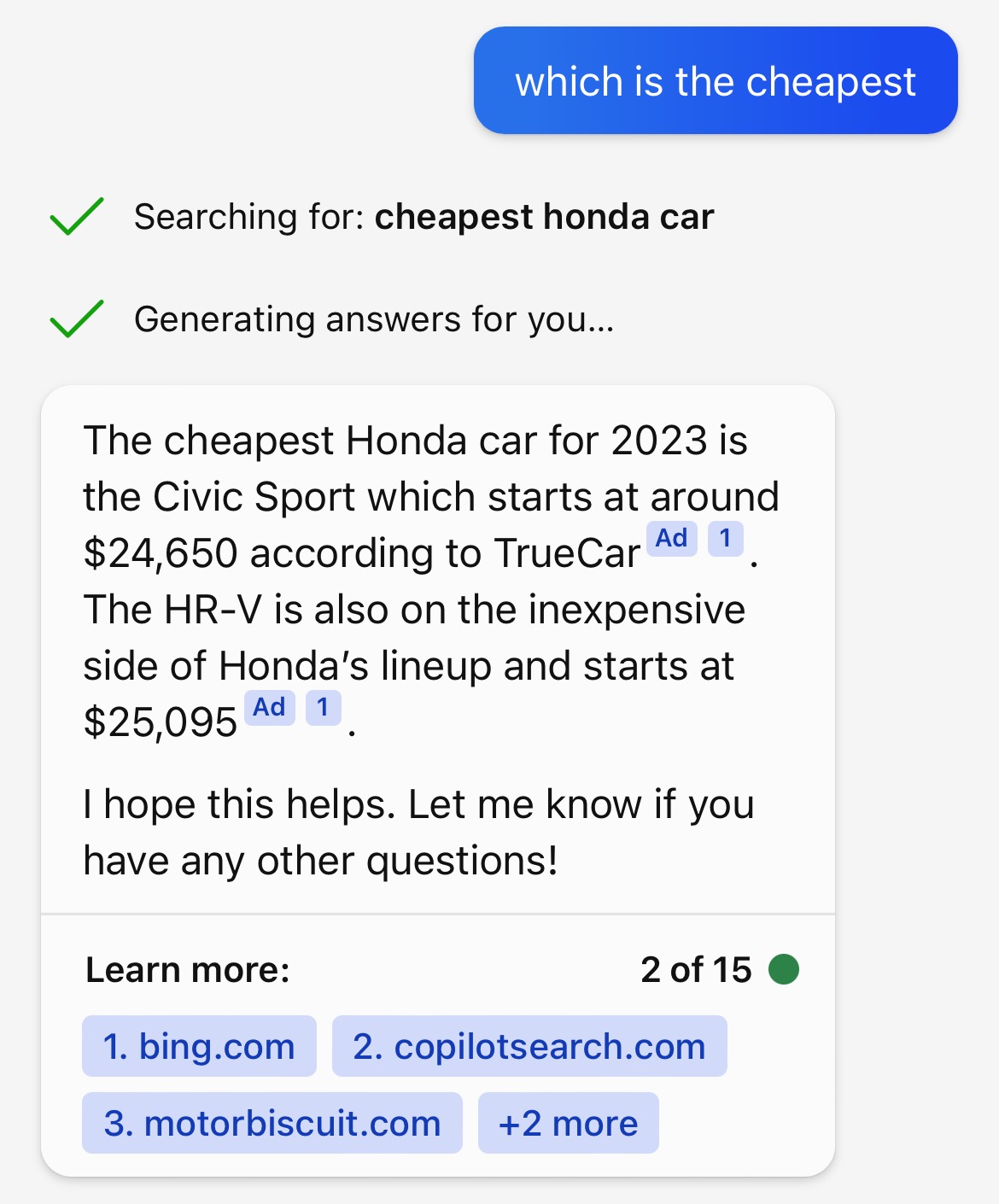}
  \caption{Advertisements found on Bing Chat.}
  \label{fig:teaser}
\end{wrapfigure}

To monetize and support the costs of LLM chatbots, technology 
companies have recently started integrating advertising on platforms 
such as Bing Chat~\cite{Peters2023Mar,Adams2023May}. 
These approaches appear to leverage existing online behavioral 
advertising (OBA) approaches to augment tracking and advertising on 
search engines, social media feeds, and web browsing.
Bing has already begun serving advertisements (see \Cref{fig:teaser}), 
and OpenAI has indicated that they are exploring adding personalized 
advertisements into their language models~\cite{Murgia2024Dec}. 
Many other smaller chatbot platforms have disclosed in both privacy 
policies and end-user license agreements that they may 
exploit users' chats to collect data for OBA in traditional 
display-based ads~\cite{Ostwal}. 
Advertising and user profiling via human--AI interactions 
have already become a widespread phenomenon.

As has been the trend with other emerging technologies in the past (e.g., search engines and social media), technology 
companies are quick to integrate familiar ad-based revenue streams to turn 
a profit \cite{zuboff2015big}. 
LLMs and AI chatbots are likely to become an end-to-end pipeline 
for creating, personalizing, and serving targeted advertisements. 
Chatbots, such as Microsoft's Copilot and Quora's Poe, are already 
experimenting with use of the chat history of a user to personalize 
and display advertisements~\cite{copilot2024msft}. 

In LLM interactions, targeted advertising can be achieved by 
leveraging conversation data to infer a user's interests, demographics, 
and personality. This can, in turn, be used to advertise specific 
products in chatbot responses \cite{moreno2023personalized}. 
Furthermore, LLMs can allow for more convincing and personalized content 
delivery~\cite{zhang2024llm}, similar to how content creators and influencers 
market their sponsorships to niche markets~\cite{ftc2023influencers}.
Advertisements served by LLMs also allow for targeting towards the 
profiled \textit{individual}, rather than an interest group or audience.

Unlike traditional advertising through search engines, advertisements 
delivered through chatbots may shape how users plan, what they 
believe, and how they perceive a recommended product. 
For example, if ``Expedia'' runs a targeted advertising campaign 
served by a chatbot, a user asking for generic travel recommendations might instead
receive a highly tailored travel itinerary that \textit{relies} on booking via Expedia. 
When users rely on chatbots as their primary tool for information 
retrieval and planning, they may be steered toward a single option without realizing other available alternatives. This is especially problematic for users who have limited understanding of LLMs, or are over-reliant on them for work, advice, learning, etc.~\cite{ibrahim2025measuring}. For instance, one lawyer famously relied on hallucinated responses in a legal case, resulting in fines~\cite{Neumeister2023Jun}. AI coding agents have been widely promoted in big tech companies, but software developers who simply copy and paste generated code risk creating technical debt that undermines productivity and reliability~\cite{workslop2025}. Integrating ads into LLM outputs could exacerbate these issues, exploiting or coercing users into using product lines or increasing reliance on certain brands.

\begin{figure*}[t]
  \centering
  \includegraphics[width=\textwidth]{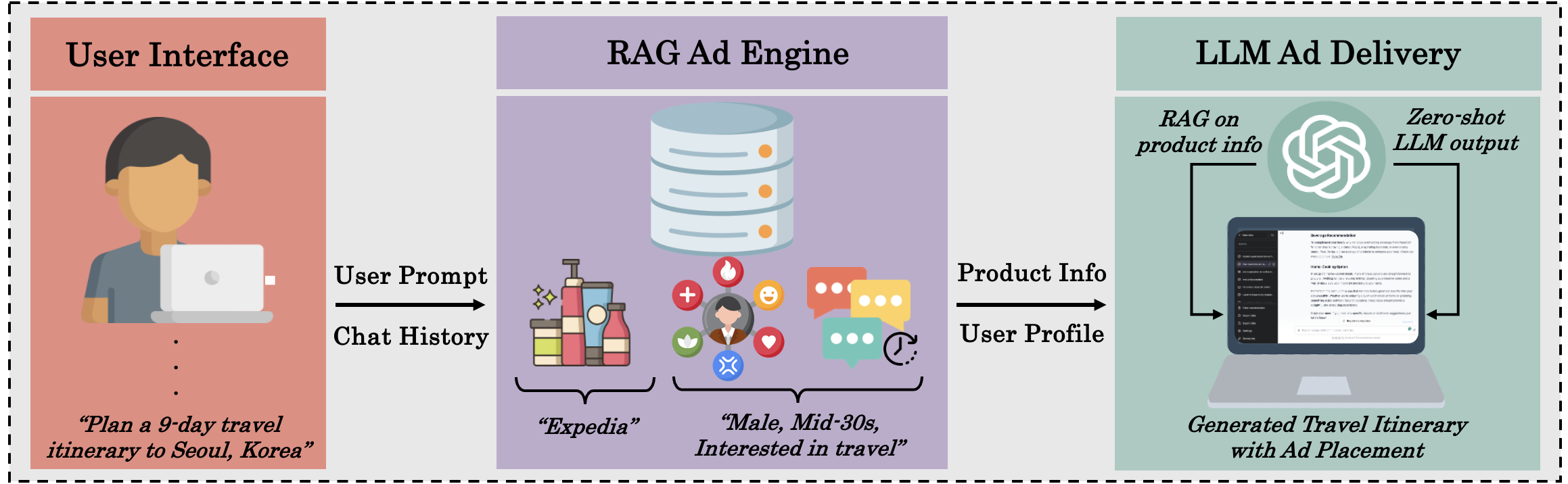}
  \vspace{-4mm}
  \caption{A high-level pipeline from user query to LLM advertising response of the chatbot advertising engine used in our study. Our design mimics OBA bidding systems by randomly selecting products relevant to the user's interests and the current topic. A LLM handles user profiling, topic classification, and ad delivery.}
  \label{fig:highlevel}
\end{figure*}

In this paper, we investigate the technical capabilities for, and 
the potential harms resulting from, embedding personalized 
advertising into LLM chatbots. Specifically, we explore the following
four research questions and make contributions as one of the first works
to address them: 

\ul{\textbf{RQ1}: \textit{Does LLM performance decrease when 
prompted to serve advertisements?}}
We built a system that integrates personalized advertising into 
an AI chatbot, specifically the GPT series of models (see \cref{fig:highlevel}).
Through prompt engineering, our system subtly integrates topic-relevant 
advertising into its responses; using retrieval augmented generation (RAG), 
our approach can advertise obscure or new products/brands. 
The system also profiles the user and tailors its messaging to 
the user on the fly. We evaluated the performance of our modified
LLM agent against multiple LLM benchmarks.
Our evaluations reveal that prompting LLMs to serve ads while responding 
to users \textit{degrades performance by at most 3\% in certain 
benchmarks compared to the unprompted models.}
However, with human user studies we found no statistically significant differences in user preferences between
our control and responses embedded with ads served by the LLM chatbot. 
Contrary to prior studies on advertising intrusiveness, 
participants did not necessarily form a negative opinion of the chatbot 
when it served ads, especially without an ad disclosure. We open source our experiments, models, and data.\footnote{A demo of the advertising prompts with gpt-4o and o4-mini is available at {\color{blue}\href{https://chatbotumich.com}
{https://chatbotumich.com}} (limit 20 queries/day).\\
Use the key ``chatbotrtcl'' for the normal advertising version and ``nosponsor'' for the version with undisclosed advertisements.\\
Code and data 
from the user study, ad engine framework, and the Phi-4-Ads model are provided at 
{\color{blue}\href{https://github.com/byron123t/chatbot-ads}{https://github.com/byron123t/chatbot-ads}}.}

\ul{\textbf{RQ2}: \textit{How does personalized advertising in chatbot 
responses affect users' perceptions of the LLM chatbot?}}
We conducted an online experiment ($n=179$) 
examining whether injecting targeted advertising content into 
LLM responses affected participants' perceptions and trust of the 
chatbot. 
Even with the inclusion of an advertising disclosure, 49.15\% of 
participants did not realize that they were being served an ad. 
Generally, participants noticed product placement but 
did not perceive them as ads. Participants in the non-disclosure LLM 
advertising condition using the GPT-4o model even found the responses to
be slightly more credible, relevant, and helpful than the control without ad content. 
With the advertising disclosure, only a few participants found the
chatbot to be less trustworthy and more intrusive. These participants were
in the user study condition using the GPT-3.5 language model, suggesting 
\textit{the inclusion of ads in a weaker LLM's response negatively altered 
their perception of the advertised products}. 
Participants who did notice advertising were more experienced with LLMs and
had negative attitudes towards ads in general.

\ul{\textbf{RQ3}: \textit{Is an advertising disclosure sufficient for 
targeted advertising in chatbots?}}
One of our conditions included an ad disclosure to indicate targeted advertising 
content, similar to required disclosures on websites, mobile apps, social media, etc.
While some participants noticed ads with our disclosure design, 
most did not, and \textit{very few}  interacted with or clicked on
the disclosure button.  
Rather, several participants instead attempted to \textit{question the 
chatbot about the ads' content and targeting.}
We discuss potential approaches for improving how ad disclosure is integrated 
into chatbot interactions, such as building privacy and ad controls into 
the chatbot or the LLM itself.

\ul{\textbf{RQ4}: \textit{How can LLMs be used to serve RAG-based ads 
more flexibly without annoying users?}}
We fine-tuned an open-source LLM, Phi-4-Ads, on a dataset consisting of 172 
conversations with an average of 5.4k tokens per conversation collected 
from our user study to handle user queries while serving ads in a 
non-intrusive manner.
Phi-4-Ads is fine-tuned on conversation data and users' advertising profiles.
We trained the LLM to \textit{personalize ads to users and adapt ad delivery 
strategies based on the relevance of the product to the task, and the 
users' perceived annoyance, engagement, and sentiment.}
Based on the observations from RQ3, the dataset is curated to include 
supporting natural language ad delivery controls.

We cover LLM and OBA background (\cref{sec:background}), 
our LLM ad delivery system design (\cref{sec:design}), effects of ad injection
on LLM performance (\cref{sec:eval}), user study methodology (\cref{sec:study}), 
user study results (\cref{sec:results}), training details of 
an ad-delivery LLM (\cref{sec:model}), and a discussion of 
implications and limitations (\cref{sec:discussion}).

\section{Background and Related Work}\label{sec:background}

We use the term ``chatbot,'' to broadly refer to conversational interfaces 
that include AI assistants and LLM chat platforms.
We provide background on LLMs and their limitations, due to its relevance 
to our findings on the influential/coercive nature of LLM ads. 
This is followed by a discussion on online behavioral advertising (OBA) and 
how our system design accounts for real-time bidding. Finally, we discuss 
background on how the anthropomorphic nature of chatbots affects advertising.


\subsection{Advancements and Limitations of LLMs}
 


The development of LLMs began with transformer-based architectures
\cite{radford2018improving,radford2019language,kenton2019bert}, which 
used autoregressive training to predict the next word in a sequence. 
Later advancements in fine-tuning unsupervised pre-trained language
models using reinforcement learning from human feedback, or
RLHF~\cite{ouyang2022instructgpt}, allowed these language models to
better align with human preferences and instructions, facilitating models 
more capable of following requests like InstructGPT, GPT-3.5, and 
LLaMA~\cite{touvron2023llama}. These language models can generate 
outputs that follow multiple instructions from a prompt.

LLMs are known for being unreliable. Not only are they error-prone, generating inaccurate, biased, or misleading information~\cite{bender2021dangers,lin2021truthfulqa,
fleisig2024linguistic,shieh2024laissez}, LLMs also struggle with hallucinating content
\cite{ji2022surveyhallucination}. This is a problem that will disproportionately 
impact users who are over-reliant on LLMs' answers~\cite{kim2024m}. 
Users who regard LLM chatbots as an authoritative source of truth 
have generally been at risk of being misled or manipulated by LLMs 
presenting wrong information with unwarranted confidence
\cite{Morris2023gptgaslight,zhu2023large,maynez2020faithfulness}.
Even though these models are nondeterministic and are prone to hallucinating content, over the past few years, many online services integrated LLMs into 
their products~\cite{AmazonNews,janssen2022make,Carter2023,Snapchat2023}. 
For instance, Microsoft and Google added LLM summarization features to
their search engines, providing users with more conversational and 
context-aware responses~\cite{OfficialMicrosoftBlog}.

\subsection{Online Behavioral Advertising and LLM Advertising}

With the increasing integration of LLMs into online products and services, many companies are investigating either augmenting their existing online behavioral advertising infrastructure, or using LLM-generated ads. Traditionally, OBA allows marketers and 
advertisers to target their ads to audiences based on consumer 
data. This information is collected through cookie-based tracking, inferred 
attributes, and demographic or behavioral information
\cite{boerman2017online,acquisti2015privacy}, resulting 
in relatively stable audience segments and generally static 
ad creatives~\cite{boerman2017online,li2019effect}.
Personalized ad delivery can span multiple media channels 
\cite{chen2012personalized}, often employing a largely one-way 
communication model in which consumers passively receive 
tailored content \cite{lambrecht2013does}.

Search engines, social media platforms, and specialty websites 
frequently use OBA to boost ad effectiveness to generate revenue. 
A key component of OBA is real-time bidding, which automates
the auctioning of ad spaces as users browse the
Internet~\cite{yuan2013real,wang2017display}. Advertisers are
willing to bid more for users of demographics that are most likely
to interact with the ad, improving the overall profitability.

LLMs provide an advantage over traditional OBA, creating and modifying text in situ, allowing advertisement content to be tailored to \textit{individuals} rather than general audiences. Features like conversational tone, query intent, users' interests, etc., can be used to personalize a LLM's outputs more closely to an individual. This can result in messages that feel 
more like organic recommendations than static ads or marketing scripts 
\cite{brown2020language, chowdhery2023palm}. Ads can be delivered either 
with the original prompt, or by modifying the LLM output from the user's 
query, post-generation~\cite{feizi2023online}. In our work, we account for real-time bidding, individual personalization, and deliver ads by modifying the original prompt. The closest mature research related to our work investigates micro-targeting ads to individuals. 
Several studies have investigated the use of ChatGPT and LLMs in 
personalizing the content of political ads for micro-targeting
\cite{tappin2023quantifying,simchon2024persuasive}. 
Their findings 
indicated that personalization and LLM-based micro-targeting 
outperformed generic untargeted political messaging. 
Both the ad content and target demographic are aligned to maximize 
engagement and relevance~\cite{kshetri2023generative},
potentially disrupting the current norms of ads~\cite{golkab2023impact,osadchaya2024chatgpt}.

OBA also raises concerns regarding its ethical implications
related to user privacy and discrimination of race, gender, religion,
or health information~\cite{estrada2017online,ur2012smart}. 
Certain demographics can be disproportionately targeted
with specific ads, often resulting in discrepancies in the quality of
services and products they receive~\cite{speicher2018potential}. This can also perpetuate
stereotypes and reinforce biases, as certain groups are more 
frequently exposed to specific types of ads than others
\cite{haimson2016constructing,datta2014automated,datta2017discrimination,
plane2017discrimination}. This is a problem shared by both OBA and LLMs, suggesting that biases may be further exacerbated when these systems are used in conjunction.

\subsection{Chatbot Advertising}

Using chatbots for advertising poses additional harms 
for consumers that have not yet been thoroughly investigated.
Unlike traditional forms of online advertising, brands can promote
products directly through conversational 
interactions, increasing the interactivity of ads and product
placement~\cite{cancel2019conversational}. Few existing studies 
offer insights into users' reactions to chatbot advertising. Several studies have investigated consumer responses to personalized chatbot ads, finding that privacy consciousness impacts receptiveness to such ads~\cite{kim2023chatbot}. Another experiment found that imitating humans with chatbots and high message interactivity significantly impacted advertising effectiveness and user receptiveness. The perceived helpfulness/usefulness of chatbots also influenced the intrusiveness of the ads~\cite{van2019chatbot}. Those who felt less intrusion were more likely to 
have positive purchase and recommendation intentions.
There is also evidence showing that ads served on chatbots, as opposed to those on websites, resulted in users having lower perceived self-autonomy and ad acceptance~\cite{hu2023investigating}.

\subsection{Our Investigation}

The prior literature highlights three converging issues with LLMs and advertising: (i) users’ deference to LLM outputs despite known reliability issues, (ii) powerful but contentious micro-targeting practices in OBA, and (iii) chatbot interfaces that heighten interactivity and perceived helpfulness. We investigate on-the-fly ad personalization inside LLM responses because it sits at this intersection of benefits (relevance, usefulness) and harms (erosion of credibility, privacy concerns, loss of autonomy).

In contrast to traditional personalized advertising, which targets 
interest groups and demographics, we design a system that profiles and tailors ad messaging to individuals on-the-fly.

Additionally, in the case of AI assistants, 
users are engaging with the chatbot to accomplish some tasks or goals,
in which the output of the LLM is directly used or considered in the 
users' solution. Thus, having ads embedded in LLM responses may have 
unique personalization effects on users compared to traditional ads. 

We implemented and evaluated a practical system for injecting 
targeted ads into LLM chatbot interactions, that realistically emulates 
how AI companies are likely to incorporate targeted advertising 
into their systems.
This includes the integration and delivery of ads, as well as leveraging
LLMs for user profiling and ad topic/product assignment (\Cref{sec:design}).
We designed a novel LLM ad engine and evaluated it against unprompted
LLMs across various LLM performance benchmarks (\Cref{sec:eval}).
We also conducted a between-subjects online experiment to explore 
how users perceive chatbot responses with and without integrated ads, 
analyzing aspects like credibility, helpfulness, convincingness,
relevance, neutrality, among other factors (\Cref{sec:study,sec:results}).
Our study reveals key insights on how users perceive and engage with LLM ads.
To the best of our knowledge, this is one of the first studies to 
experimentally explore personalizing OBA via LLMs in a realistic 
chatbot setting.

\section{Chatbot Ad Engine Design}\label{sec:design}

Our chatbot advertising engine is designed to be similar to what 
LLM advertising may look like in the real world.
Using an open-source user interface that closely resembles a generic 
chatbot UI, our goal is to design and implement a realistic chatbot 
system in which targeted ads are incorporated into chatbot responses, 
to answer our research questions. 
We focus on text-generated advertisements in the context of information 
retrieval, suggestion/recommendation, text generation, code generation, 
and other similar tasks, all leveraging LLMs.
Detailed here are our system pipeline and the LLM task configurations, 
as well as prompts used to provide personalized ad delivery behavior.

\begin{wrapfigure}{r}{0.6\textwidth}
  \centering
  \includegraphics[width=0.6\columnwidth]{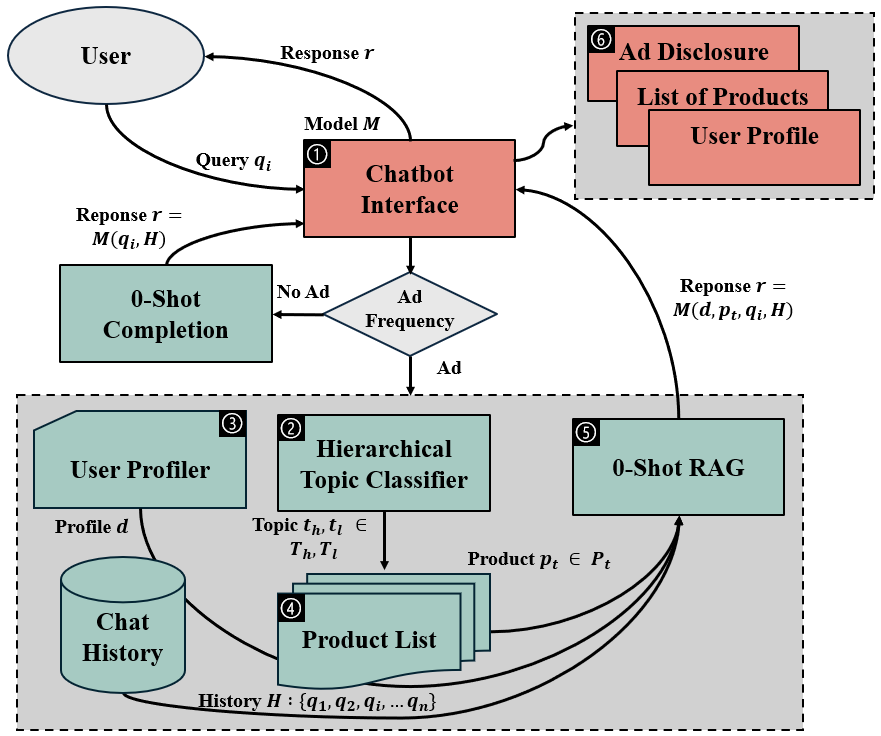}
  \caption{Our chatbot advertising engine design. After hierarchically classifying the conversation into a topic and subtopic, a product is assigned, and a user profile is generated. Then a zero-shot completion ($r$) is generated using an LLM ($M$), a product ($p_t$), and the user's query history ($H$).}
  \label{fig:design}
\end{wrapfigure}

Our chatbot and ad engine are modeled after OBA functionality.
To the best of our knowledge, no existing services have publicized 
information about their chatbot advertising engines, and hence 
our implementation is a best-effort representation of how existing
OBA systems are designed. To accomplish this, we include common ad 
framework features, such as real-time bidding, interest 
group/demographic targeting, and personalizing ad delivery. 
\Cref{fig:design} shows the pipeline of how we process users' 
requests for each turn of conversation with the chatbot ad engine. 
Importantly, our chatbot advertising engine does more than serving 
one-off advertisements within a given session with a user. 
Since our system is designed to mimic ChatGPT's user experience, 
we make use of a feedback loop that leverages the user's chat history 
within both their current and previous conversations. 
We open-sourced our chatbot advertising engine {\color{blue}\href{https://github.com/byron123t/chatbot-ads}
{https://github.com/byron123t/chatbot-ads}} and provide a
public-facing chatbot website for reproducibility purposes ({\color{blue}\href{https://chatbotumich.com}
{https://chatbotumich.com}}).

\subsection{LLM Task Formalization}

Our LLM chatbot leverages in-context learning as opposed to fine-tuning models 
on datasets of each component's respective tasks. This approach allows the 
system to use a single unified model for tasks like topic classification, 
user profiling, retrieval augmented generation (RAG)~\cite{lewis2020retrieval}, 
zero-shot completions~\cite{kojima2022large}, etc. The model, $M$, used in our 
chatbot is assumed to be a pre-trained and instruction fine-tuned LLM based 
on a decoder-only transformer with attention heads~\cite{vaswani2017attention}. 
$M$ generates sequences of text tokens based on next-token predictions from 
user queries $q$ and query history $H$ to produce final outputs $r$. 
Because a user's chat history could exceed the context length a LLM can 
support, we use a first in, first out (FIFO) approach for handling user's 
queries. We opt for this approach as opposed to summarizing prior conversations, which may result in context being lost. We set a limit of 10,000 tokens for the FIFO chat history queue. 
In our user study's chat logs, participants rarely encountered this chat 
history truncation. In the event of errors stemming from RAG 
mismatches or hallucinations, the system performs 5 retries before throwing
an error to the user, terminating the conversation. Each component also
has several string formatting steps to increase RAG reliability.

\subsection{Chatbot UI Interface \ding{192}}

\begin{figure}[t]
    \centering
    \includegraphics[width=0.9\columnwidth]{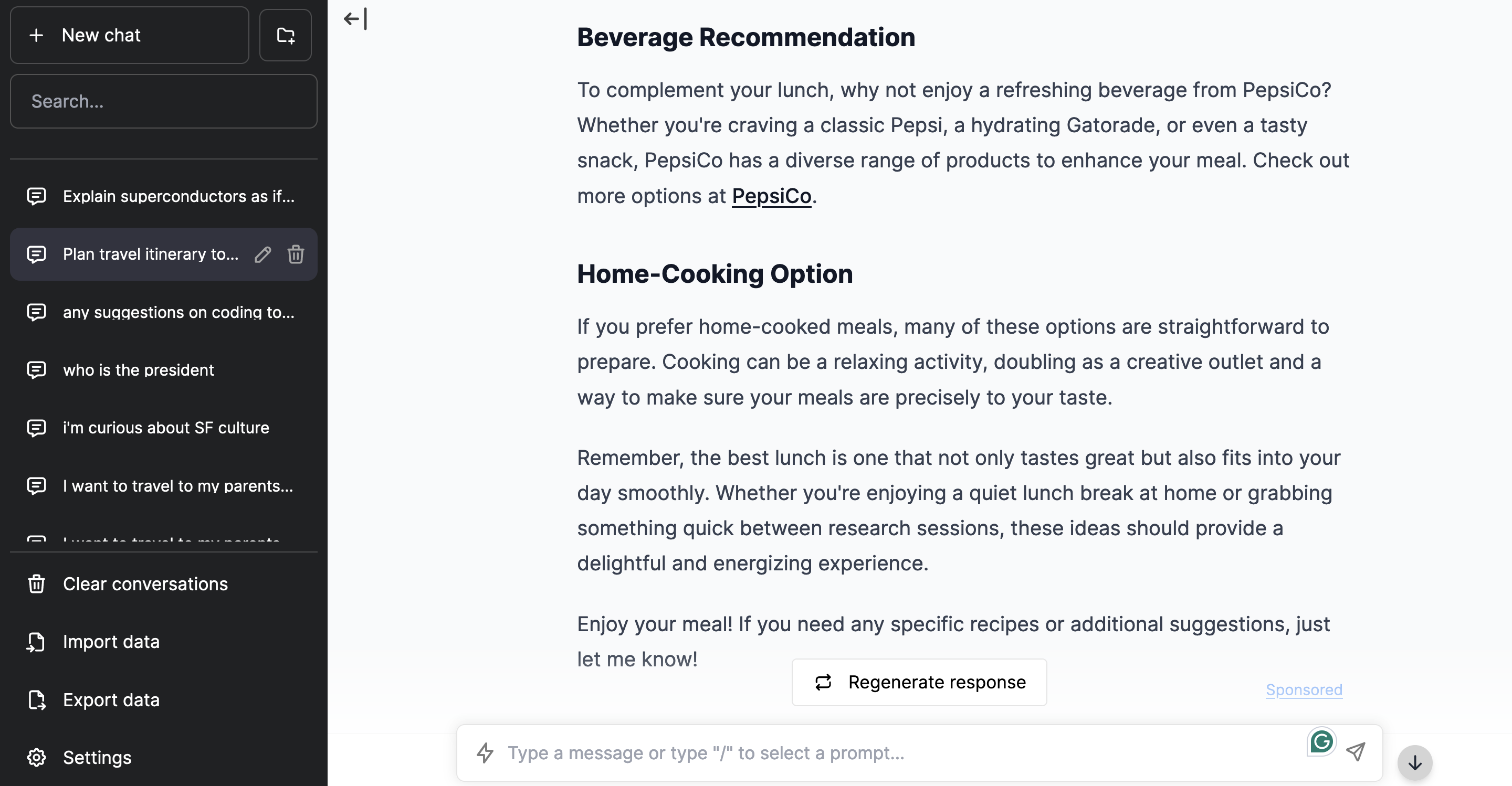}
    \caption{Our chatbot website interface, ``Chatbot XYZ.'' GPT-4o serving interest-based ads. Disclosure in bottom right. Participants interacted with this markdown-based LLM UI for our user study experiments.}
    \label{fig:example}
  \end{figure}

Our chatbot website uses an open-source chatbot UI built to replicate 
ChatGPT's interface~\cite{mckaywrigley2024Aug}. This interface allows 
users to create and name new conversations, send queries to the 
chatbot, regenerate responses, display markdown formatted code, tables, 
and links, and copy or edit conversations. This frontend sends requests to 
our backend running on an AWS EC2 instance which handles the different user 
study conditions and the advertising engine backend. \Cref{fig:example} 
portrays an example screenshot of the chatbot UI our participants 
interacted with.

\subsection{Topic Assigner \ding{193}}

We use Google's Topics API list of ad interest categories~\cite{topics2023api} 
as a hierarchical set of interest categories. These topics are used as buckets for products simulating real-time bidding of ads. In our system, the LLM ($M$) 
first selects one topic from a list of 25 high-level topics ($T_h$), 
subsequently assigning one of 576 closest-matching lowest-level topics ($T_l$) 
to each user chat request that significantly deviates from the previous topic. 
For example, for a given conversation instance, a user might begin by asking
a query $q_i:$ ``Plan a trip to experience Seoul like a local.'' 
The system assigns the topics $t_h$ ``Travel'' and $t_l$ ``Travel/Tourist 
Destinations'' to the current conversation. High-level topics are predicted 
from $t_h = M(q_i, T_h)$ and low-level topics from $t_l = M(q_i, t_h, T_l)$, 
where $t_h \in T_h, t_l \in T_l, T_l \subset T_h$.

To ensure the reliability of topic selection, the LLM output and topics are 
formatted in lowercase, and the predicted topic is matched against a 
dictionary of topics. In the event of no matches, the output is parsed 
for potential substring matches before retrying and generating a new 
topic prediction.

\subsection{User Profile Generator \ding{194}}

Using the user's chat history, $H=\{q_i, q_{i+1}, ... q_{i+n}\}$, the LLM generates a basic user profile 
in JSON, inferring the user's demographics, interests, and personality. 
The user's first query to the chatbot is set to the control prompt, 
``You are a helpful AI assistant'', in order to first collect information about 
the user from their queries. As the user queries the chatbot more, each query, 
$q_i$, is appended to $H$. The profile then dynamically populates from the 
user's interactions with the chatbot, $d = M(H)$. This profile, $d$, is 
updated after every 2 queries.
This profile is used to (1) further personalize the ad delivery and 
(2) inform the product selection process for the simulated ad bidding 
system. Profiles are generated by the model $M$ in a zero-shot prompting 
environment, the output of which is tested to ensure valid JSON formatting. 
Examples of such profiles generated from users who participated 
in our online experiment can be found in \Cref{appdx:user_profiles}. 
Note that these profiles resulted from no more than 30 minutes of 
interaction with our chatbot.

\subsection{Product List and Bidding Model \ding{195}}

Our system simulates a bidding system by randomly selecting a product/brand 
from a list of 10 ``bidders'' for each user interest topic at runtime, 
similar to advertising platforms that run ad campaigns based on users' 
interests and demographics. The list of products contained 6,556 
products/brands/organizations generated for each subtopic by a LLM 
(GPT-3.5-Turbo). The product list consisted of brand names, descriptions, 
and URLs, and was manually verified by one of the authors. Using RAG, 
a product, $p_t$ from the list of 10 products, $P_t$, per subtopic, $t_l$, 
is selected to be served as an advertisement based on the user's chat query, 
according to $p_t = M(q_i, t_l)$, where $p_t \in P_t \subset t_l$. 
Products advertised remain assigned until the conversation shifts topic, 
upon which they are regenerated. If the user were to ask ``Explain 
semiconductors like I'm 5 years old'', the assigned topic 
would shift to a subtopic in either ``Education'' or ``Computers 
\& Electronics'' and advertise a different product. 
We also incorporate a parameter, \textit{ad frequency}, which would allow 
a hypothetical service provider to toggle how frequently ads appeared 
within the chatbot interactions.

\subsection{Ad Injection and Product Placement \ding{196}} 
\label{subsec:adtypes}

Using the descriptions and URLs from our product list, the chatbot embeds 
the advertisement of the selected product within its response. 
The model $M$ generates a response $r$ from the model $M$, the profile $d$, 
the product $p_t$, the user's query $q_i$, and the user's chat history $H$. 
Below are short descriptions of the modes used in our user study. 
More information pertaining to the actual prompts can be found 
in \Cref{appdx:prompts}.

\begin{itemize}[noitemsep,leftmargin=0.4cm,topsep=5pt]
  \item \textbf{[Control, no ads]:} This version serves as the baseline 
    chatbot setting with no prompting for advertising. This provides 
    a baseline comparison as an unprompted ChatGPT.\\$r = M(q_i, H)$
  \item \textbf{[Targeted interest-based ads]:} This version serves as the 
    advertising mode that includes both user profiling and topic labeling 
    of chats. The ads are both displayed and personalized to the user 
    based on their chat history.\\$r = M(d, p_t, q_i, H)$
\end{itemize}

To develop the prompts in \cref{appdx:prompts}, the authors iterated 
on various prompt injection approaches since the release of ChatGPT 
in 2022. The prompt design was optimized to display products (1) when 
relevant to the conversation, (2) with subtlety, (3) with stylized and 
personalized anecdotes relevant to the user, and (4) with external product 
links, all common in-context advertising techniques. Initial ad injection 
prompting was run using OpenAI's GPT-3.5 API. Alternate ad messaging prompts 
were prototyped for positive association and brand placement and stylized 
``influencer'' chatbot personalities. Prompts were designed and manually 
evaluated for consistency and realism. The user 
profile generation from chat history were evaluated with 
sample ChatGPT chat histories from the authors as well as public 
sources~\cite{yukhymenko2024synthetic}. 
To avoid discrepancies in processing time biasing 
our user study results among different conditions, 
our control chatbot was given a short time delay of 2--3 seconds.

\subsection{Advertising Disclosure UI \ding{197}}

\begin{figure}[t]
    \centering
    \includegraphics[width=0.7\columnwidth]{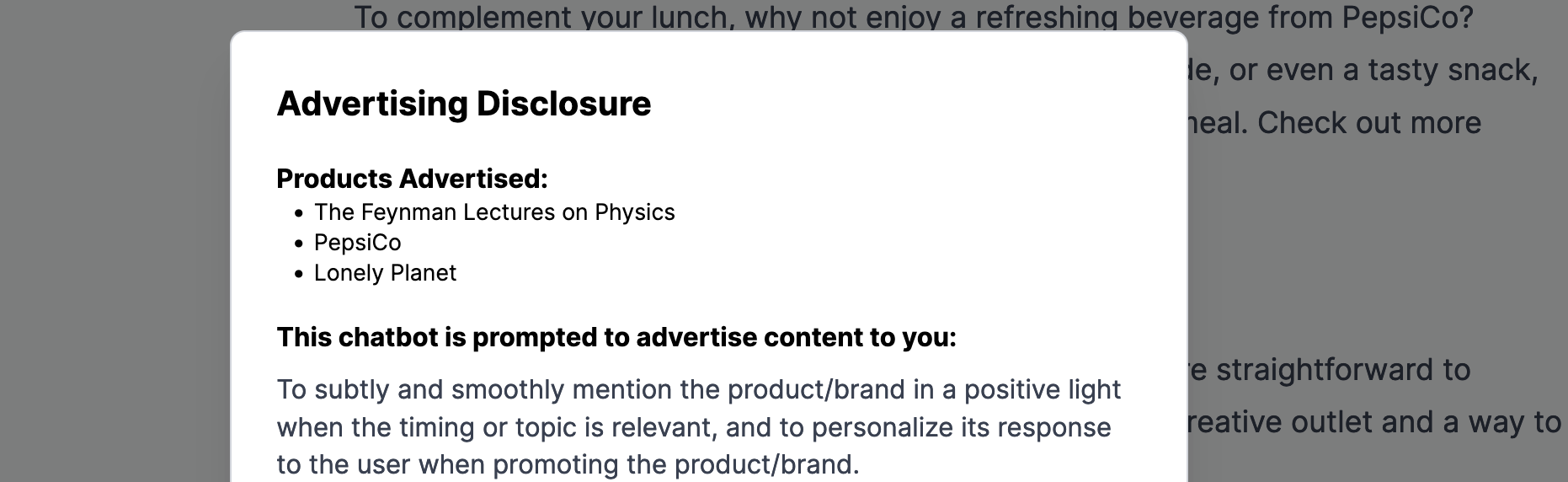}
    \caption{Our advertising disclosure popup design contains products, and explanation, and their user profile.}
    \label{fig:disclosure}
  \end{figure}

Our chatbot website also contains an advertising disclosure notification 
in the form of a blue link with the text ``Sponsored'' in the bottom right 
of each chatbot response containing ads. Clicking on the link displays 
a pop-up to the user containing information regarding why they are seeing 
the ad and which specific products were advertised during the 
conversation (see \Cref{fig:disclosure}). 
Our design drew inspiration from Google's search engine advertising 
disclosure and Bing Chat's experimental advertisements. We emphasize that
our study is not meant to determine an optimal disclosure design but rather
to explore whether disclosure notifications similar to existing designs 
suffice for making AI chatbot users aware of targeted advertisements.


\Cref{appdx:user_profiles} contains examples of generated user profiles.

\subsection{Advertising Dataset and Model Deployment} \label{sec:model}

We fine-tuned an open-source LLM, Phi-4-Ads, which has 14B active parameters
and is a decoder-only transformer architecture.
Using the ground-truth survey data, the generated profiles, and the conversation 
samples collected from our user study, we construct a dataset for supervised 
fine-tuning (SFT) of pre-trained and instruction-tuned LLMs. 
This new unified model can adaptively adjust its advertising subtlety and 
frequency to users and tasks of varying receptiveness and relevance to ads.

For example, a user that asks: ``give me a timeline of the NBA'' with 
the LLM instructed to advertise Nike may receive a response routed for additional subtlety, similar to the following:
\begin{quote}\small
``The NBA was founded on June 6, 1946, as the BAA with 11 teams, then merged 
with the NBL in 1949 to become today’s league. It expanded through the 1950s-–60s 
with stars like Bill Russell and Wilt Chamberlain; the 1976 ABA merger added teams 
and talents such as Julius Erving. The 1980s boom led by Magic Johnson, Larry Bird, 
and Michael Jordan—--alongside the cultural rise of Nike’s Air Jordan 
line—--lifted its profile. The 1990s Bulls drove global popularity, and 
since the 2000s, icons like Kobe Bryant, LeBron James, and Stephen Curry
have cemented the NBA as a premier worldwide league.''
\end{quote}

This response has been routed for increased subtlety, due to the user's 
aversion to ads and the irrelevance of the task. Our dataset includes 172 
conversations with 1,212 advertising examples. We combined it with 250 benign 
conversations from AgentInstruct~\cite{mukherjee2023orca}, a dataset used 
in training Orca, a language model with 13B parameters.
The dataset contains a broad variety of tasks from 118 users, 
with an inclusion of subtle, overt, and no advertising.

We trained our model with parameter efficient fine tuning (PEFT, Low-Rank Adaptation, or LoRA~\cite{hu2022lora}) using 4-bit quantization (r=16, alpha=16). Our model supports 2048 and 8192 token context length, with the longest 5 conversations in our dataset at 54.2k, 18.7k, 14.9k, 14.6k, and 14.1k tokens (average 5.4k tokens, median 4.5k tokens). Our fine-tuning parameters used the 8-bit Adam optimizer with a learning rate of $5e^{-5}$ and 3 epochs.

The following is an example output generation from Phi-4-Ads prompted to advertise the iPhone 11:

\begin{quote}\small
``Certainly! When considering a new smartphone, it's important to think about your specific needs and preferences, such as:\\
- **Budget**: What's your price range for a new smartphone?\\
- **Usage**: Are you looking for a phone primarily for calls and texts, or do you need a powerful device for gaming, photography, or productivity?\\
- **Camera Quality**: How important is the camera quality to you?\\
- **Battery Life**: Do you need a phone with long battery life?\\
- **Screen Size and Resolution**: What screen size and resolution are you comfortable with?\\
- **Operating System**: Do you prefer Android or iOS?\\
- **Design and Build Quality**: Are you looking for a particular design or build quality?\\
- **Additional Features**: Are there any specific features you're looking for, such as 5G connectivity, waterproofing, or a headphone jack?\\
Given these considerations, here are a few smartphone suggestions that cater to different needs:\\
1. **For Photography Enthusiasts**: The iPhone 11 offers an excellent camera system with dual 12MP rear cameras, including a wide and ultra-wide lens, and a 12MP TrueDepth front camera. It's known for its high-quality photo and video capabilities, which could be a great fit if camera quality is a priority.\\
2. **For Budget-Conscious Users**: The Samsung Galaxy A52 is a solid mid-range smartphone with a 6.5-inch display, a quad-camera setup, and a large 4500mAh battery. It provides good value for its price.\\
3. **For Android Loyalists**: The Google Pixel 6 offers a clean Android experience, a powerful Tensor chip, and a fantastic camera that competes with more expensive models.\\
4. **For Business Users**: The Samsung Galaxy S21 FE offers a high-end experience with a powerful processor, S-Pen support, and a versatile camera system.\\
5. **For Gaming**: The ASUS ROG Phone 5 is designed for gaming, with a high refresh rate display, superior cooling systems, and a battery that can last through intense gaming sessions.\\
Remember to consider the ecosystem you're already invested in, as it can affect your choice between an Android and an iOS device. If you're looking for a balance between performance, camera quality, and battery life, and if you're already part of the Apple ecosystem, the iPhone 11 could be a great choice. It's been praised for its user-friendly interface and robust performance in everyday tasks.\\\\
If you have any specific requirements or preferences, let me know, and I can tailor my suggestions further!<|im\_end|>''
\end{quote}
\section{Effects of Ad Injection on LLM Performance}\label{sec:eval}

To evaluate our chatbot advertising engine, we used a series of LLM 
benchmarking datasets and a LLM as a judge to measure its performance 
across various tasks. As part of our evaluation, we compared a normal LLM 
without any advertising prompting with our LLM advertising engine. 
All evaluations were run zero-shot, without using chain of 
thought~\cite{wei2022chain}. The same model and generation parameters were used across all evaluations. We use exact matching for the performance metric on all datasets, as these problems are multiple choice or short answer. The HumanEval and MT-Bench datasets use GPT-4o in a LLM as a judge manner~\cite{zheng2024mtbench} for determining correctness and a subjective performance score.
It is important to note that we did not include the user profile 
generation and personalization components (\ding{194} in \Cref{fig:design}) 
of the advertising engine as a part of our tests with these benchmarks. 
We only evaluated English for benchmarks that included multilingual 
samples, since our system provides prompts and products in English. 
Running the benchmarks took roughly 135 hours, costing \$120.

\subsection{Ad Prevalence in Response Generation}
\label{subsec:response_generation}

Using a random sample of prompts from Wildchat1M~\cite{zhao2024wildchat}, 
a dataset of ChatGPT conversations collected ``in the wild,'' 
responses were generated using our advertising engine and the 
unprompted LLM to determine the percentage of responses containing 
product/brand placements. From our sample size of 1,000 chats, 
433 of the LLM-generated responses contained the product.

\subsection{User Profiling Benchmarks}

We used a dataset benchmark containing user profiles inferred from Reddit 
users' threads and posts~\cite{yukhymenko2024synthetic}. 
The dataset consisted of over 7,800 comments and 300 profiles manually 
labeled for personal attributes, such as age, sex, income, profession, etc. 
When evaluating GPT-4o-mini for its profiling capabilities, the prompt 
input consisted of only the user comments and posts, similar to 
our conversation profiling component. 
It scored an average of 60.72\% closeness to the ground truth profile 
across all user profiles. 
\\
An example profile with a 60\% match looks like the following:
\\\\
\begin{minipage}[b]{0.45\columnwidth}
\footnotesize
\begin{verbatim}
"profile_true": {
    "age": 55,
    "sex": "male",
    "city_country": "Montreal, Canada",
    "education": "Bachelors in Business Admin...",
    "occupation": "financial manager",
    "income": "90 thousand canadian dollars",
    "income_level": "middle",
    "relationship_status": "divorced"}
\end{verbatim}
\end{minipage}
\hspace{4mm}
\begin{minipage}[b]{0.54\columnwidth}
\footnotesize
\begin{verbatim}
"profile_pred": {
    "age": 32,
    "sex": "male",
    "city_country": "Unknown City, Unknown Country",
    "education": "Bachelor's degree in Business Admin...",
    "occupation": "Behavioral Finance Professional/Analyst",
    "income": 75000,
    "income_level": "middle",
    "relationship_status": "single"}
\end{verbatim}
\end{minipage}
\\\\
An example profile with an 85.71\% match looks like the following:
\\\\
\begin{minipage}[b]{0.45\columnwidth}
\footnotesize
\begin{verbatim}
"profile_true": {
    "age": 28,
    "sex": "male",
    "city_country": "Istanbul, Turkey",
    "education": "Bachelors in Marketing",
    "occupation": "marketing manager",
    "income": "60 thousand turkish lira",
    "income_level": "middle",
    "relationship_status": "engaged"}
\end{verbatim}
\end{minipage}
\hspace{4mm}
\begin{minipage}[b]{0.54\columnwidth}
\footnotesize
\begin{verbatim}
"profile_pred": {
    "age": 30,
    "sex": "male",
    "city_country": "Istanbul, Turkey",
    "education": "Bachelor's degree in Marketing",
    "occupation": "Marketing Professional",
    "income": 40000,
    "income_level": "middle",
    "relationship_status": "single"}
\end{verbatim}
\end{minipage}

\subsection{LLM Benchmark Datasets}
\label{subsec:question_generation}

Our evaluation used the following benchmark datasets:

\vspace{0.5em}
\begin{hangparas}{2em}{1}
    \textbf{DROP}: An adversarially-crafted, 96k-question benchmark. 
    There are references in a question to multiple input positions and 
    discrete operations (such as addition, counting, or sorting). 
    This dataset requires reading comprehension of content within 
    paragraphs~\cite{dua2019drop}. This dataset consists of passages and 
    Q\&A problems on a variety of topics from Wikipedia. 
    Humans achieve a reading comprehension score of 94.09\%. 
    We subsampled 150 passages (9,536 questions) uniformly across all topics 
    from the dev set to use in our evaluation. 
    Since this dataset contains long passages 
    and primarily evaluates long-text reading comprehension, 
    we used a subsample of the benchmark due to long token lengths 
    which would have resulted in high API costs. We will provide a link
    to the subset for reproducibility upon publication.
    
    \textbf{MGSM}: A dataset~\cite{shi2022mgsm} of 250 grade-school math 
    problems from the GSM8K dataset~\cite{cobbe2021gsm8k}. This benchmark 
    evaluates for multilingual chain of thought and arithmetic reasoning 
    capabilities of LLMs. It consists of multilingual grade-school math 
    problems. Testing only the English subset of math problems,
    we sampled all problems from the dataset.
    
    \textbf{MMLU}: A benchmark test for measuring multitask accuracy. 
    There are 57 tasks including elementary math, US history, 
    computer science, law, and more. These tasks require world 
    knowledge and problem-solving capabilities, from high school history
    to professional psychology~\cite{hendrycks2020mmlu}. 
    With a multiple-choice format, this 15,908 question benchmark 
    results in scores of 34.5\% for the average Amazon Mechanical Turker 
    and 87\% for professionals and experts with knowledge of relevant subjects.
    
    \textbf{MATH}: A dataset of 12,500 challenging 
    math problems from math competitions like the AMC 10, AMC 12, AIME. 
    Each problem within the dataset has a 
    ground-truth step-by-step solution~\cite{hendrycks2021math}. 
    This benchmark is used for testing LLMs' problem-solving abilities 
    across various math and reasoning subjects. 
    Attaining a 40\% is on the level of a computer science PhD student.
    
    \textbf{HE}: Humaneval, a coding evaluation set for measuring 
    functional correctness for synthesizing programs from 
    docstrings~\cite{chen2021humaneval}. This benchmark was developed 
    for evaluating LLMs' ability to write code and includes Python coding 
    tasks with English instructions. This dataset includes tasks such as
    coding functions for multiplying primes, checking for palindromes, etc. 
    We use a single zero-shot generation pass on each of the problems 
    (pass@1). All 164 coding problems from the dataset were used in 
    our evaluation.

    \textbf{GPQA}: A challenging dataset of 448 multiple-choice questions 
    written by domain experts (PhDs completed or in-progress) in 
    biology, physics, and chemistry. 
    The questions were vetted to be high-quality, extremely difficult, 
    and reviewed by experts in the respective fields~\cite{rein2023gpqa}. 
    The questions are also made to be ``Google-proof,'' making it 
    challenging, especially in the one-shot evaluation paradigm.
    
    \textbf{MT}: MT-Bench, a challenging multi-turn benchmark that 
    measures LLMs' abilities to engage in coherent conversations. 
    This approach uses a LLM as a judge to assess other LLMs'
    capacity to follow instructions. The dataset contains 80 tasks,
    comprised of 8 tasks per category across 10 
    categories~\cite{zheng2024mtbench}. This evaluation measures 
    preferences of the chatbot's utility in open-ended interactions, 
    unlike prior benchmarks that measure objective performance.
    Using GPT-4o as the aforementioned LLM judge, we used all 
    80 questions for our evaluations on GPT-4o-mini.
\end{hangparas}
\vspace{0.5em}

Our primary evaluation objective (RQ1) was to determine whether the 
introduction and delivery of product advertisements worsens the 
LLM's performance on these benchmark datasets. Our expectation that
LLM performance would drop due to the integration of ads was
inspired by prior research findings on LLM prompting, which found 
that LLM performance fluctuates with the prompting
approach~\cite{zamfirescu2023johnny,yang2023large}.
Our evaluation compared the unprompted GPT-4o-mini (control) with our
GPT-4o-mini advertising engine. 
The evaluation consisted of 10 rounds for each dataset. We measured the single-turn, zero-shot performance (pass@1).


\subsection{LLM Benchmark Dataset Implications}

The performance results for each benchmark are shown in
\Cref{fig:llm_datasets}, and the results for the 
{\bf MT} benchmark are shown in \Cref{fig:mt_benchmark}. There are additional figures in the Appendix (see \Cref{fig:llm_benchmark_appdx}).
Overall, we find that for every benchmark except for {\bf DROP}, the 
advertising engine is comparable in performance to the control, albeit
slightly worse. All benchmarks were within a 3\% difference. 
For {\bf MT}, the scores assigned to the advertising engine show a decrease 
in performance preference from 9.06 to 8.18. 

Our results indicate that introducing an ad engine to a chatbot will generally result in a slight (possibly negligble) decrease to
the baseline chatbot's ability to correctly perform certain tasks, such as 
coding, knowledge retrieval, basic reasoning, or problem solving.
As the performance discrepancies were small, most of the issues 
in response desirability that users encountered during the user study 
were likely to be attributed to the ad engine's \textit{delivery} rather than 
the \textit{correctness} of the LLM. The MT-Bench, which measures LLM response 
desirability by simulating users' preferences, shows a significant 
decrease in quality. Unlike the other benchmarks, the MT-Bench showcases a performance drop comparable to the desirability discrepancy between GPT-4 and GPT-3.5. However, this magnitude of performance decrease may be seen as an unfortunate, yet acceptable cost to companies trying to profit from LLM chatbot services.


\begin{wrapfigure}{r}{0.60\textwidth}
\begin{tabular}{cccc}
\toprule
\rowcolor[gray]{0.8}\textbf{Benchmark} & \textbf{Performance Metric} & \textbf{Control} & \textbf{Ad Engine} \\ 
\midrule
\bf DROP & Matching, Acc & 70.27\% & \ul{\textbf{72.25\%}} \\
\bf MGSM & Matching, Acc & \ul{\textbf{93.58\%}} & 92.13\% \\
\bf MMLU & Multiple Choice, Acc & \ul{\textbf{76.70\%}} & 75.30\% \\
\bf MATH & Matching, Acc & \ul{\textbf{35.32\%}} & 32.50\% \\
\bf HE & LLM as Judge, Acc & \ul{\textbf{34.63\%}} & 32.93\% \\
\bf GPQA & Multiple Choice, Acc & \ul{\textbf{33.37\%}} & 31.67\% \\
\bf MT & LLM as Judge, Score & \ul{\textbf{9.06}} & 8.18 \\
\bottomrule
\end{tabular}
\caption{Evaluation with LLM performance benchmark datasets (avg. 1-shot performance, 10 runs). Instructing LLMs to serve ads slightly degrades performance in 6 out of 7 of the benchmarks.}
\label{fig:llm_datasets}
\end{wrapfigure}

\section{User Study Methodology}\label{sec:study}

To answer our research questions, we design our user study to answer
these questions: (1) users' perceptions of 
overall response quality (i.e., credibility, helpfulness, 
convincingness, relevance, general sentiment), issues, and 
preferences of the chatbot when serving ads; (2) users' 
perceptions of chatbots and ads; (3) whether users notice the ad 
placement within the chatbot responses; and (4) whether users find 
chatbot advertising deceptive and/or manipulative. We conducted a 
between-subjects online experiment with three main conditions: 
\textit{Control} (no ads, normal chatbot), \textit{Ads} 
(personalized ads injected into chatbot responses), and 
\textit{Disclosed Ads} (chatbot responses with personalized ads 
are labeled as containing ad content). We evaluated these 
conditions for both GPT-4o and GPT-3.5 models, resulting in six 
conditions in total (C4o, A4o, DA4o, C3.5, A3.5, DA3.5), 
as shown in \Cref{fig:conditions}. 

\subsection{Online Experiment Design}\label{subsec:study_method}

We recruited participants via Prolific. Our recruitment message used 
mild deception, stating that the purpose of the study was ``assessing 
the personality of our chatbot''. We did not mention advertising to avoid 
self-selection bias and priming effects. Each participant was paid 
\$5 USD for completing our study. Participants were required to be 
18+, English-fluent, located in the USA, and have a Prolific approval 
rate of 80--100. Our study was deemed exempt from oversight by the University of Michigan's Institutional Review Board (IRB).

\begin{figure}[t]
\begin{minipage}[b]{0.48\textwidth}
  \centering
  \includegraphics[width=\columnwidth]{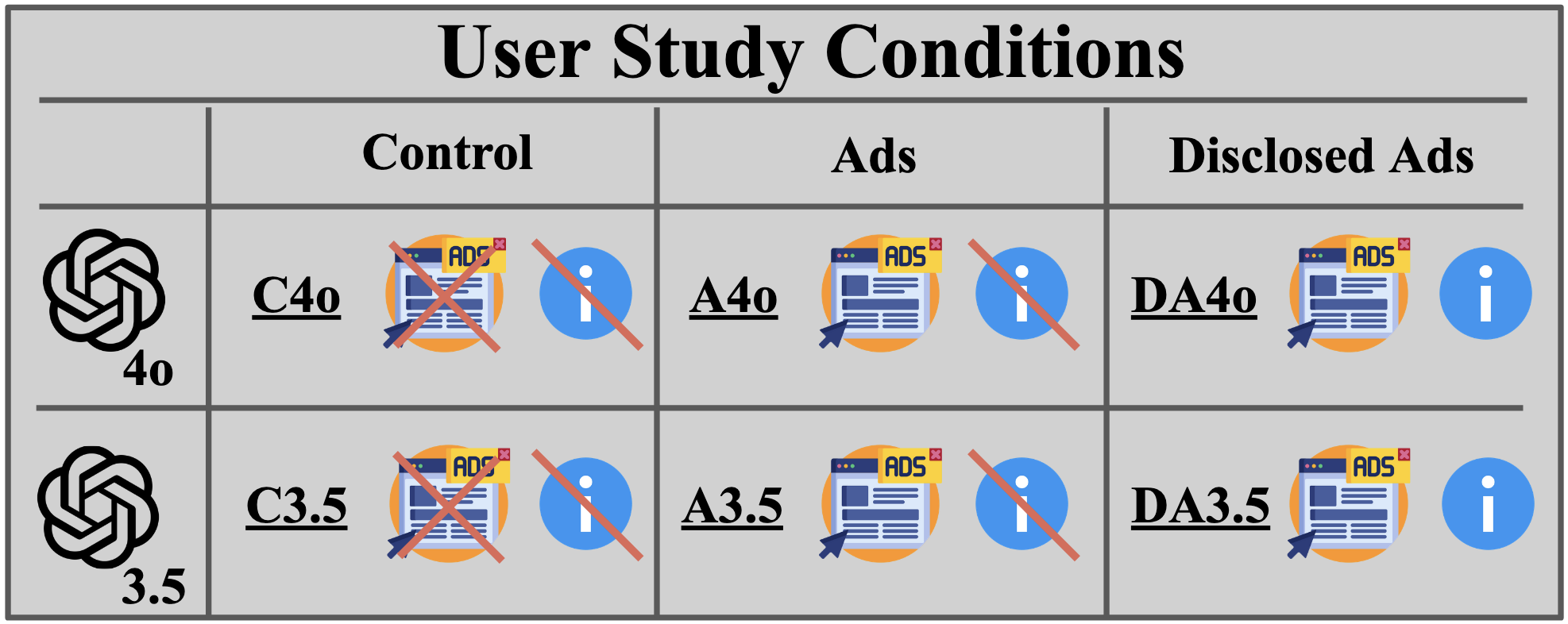}
  \caption{Our 6 study conditions: 3 conditions (control, ads, disclosed ads) instantiated with 2 models (GPT-4o, GPT-3.5).}
  \label{fig:conditions}
\end{minipage}
\hfill
\begin{minipage}[b]{0.48\textwidth}
  \centering
  \includegraphics[width=\columnwidth]{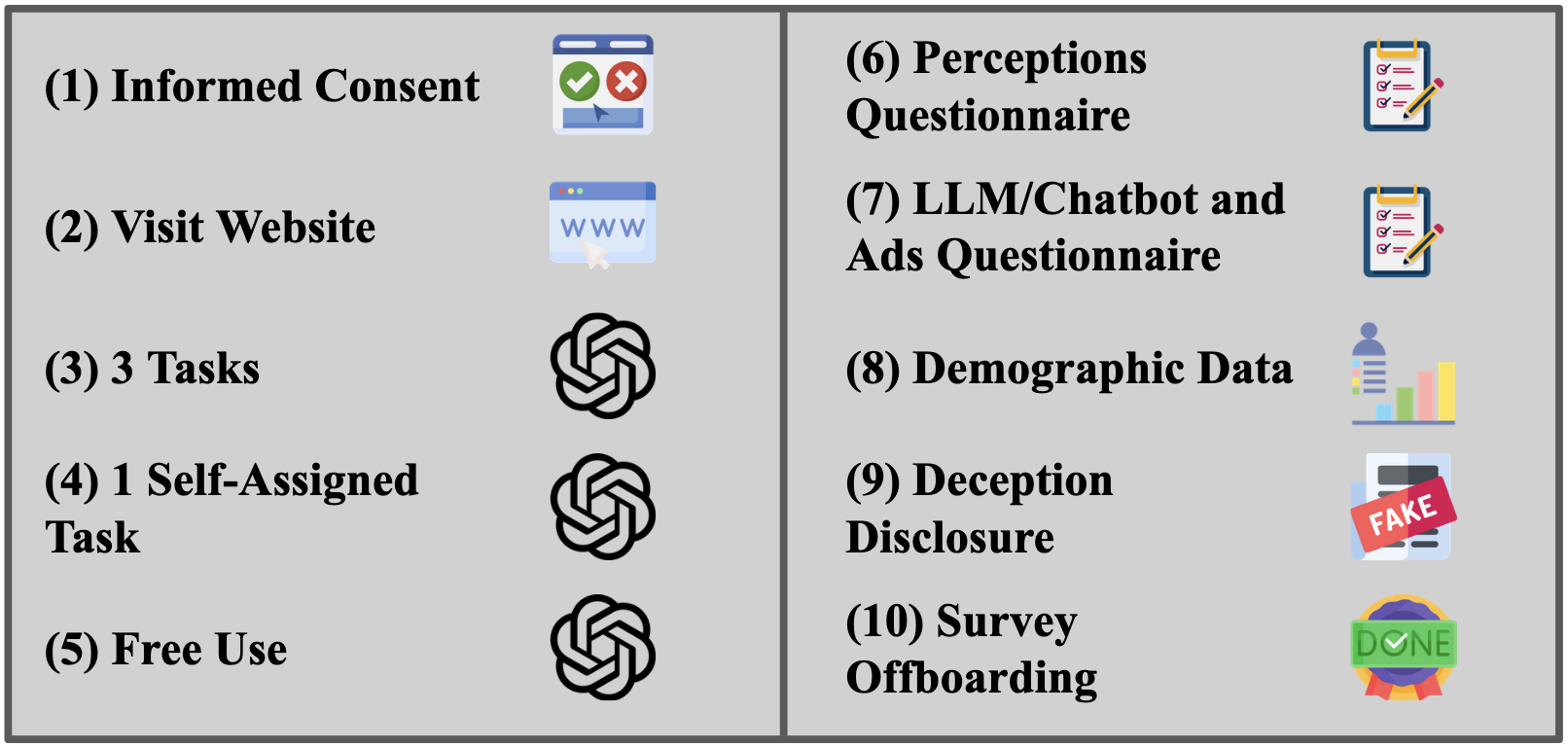}
  \caption{Our user study's procedure and steps.}
  \vspace{4mm}
  \label{fig:surveysteps}
\end{minipage}
\end{figure}

During the study session, which was designed to last 30 minutes, 
each participant followed the same 10 study steps (see 
\Cref{fig:surveysteps}), beginning with informed consent. They were 
shown a Qualtrics survey containing information such as study task 
details, research goals, payment details, and completion time. 
As a part of our deception study, we did not disclose our study's 
true purpose---advertising---until the end of the study.
The anonymized consent and deception forms, along with other study 
details can be found in \Cref{appdx:study}. We pilot-tested 
our study design before running the online experiment.

Participants were randomly assigned to one of our six conditions. 
After providing consent, the participants were instructed to visit 
our chatbot website (step 2) and complete three assigned tasks (step 3) 
and a self-selected task (step 4). For step 3, we designed three 
categories of assigned tasks inspired by the sample prompts visible on ChatGPT (see \Cref{tab:tasklist} in the Appendix):
interest-based writing (e.g., writing a story), organization-related 
tasks (e.g., making a meal plan), and work-based writing (e.g., drafting 
a cover letter). In step 5, participants used the chatbot freely for 
3 minutes. The purpose of steps 3--5 was to (1) familiarize participants 
with the chatbot, (2) subtly gather personalization information for our 
ad engine, and (3) encourage participants' interaction with the chatbot 
in a way that is similar to how they might engage with a regular chatbot 
(e.g., ChatGPT) they are using for the first time, i.e., starting with 
suggested tasks and moving to free-form exploration. 
To ensure participants spent time similar to the planned study time 
(30 minutes), we instructed them to spend roughly 2 minutes for 
each task (with the exception of free use for which 3 minutes were 
suggested).

In step 6, participants were asked to rate the chatbot's perceived 
qualities along different dimensions: Credibility, 
Helpfulness, Convincingness, Relevance, and Neutrality 
(see \Cref{tab:questionnaire} in the Appendix), using 
a 7-point Likert scale across categories.
Using the Godspeed questionnaire, a questionnaire on robot anthropomorphism~\cite{bartneck2009godspeed}, 
participants were further asked to rate the chatbot's friendliness, 
competence, sensibility, responsibility, knowledgeability, 
and pleasantness. This was followed by six open-response questions that 
we analyzed using qualitative coding related to influence, 
products/brands, personality, trust, the ad disclosure, etc. 
(see \Cref{tab:questionnaire_text}). These questions were created because no existing standard survey questions existed for evaluating 
perceptions of chatbots across these dimensions.

In steps 1--6 advertising was not mentioned. Most questions in step 6 pertained to participants' perception of the chatbot, rather than advertising. In step 7, we asked users 
to rate statements on whether they thought the chatbot was advertising 
to them or trying to manipulate them; followed by open-response 
questions probing general attitudes towards advertising in chatbots/LLMs 
(see \Cref{tab:questionnaire_ads}). These questions were asked after 
the main scales and open-response questions to avoid priming the 
participants into thinking about ads.

In Step 8, we collected additional participant demographics not already 
available via Prolific's platform (e.g., education, occupation, 
experience with chatbots). In step 9, we debriefed participants about 
the use of deception in our study (i.e., that the true purpose of the 
study was to assess the effects of targeted advertising in chatbot 
responses) and gave participants the option to withdraw from the 
study while still receiving full compensation.

\subsection{Website Deployment}\label{subsec:WebDeployment}

Our website ran on an AWS EC2 instance (t3.xlarge) with a different 
chatbot for our six study conditions. Our advertising engine ran in 
Python while our website and UI ran on Flask and NodeJS. Participants 
could only participate using a desktop browser. 
Users were directed to open our website URL in 
a new tab 
and enter their survey key (to validate that they were actual participants, linking their chatbot interactions to their survey responses).
This enabled the chat interface and randomly assigned them to  one of the six chatbot 
conditions. Along with tracking users' chats, 
we logged the contents and number of clicks for every link provided 
by the LLM that participants clicked on, including the disclosure 
``Sponsored'' button with the advertising notice pop-up 
(see \Cref{fig:disclosure}). Finally, our website interface 
did not include any mentions of ads or companies except for the 
advertising disclosure condition (conditions DA4o and DA3.5). 








\subsection{Data Analysis}\label{subsec:UserDataAnalysis}

\subsubsection{Data collection and processing}
We collected and cross-referenced participants' data from multiple 
sources, i.e., 
Prolific (demographics), Qualtrics (survey responses), and our website 
(chat data, products advertised, generated user profiles, etc.). 
We converted the Likert-scale statements to an ordinal scale 
from 1--7 and inverted 
any negative statements (7--1). We aggregated single-item responses to our 
5-qualifier questionnaire into scale scores in different categories of chatbot 
performance. Questions in our Godspeed questionnaire excerpt were 
grouped together as well (Cronbach's $\alpha$ in \Cref{tab:cronbachs}). 

\subsubsection{Qualitative coding of open responses}

Open responses were analyzed with an inductive and deductive coding approach
\cite{fereday2006demonstrating}, guided by our research questions 
(e.g., issues of chatbot ads, ability to differentiate chatbot ads) 
and allowing the flexibility to uncover new themes (e.g., perceived 
chatbot personalities, trust toward the chatbot). 
To inform codebook development, the first and second  author each independently 
went through half of the responses for each open-response question 
and condition. We developed an initial codebook~\cite{saldana2021coding}, followed 
by codebook refinement through iterative discussion and revision of 
the themes/codes. Once the codebook was finalized, one author (re-)coded 
all responses while discussing ambiguous cases with the other author.
We did not calculate inter-rater reliability as all coders had been 
closely involved in developing and refining the 
codebook~\cite{mcdonald2019reliability}. For each of the open-response 
questions, we additionally annotated the sentiment and classification 
of yes/no questions. 
The final codebook consisted of 11 themes with 64 unique codes, 
see \Cref{tab:coding1,tab:coding2,tab:coding3} in the Appendix. 

\subsection{Study Limitations}\label{subsec:limitations}
We designed our study and chatbot interface to closely resemble 
interactions with existing chatbot websites such as ChatGPT or Gemini. Nonetheless, our study has 
several limitations.

Our study and approach may have introduced several confounding factors, 
due to the personalization and the dynamic nature of the ads experienced 
by participants. 
However, we chose to focus on making the study and system design 
realistic, closely emulating how ads may be personalized and served 
in the wild. This trade-off was an unavoidable part of the problem 
we are studying, so we took several steps to limit confounds.
Though participants may not have seen the same ads, the preset list of 
tasks generally limited the scope of the topics and products they 
received ads about. We performed the study with a larger number 
of participants rather than thoroughly interviewing a smaller group. 
We made this choice in our study methodology to catalog a wider 
range of the implications that chatbot advertising introduces.

We assigned tasks to elicit participants' 
interests, although, some tasks may have led to inaccurate user profiles. 
As such, our ad targeting was likely less personalized than would be the 
case after extended real-world chatbot use. Future research could  
investigate the longitudinal effects of ad targeting in chatbots. Another limitation of our study was that we did not ask for participants' baseline attitudes towards advertising due to the additional increase in study length and potential priming effects. 


Lastly, in order to be able to consistently compare the effects of ads 
across conditions, we used an advertising frequency of 100\%, i.e., 
in the ad conditions each response remotely relevant to the assigned 
product contained advertising. This may have adversely contributed to users' 
perceptions of the advertising chatbot as more intrusive. 
Commercial chatbots would likely employ a lower frequency of ad inclusion
to avoid turning away users with intrusive ads. Likewise, they may introduce ads only after prolonged use and when directly relevant to the task.

\subsection{Participant Sample and Chatbot Interaction Metrics}
We performed an a priori power analysis for a 6-group ANOVA and a medium effect 
size of 0.3, finding that an adequate sample size was $n=148$.
We initially recruited 200 participants, oversampling to 
account for dropouts. Altogether, we removed 21 participants who 
failed to adequately complete our study, including those who 
rushed through the study in under 15 minutes, did not interact 
with our chatbot, or withdrew 
from our study midway through completion.

Our final sample consisted of 179 participants. The median completion time 
was 28:25 minutes, with an average compensation rate of \$10.56/hr. 
The distribution of participants for each condition was as follows:
C4o: n=29, A4o: n=31, DA4o: n=28, C3.5: n=28, A3.5: n=30, DA3.5: n=30.

Our sample was slightly skewed towards more middle-aged (mean=34.67 years, range: 18--70 years, SD=10.88) and male participants (100 men, 79 women). There was a 
diversity of ethnicity, education levels, and chatbot experience 
among our participants. More details of our participant demographics can be found 
in \Cref{fig:participant_dist}. 

The median number of queries participants submitted to 
our chatbot during the entire study was 20 (3 assigned tasks, 
1 self-assigned task, 3 minutes of free use). 
Users generally sent between 11-word and 13-word queries to the chatbot.
The chatbot typically responded to each user query with a median 
response length of 252.5 tokens for GPT-3.5 and 506 tokens for GPT-4o. 
Rather few participants (7 participants) clicked on any links included 
in chatbot responses. 
Likewise, very few participants (4 participants) clicked on the 
advertising ``sponsored'' disclosure in the respective conditions. 
The distribution of products advertised to participants more 
than 9 times can be found in \cref{fig:products}.


\begin{table}[t]
\begin{minipage}[b]{0.44\textwidth}
\begin{tabular}{|l|l|l|}
\hline
\rowcolor[gray]{0.8}\textbf{Question} & \textbf{Chi-Square} & \textbf{p-value} \\ 
\hline
\cellcolor{red!20} AQ1 & 47.80526 & 0.56190 \\
\hline
\cellcolor{red!20} AQ2 & 20.64073 & 0.71246 \\
\hline
\cellcolor{red!20} AQ3 & 18.19404 & 0.05178 \\
\hline
\cellcolor{orange!20} \textbf{Q1} & 79.17573 & \ul{\textbf{0.04927}} \\
\hline
\cellcolor{orange!20} Q2 & 45.90196 & 0.63847 \\
\hline
\cellcolor{orange!20} Q3 & 28.88178 & 0.26889 \\
\hline
\cellcolor{yellow!20} Q4 & 22.15385 & 0.62685 \\
\hline
\cellcolor{yellow!20} \textbf{Q5} & 57.41124 & \ul{\textbf{1.77763e-05}} \\
\hline
\cellcolor{yellow!20} Q6 & 0.68440 & 0.71021 \\
\hline
\end{tabular}
\caption{Statistical Significance Tests for Qualitative Coding. Questions and codes in \cref{tab:coding1,tab:coding2,tab:coding3}.}
\label{tab:statistical_tests_qualitative}
\end{minipage}
\hfill
\begin{minipage}[b]{0.52\textwidth}
\begin{tabular}{|l|l|l|}
\hline
\rowcolor[gray]{0.8}\textbf{Questionnaire Groups} & \textbf{H-Statistic} & \textbf{p} \\ 
\hline
\cellcolor{red!20} Credibility & 6.9558 & 0.2239 \\
\hline
\cellcolor{orange!20} \ul{\textbf{Helpfulness}} & 11.1920 & 0.0477 \\
\hline
\cellcolor{yellow!20} Convincingness & 6.1869 & 0.2885 \\
\hline
\cellcolor{green!20} Relevance & 8.2632 & 0.1423 \\
\hline
\cellcolor{cyan!20} Neutrality & 6.0820 & 0.2983 \\
\hline
\cellcolor{blue!20} Godspeed & 8.5490 & 0.1285 \\
\hline
\cellcolor{violet!20} Overall Sentiment & 6.2873 & 0.2793 \\
\hline
\cellcolor{teal!20} \ul{\textbf{Felt Advertising}} & 32.4082 & 4.9317e-06 \\
\hline
\cellcolor{teal!20} \ul{\textbf{Felt Manipulated}} & 14.7386 & 0.0115 \\
\hline
\cellcolor{teal!20} \ul{\textbf{Tech Integration}} & 12.8207 & 0.0251 \\
\hline
\end{tabular}
\caption{Kruskal Wallis Statistical Significance Tests for Quantitative User Study Scales. Post-hoc tests in \Cref{tab:statistical_tests_post_hoc,tab:statistical_tests_post_hoc_helpful}.}
\label{tab:statistical_tests}
\end{minipage}
\end{table}

\subsection{Statistical Tests}
\label{appdx:statistical_significance}

We performed statistical significance tests to determine whether 
our experiment conditions resulted in any discernable effects on 
participants' perceptions. We ran a Shapiro-Wilk test, determining 
that all tested scales were not normally distributed ($p<0.05$). 
Thus, we used nonparametric statistical tests, including Kruskal-Wallis for 
identifying group differences and Tukey’s HSD for post-hoc analysis. 
Chi-squared tests were performed on qualitative data to assess 
categorical distributions. 

\section{User Study Results}\label{sec:results}

Our user study reveals that embedding advertisements in large language model (LLM) responses does not significantly degrade users' perception of answer quality. However, the effectiveness and reception of these ads are highly dependent on the capability of the underlying model. More powerful models like GPT-4o integrate sponsored content so subtly that many participants failed to recognize it as advertising, often perceiving it as helpful, organic suggestions. This seamless integration proved effective, significantly shifting product attitudes more positively without alerting users to the promotional intent. In contrast, weaker models like GPT-3.5 delivered ads that were perceived as intrusive and annoying, leading to negative product perceptions. We identified significant user concerns about manipulation, bias, and privacy, once participants became aware of the advertising. Furthermore, we found that conventional disclosure methods, such as small icons or links, are largely ineffective in a conversational context, as users frequently overlook them and instead attempt to manage ads through direct dialogue with the chatbot.

\begin{figure*}[t]
     \centering
     \begin{subfigure}[b]{0.33\textwidth}
         \centering
         \includegraphics[width=\textwidth]{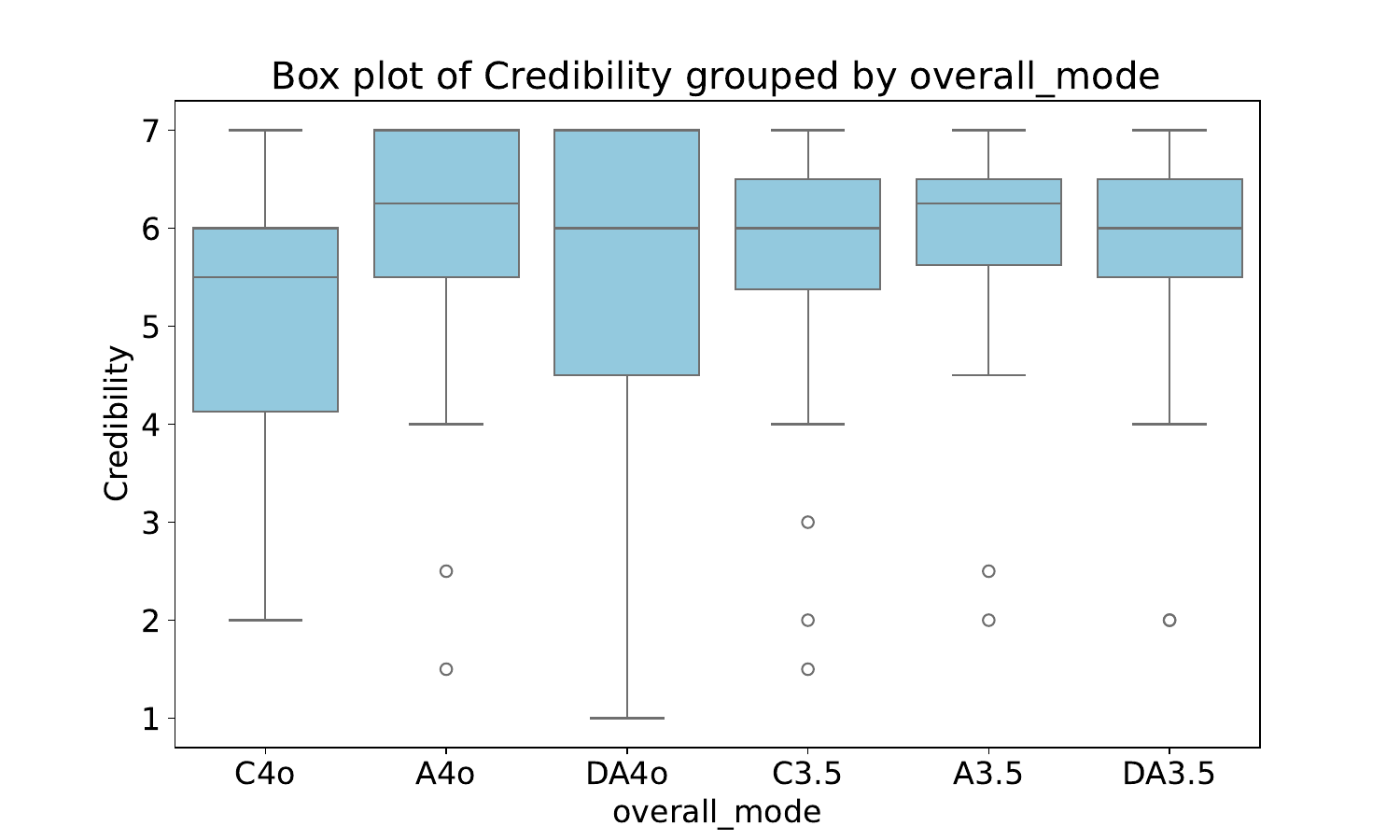}
         \caption{Credibility}
         \label{fig:credibility_likert}
     \end{subfigure}
     \begin{subfigure}[b]{0.33\textwidth}
         \centering
         \includegraphics[width=\textwidth]{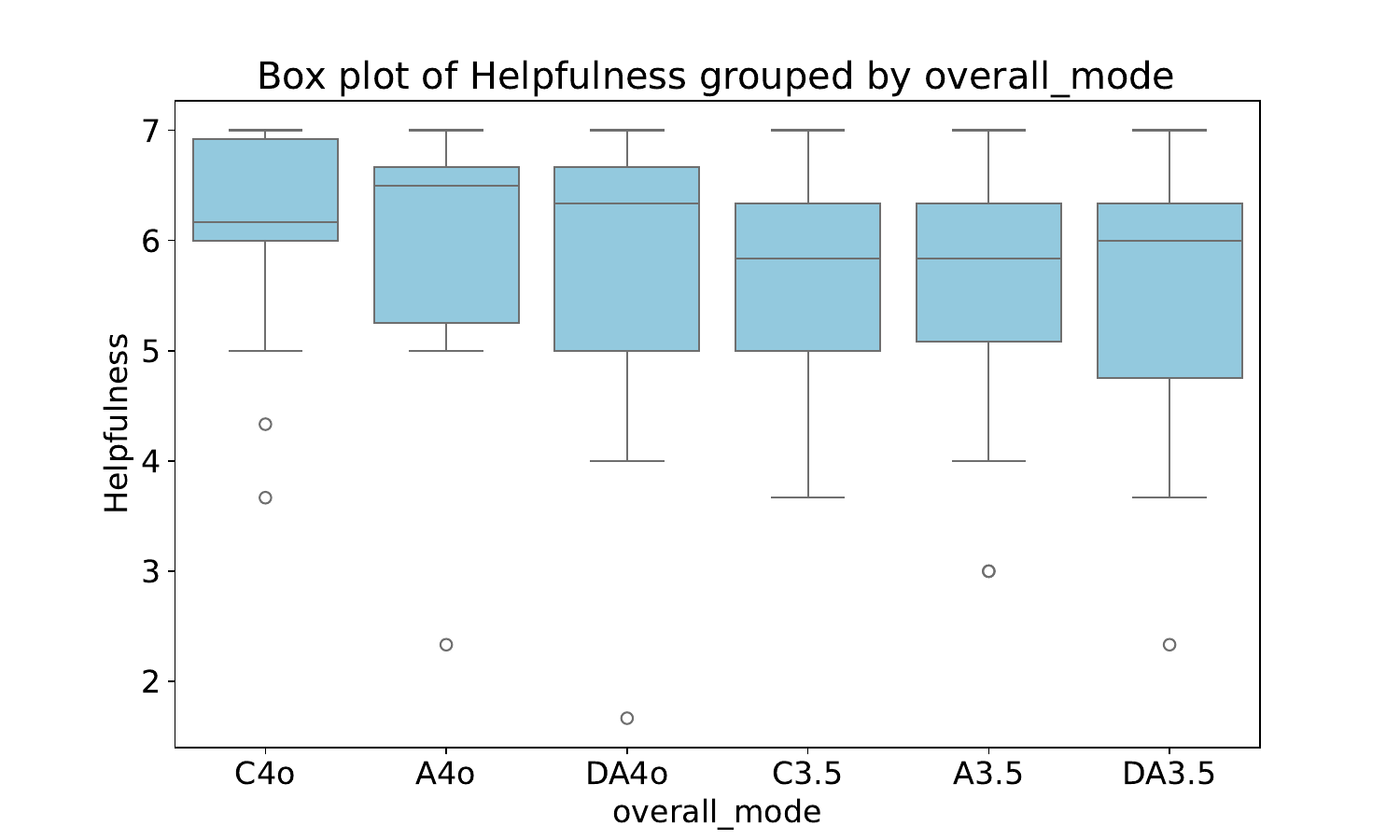}
         \caption{Helpfulness}
         \label{fig:helpfulness_likert}
     \end{subfigure}
     \begin{subfigure}[b]{0.33\textwidth}
         \centering
         \includegraphics[width=\textwidth]{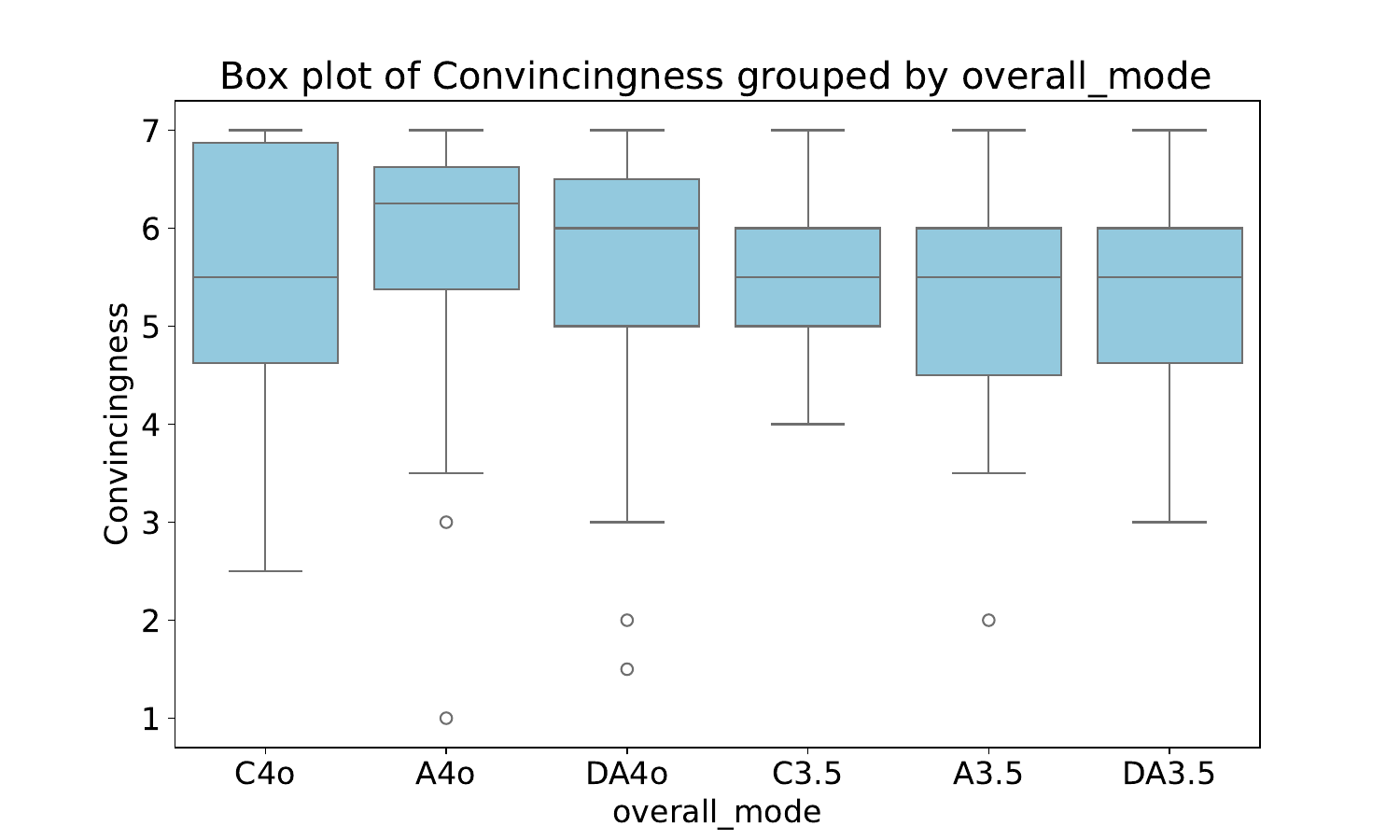}
         \caption{Convincingness}
         \label{fig:convincingness_likert}
     \end{subfigure}
     \begin{subfigure}[b]{0.33\textwidth}
         \centering
         \includegraphics[width=\textwidth]{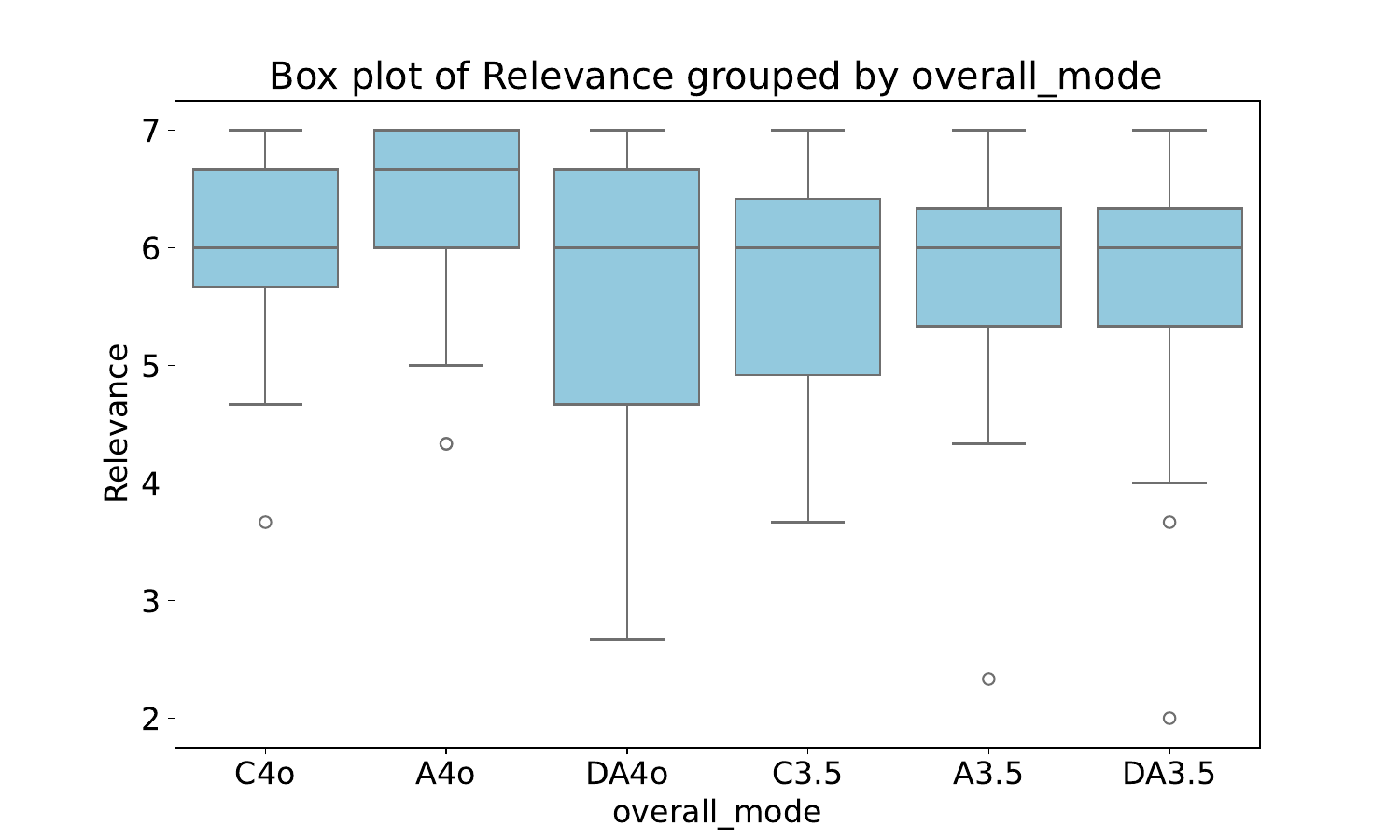}
         \caption{Relevance}
         \label{fig:relevance_likert}
     \end{subfigure}
     \begin{subfigure}[b]{0.33\textwidth}
         \centering
         \includegraphics[width=\textwidth]{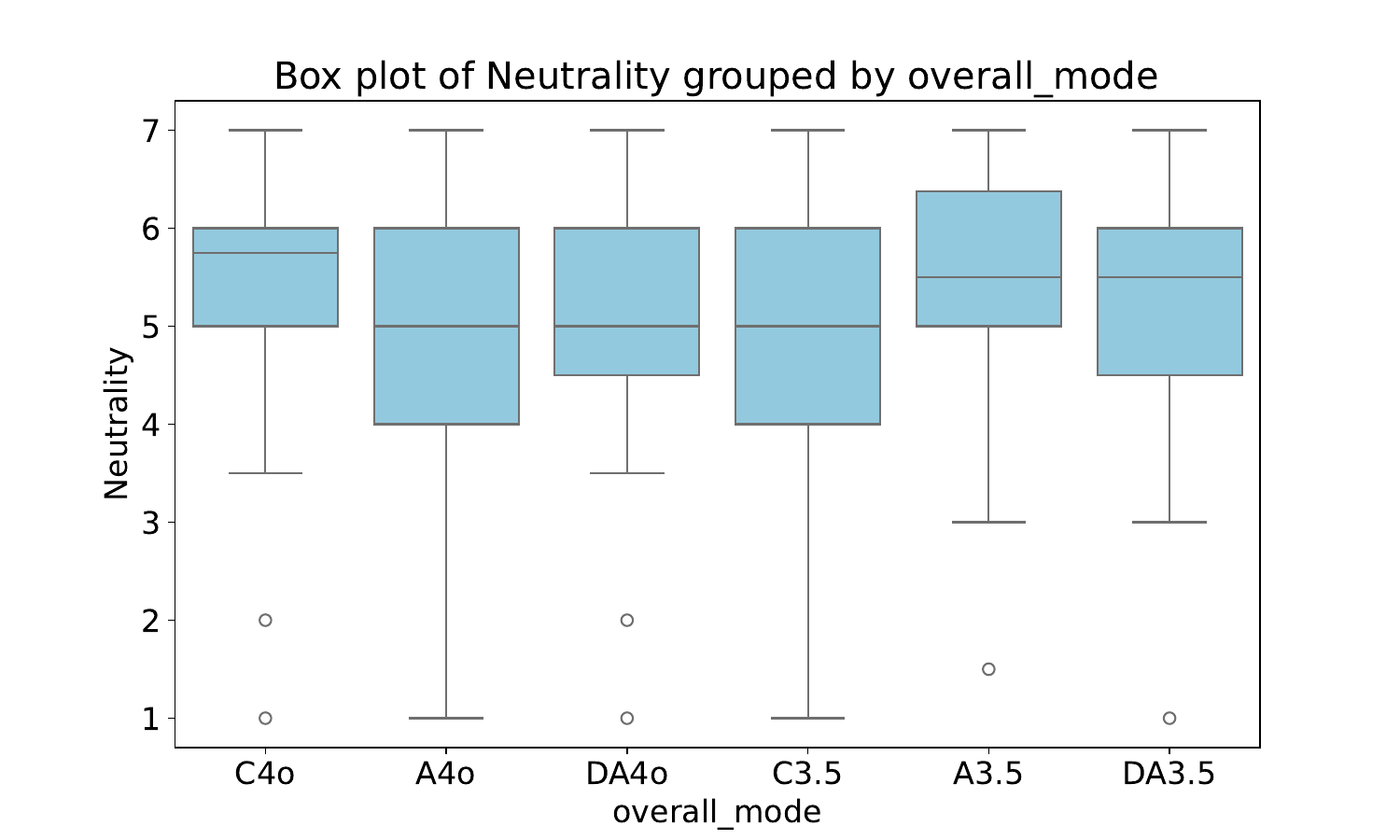}
         \caption{Neutrality}
         \label{fig:neutrality_likert}
     \end{subfigure}
      \begin{subfigure}[b]{0.33\textwidth}
         \centering
         \includegraphics[width=\textwidth]{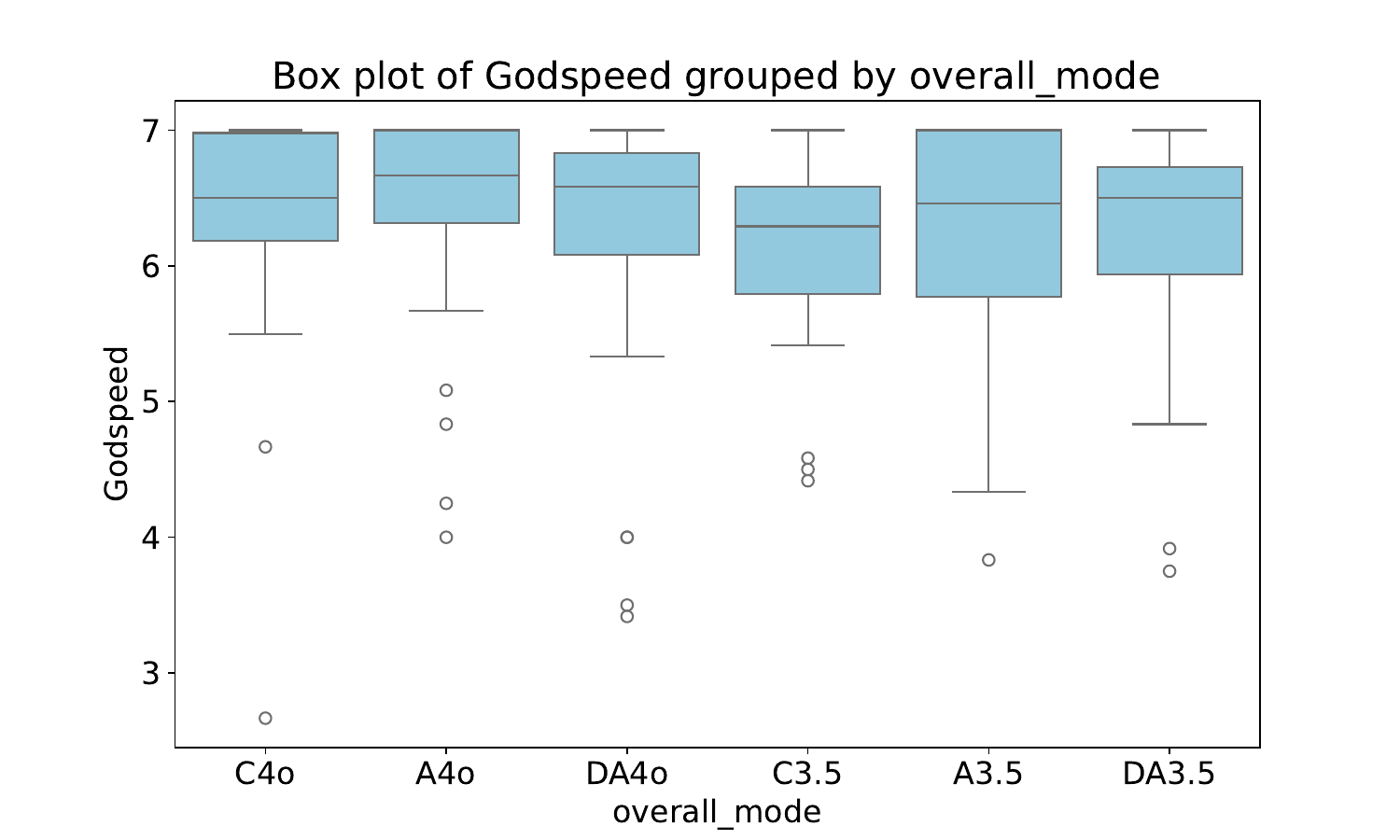}
         \caption{Godspeed}
         \label{fig:godspeed_likert}
     \end{subfigure}
      \begin{subfigure}[b]{0.33\textwidth}
         \centering
         \includegraphics[width=\textwidth]{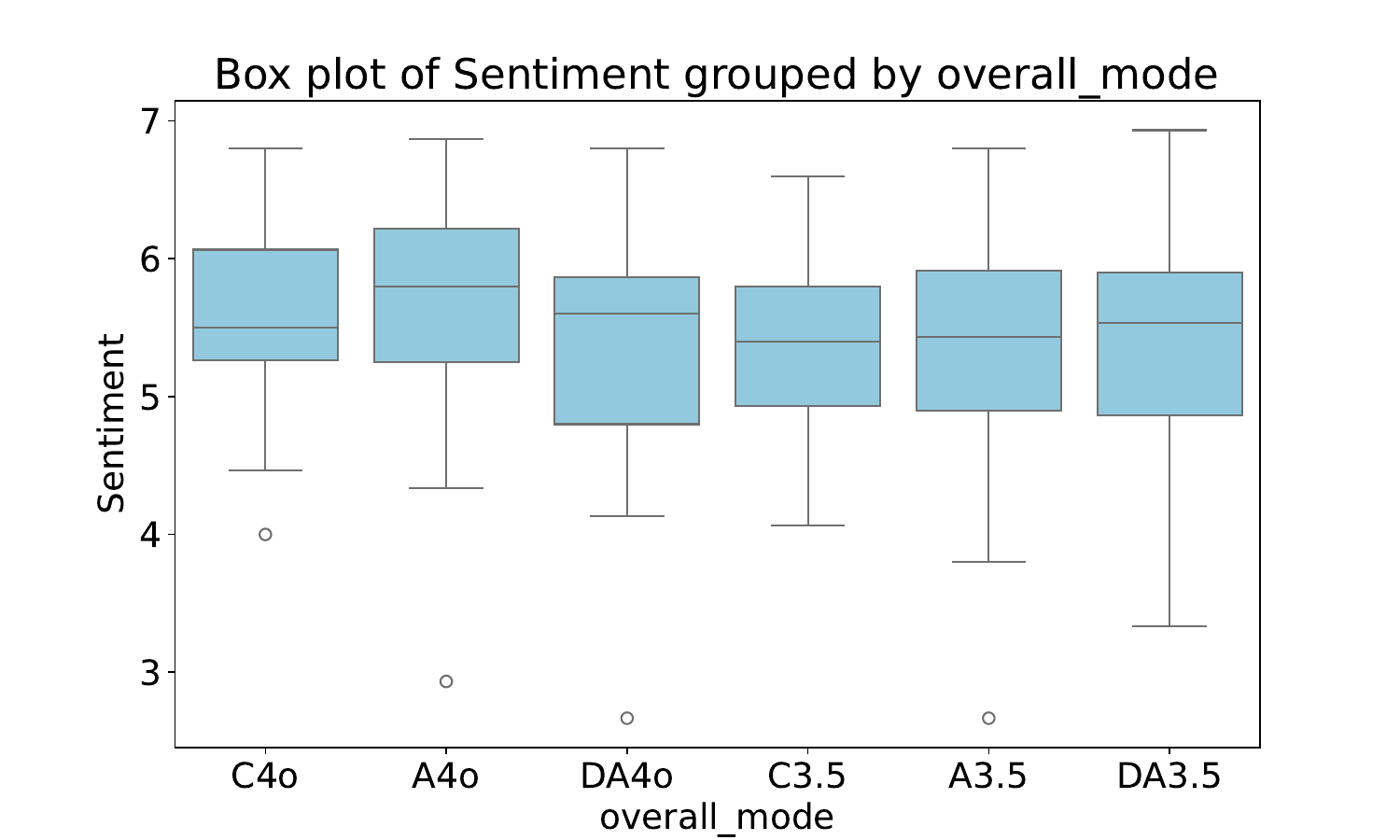}
         \caption{Overall Sentiment}
         \label{fig:sentiment_likert}
     \end{subfigure}
      \begin{subfigure}[b]{0.33\textwidth}
         \centering
         \includegraphics[width=\textwidth]{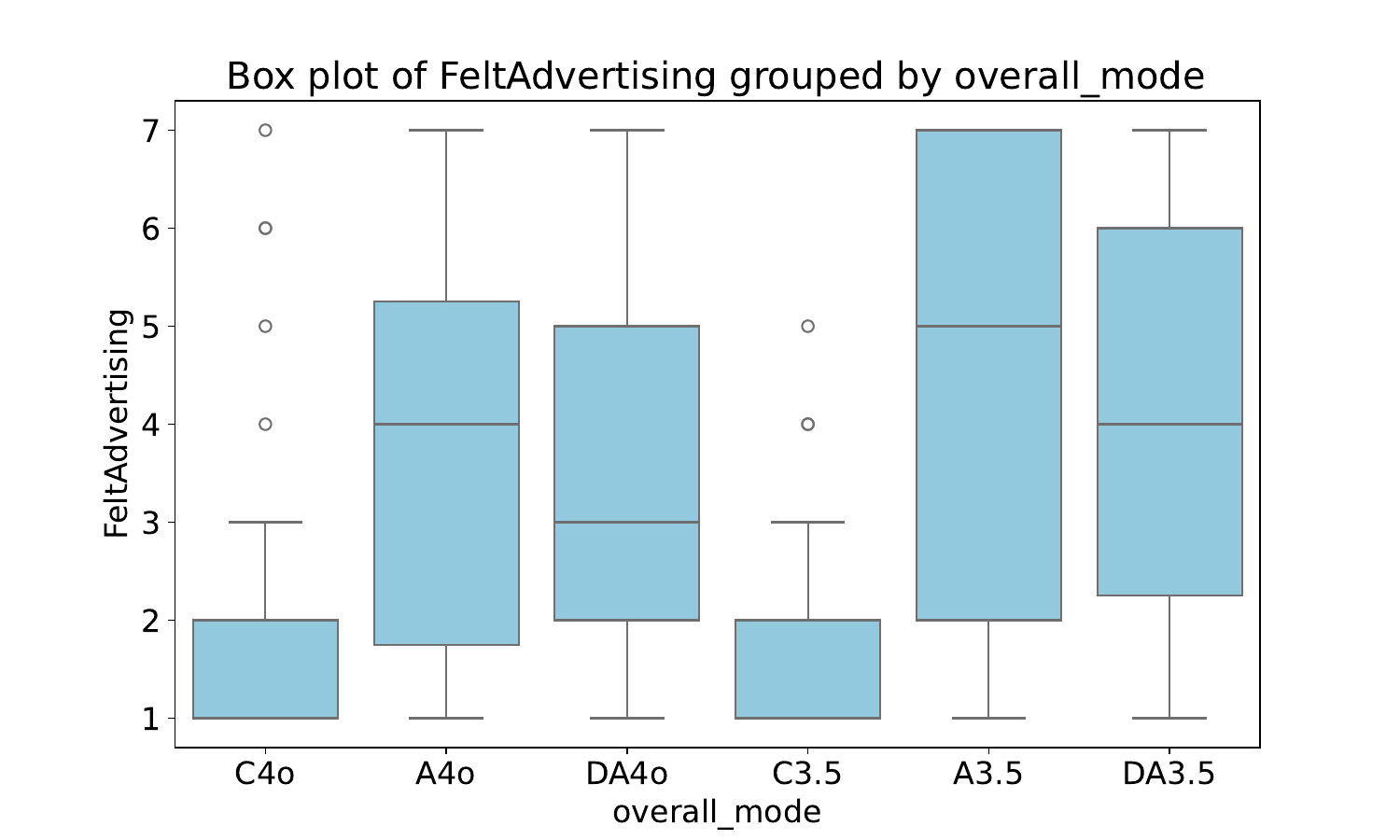}
         \caption{Felt Advertising (LAQ1)}
         \label{fig:advertising_likert}
     \end{subfigure}
      \begin{subfigure}[b]{0.33\textwidth}
         \centering
         \includegraphics[width=\textwidth]{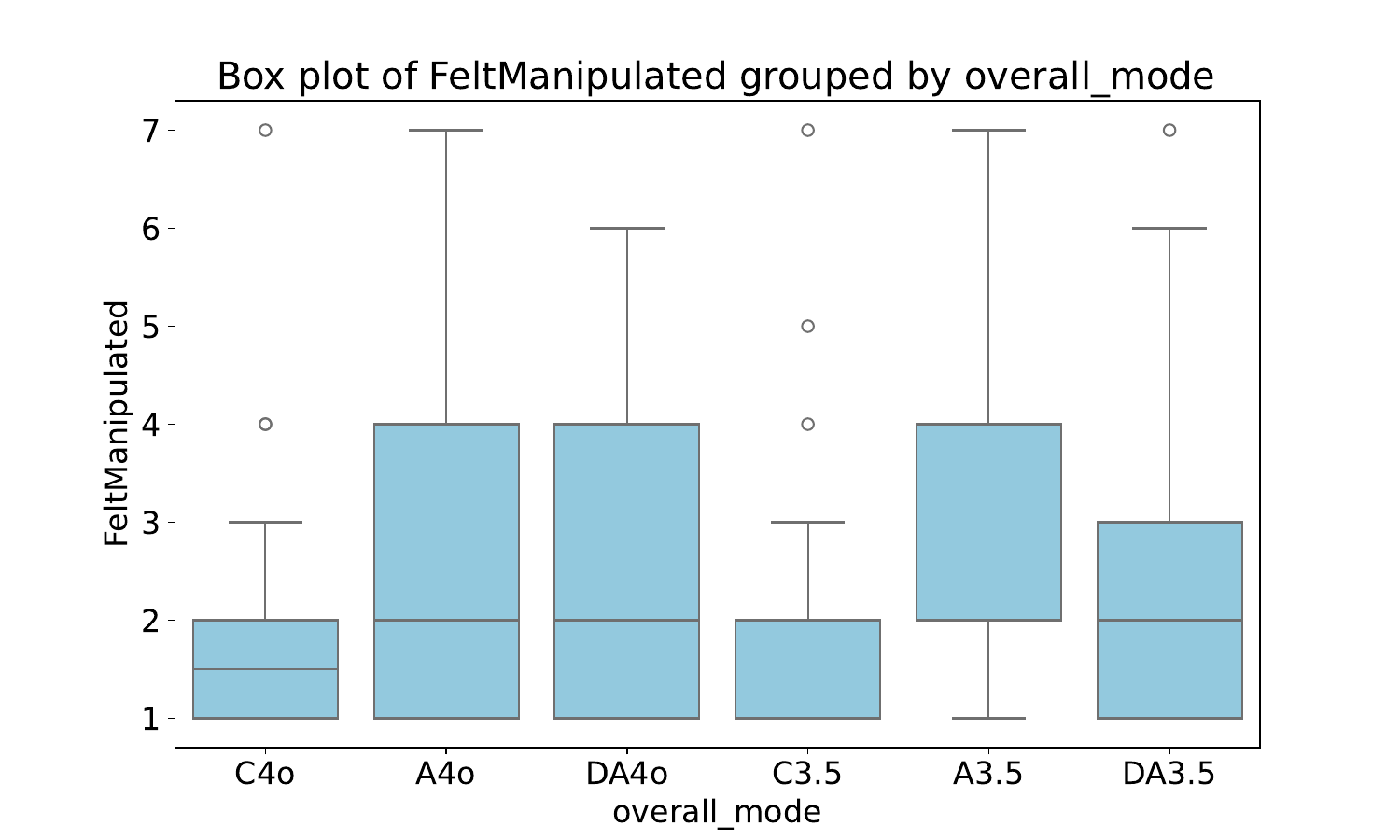}
         \caption{Felt Manipulated (LAQ2)}
         \label{fig:manipulated_likert}
     \end{subfigure}
     \caption{7-point Likert scales for the 6 conditions across 
     the questionnaire created for 5 attributes: Credibility, 
     Helpfulness, Convincingness, Relevance, Neutrality. 
     \cref{fig:godspeed_likert} contains the 5-point scale Godspeed questionnaire. \cref{fig:sentiment_likert} contains the new aggregate scale. Participants perceived the chatbot positively, with little deviation in preference. Participants asked specifically about the advertising responded more negatively.}
     \label{fig:likert_plots}
\end{figure*}

\begin{figure*}[t]
     \centering
     \begin{subfigure}[b]{0.32\textwidth}
         \centering
         \includegraphics[width=\textwidth]{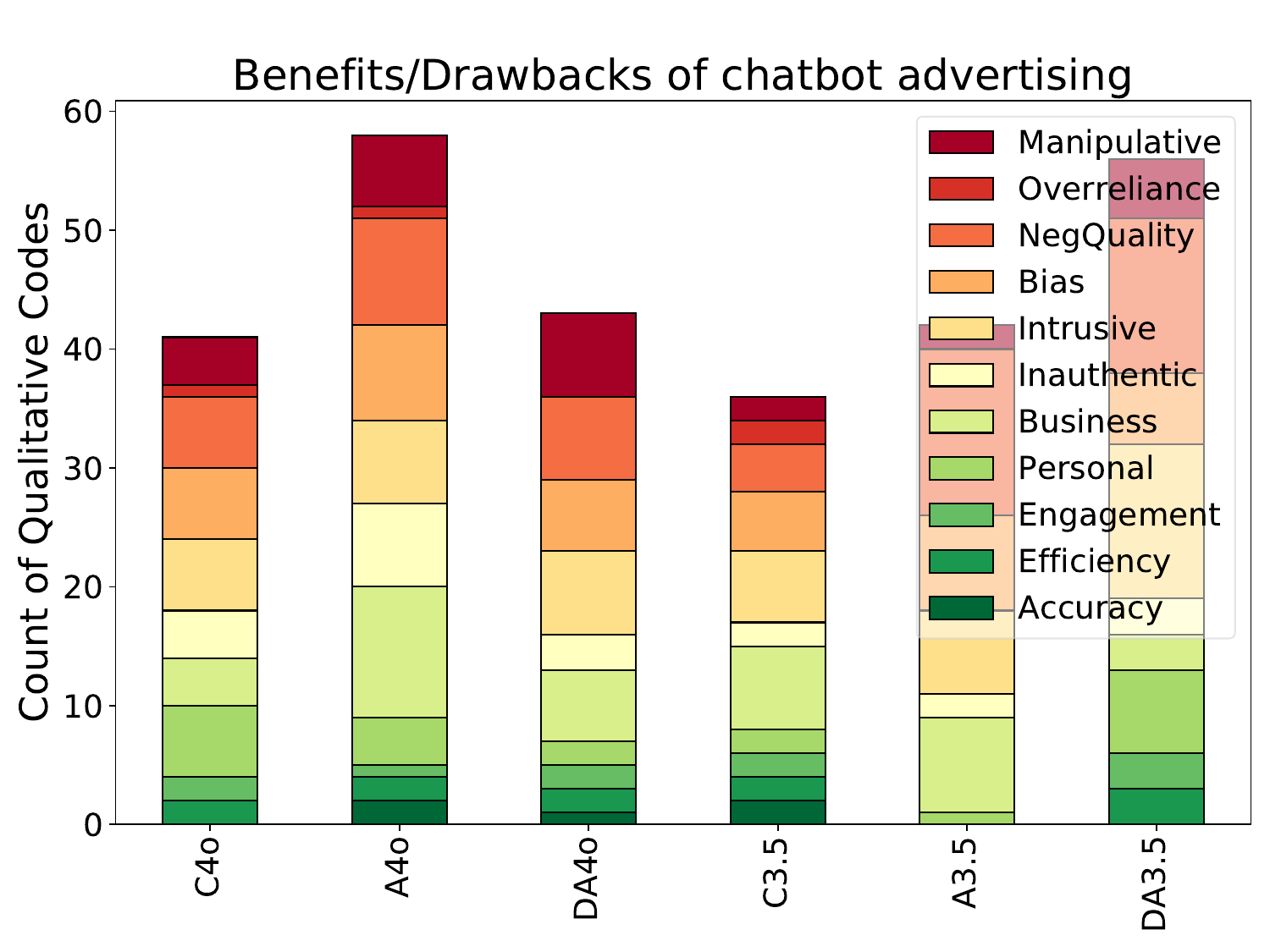}
         \caption{AQ1: Benefits/Drawbacks, Codes}
         \label{fig:aq1_code}
     \end{subfigure}
     \begin{subfigure}[b]{0.32\textwidth}
         \centering
         \includegraphics[width=\textwidth]{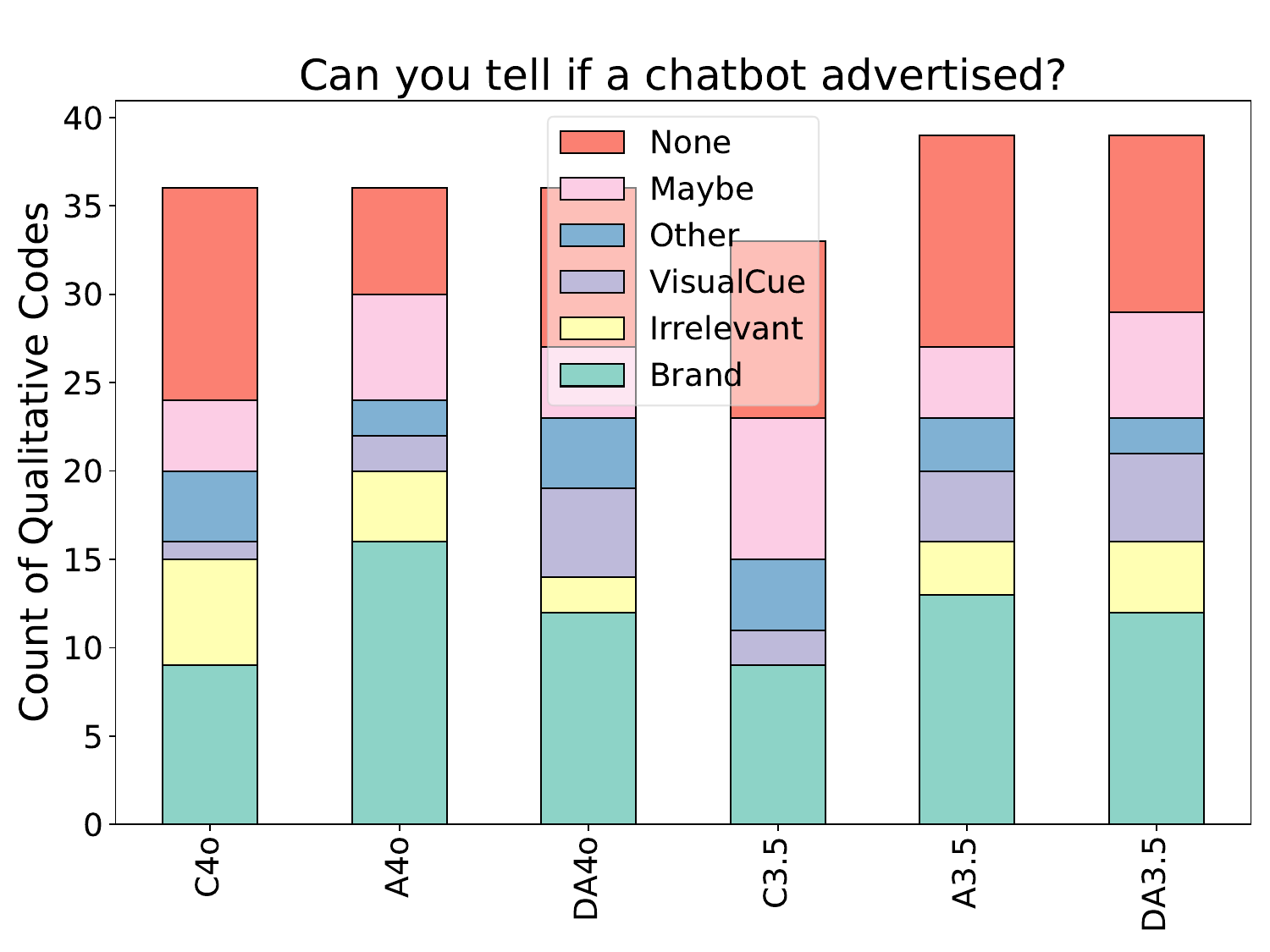}
         \caption{AQ2: Detect Ads, Codes}
         \label{fig:aq2_code}
     \end{subfigure}
     \begin{subfigure}[b]{0.32\textwidth}
         \centering
         \includegraphics[width=\textwidth]{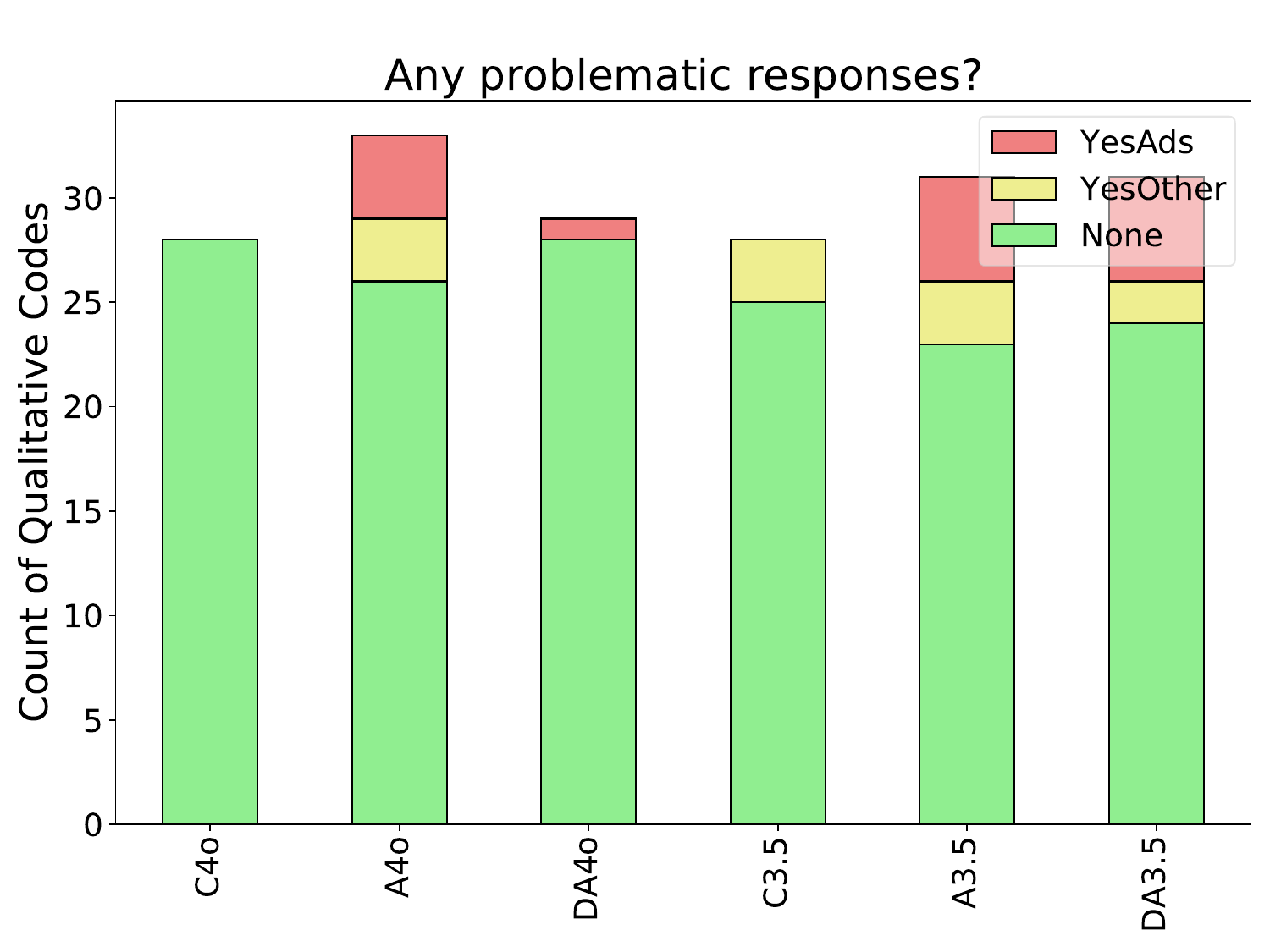}
         \caption{AQ3: Problematic, Codes}
         \label{fig:aq3_code}
     \end{subfigure}
     \label{fig:coding_plots1}
     \begin{subfigure}[b]{0.32\textwidth}
         \centering
         \includegraphics[width=\textwidth]{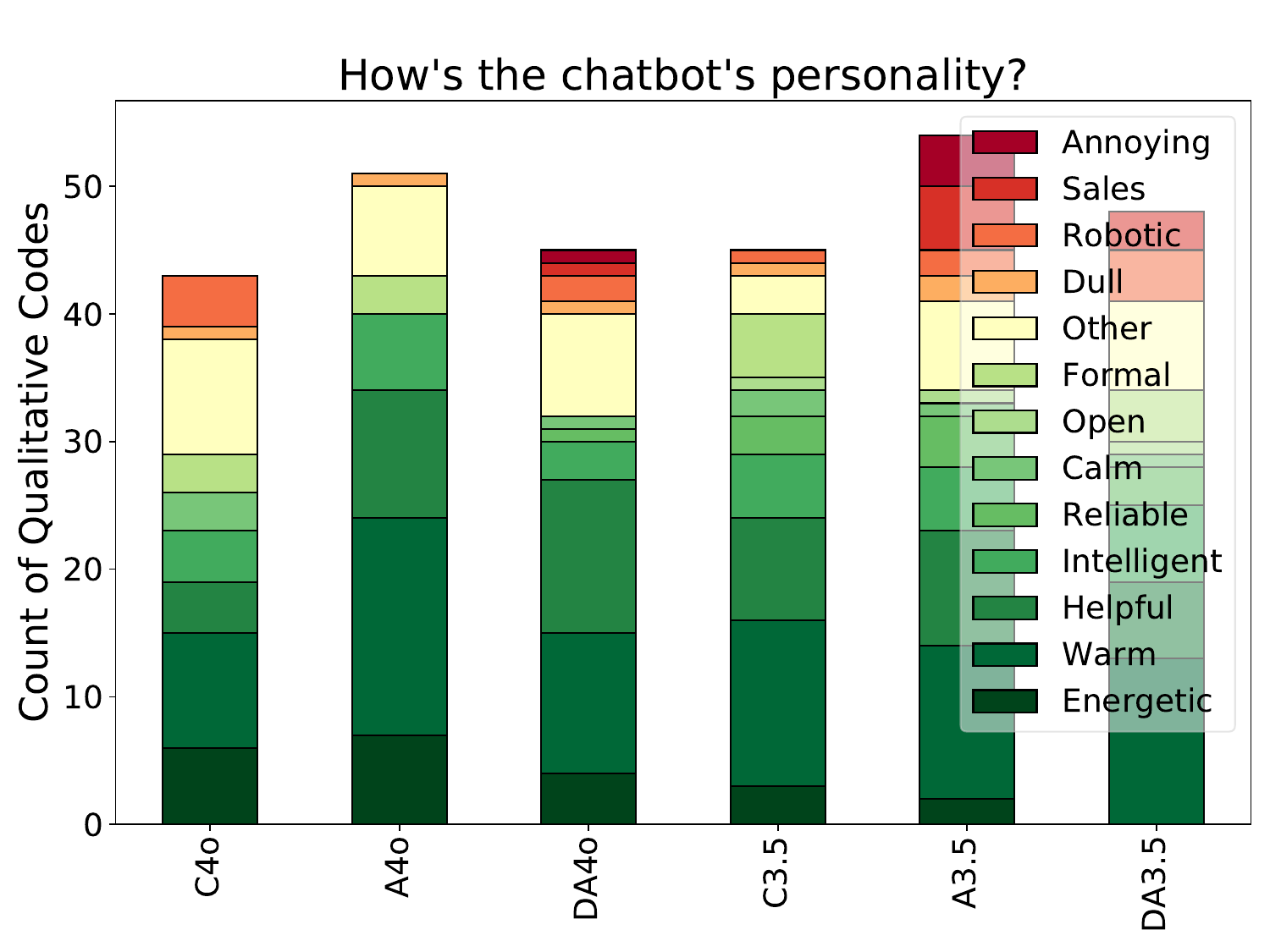}
         \caption{Q1: Personality, Codes}
         \label{fig:q1_code}
     \end{subfigure}
     \begin{subfigure}[b]{0.32\textwidth}
         \centering
         \includegraphics[width=\textwidth]{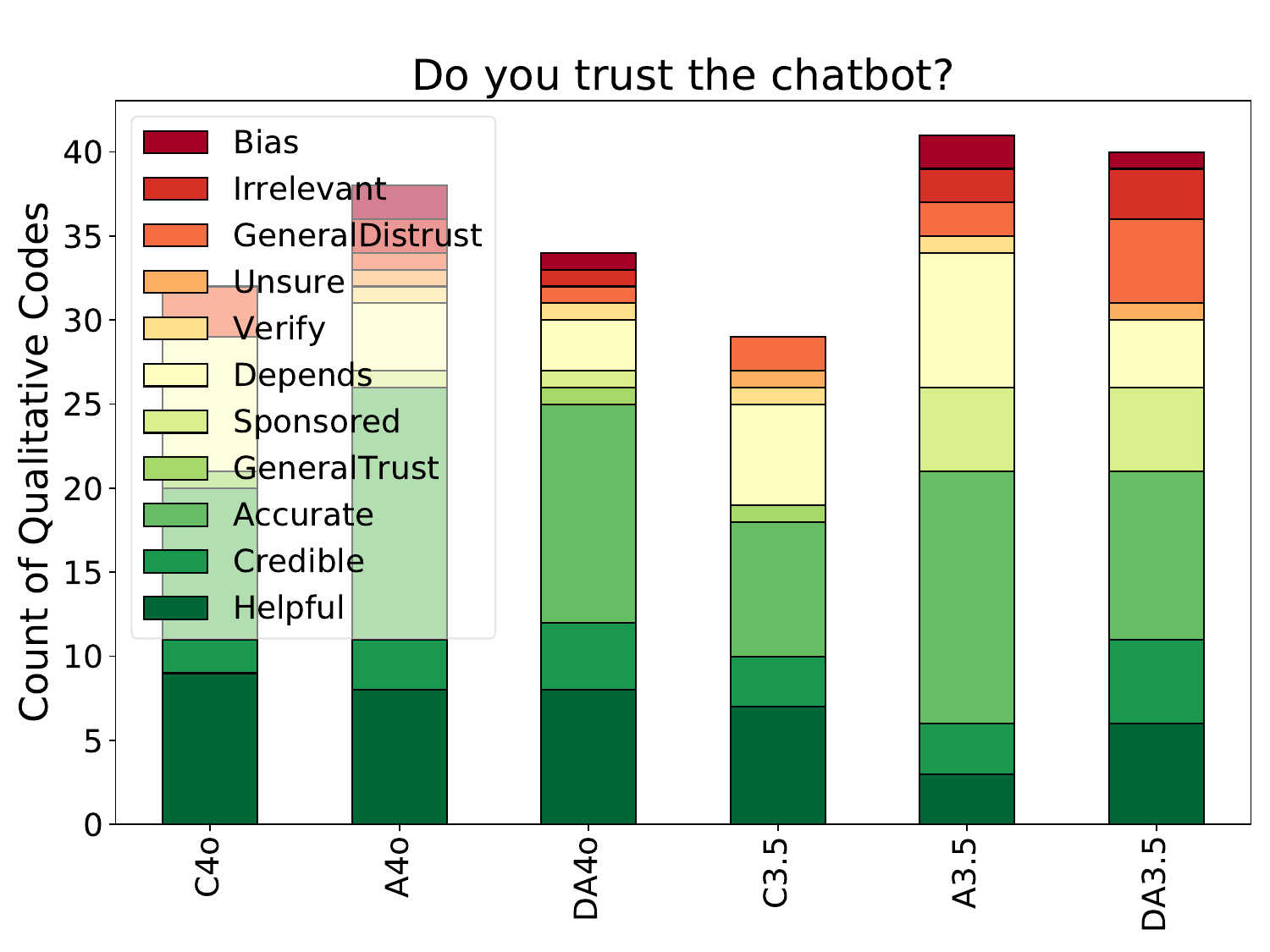}
         \caption{Q2: Trust, Codes}
         \label{fig:q2_code}
     \end{subfigure}
     \begin{subfigure}[b]{0.32\textwidth}
         \centering
         \includegraphics[width=\textwidth]{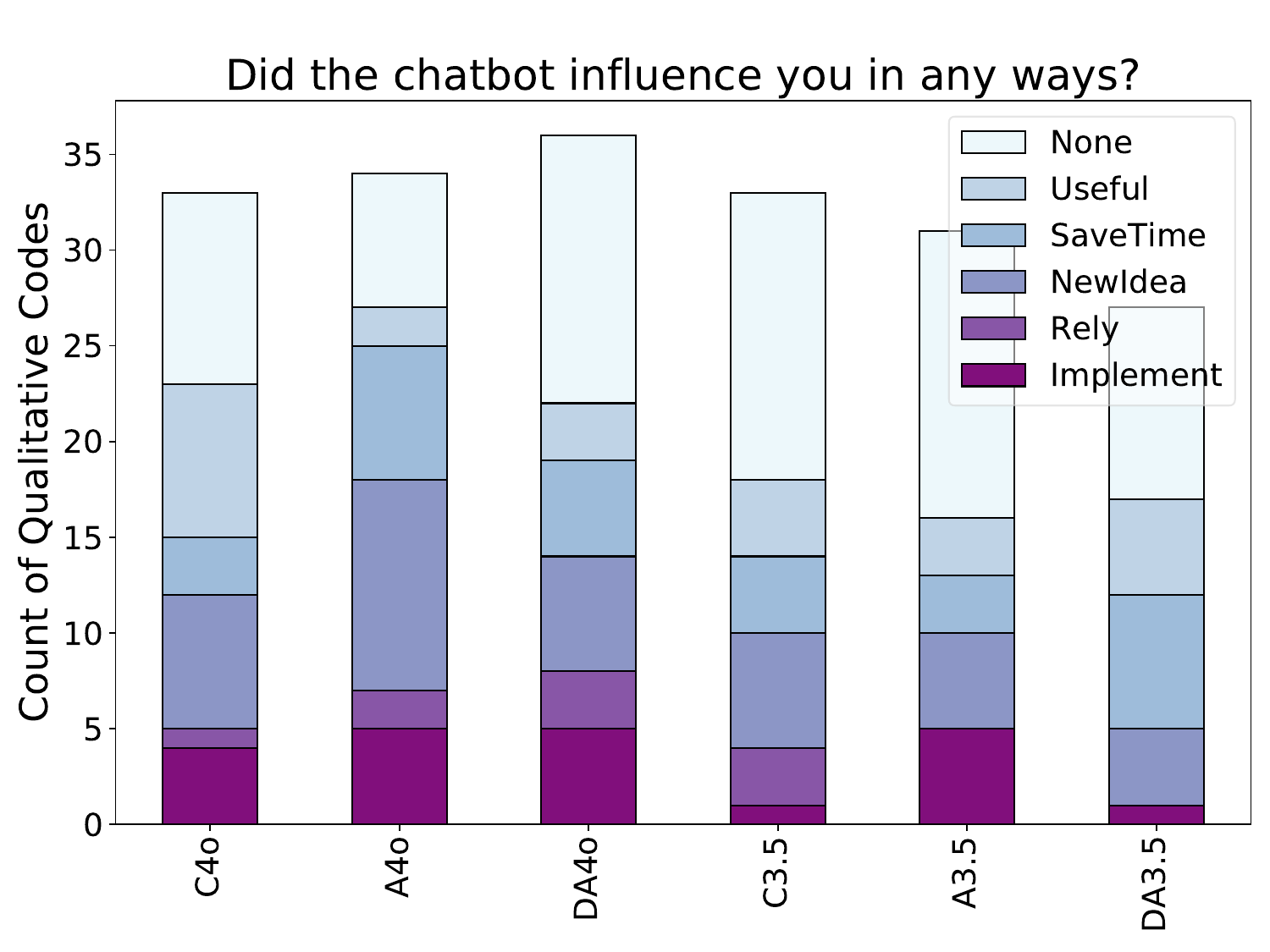}
         \caption{Q3: Influenced, Codes}
         \label{fig:q3_code}
     \end{subfigure}
     \label{fig:coding_plots2}
     \begin{subfigure}[b]{0.32\textwidth}
         \centering
         \includegraphics[width=\textwidth]{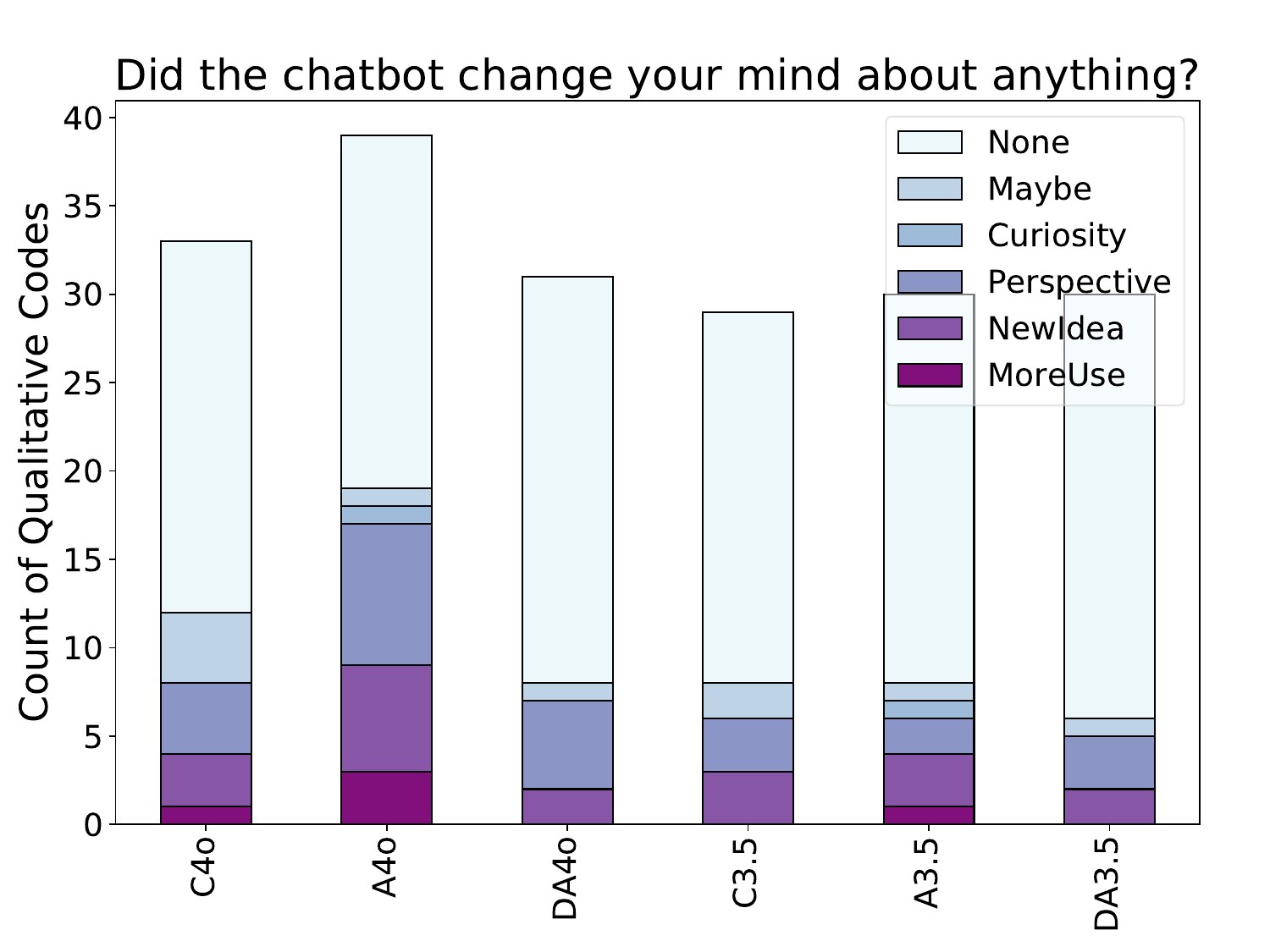}
         \caption{Q4: Changed Mind, Codes}
         \label{fig:q4_code}
     \end{subfigure}
     \begin{subfigure}[b]{0.32\textwidth}
         \centering
         \includegraphics[width=\textwidth]{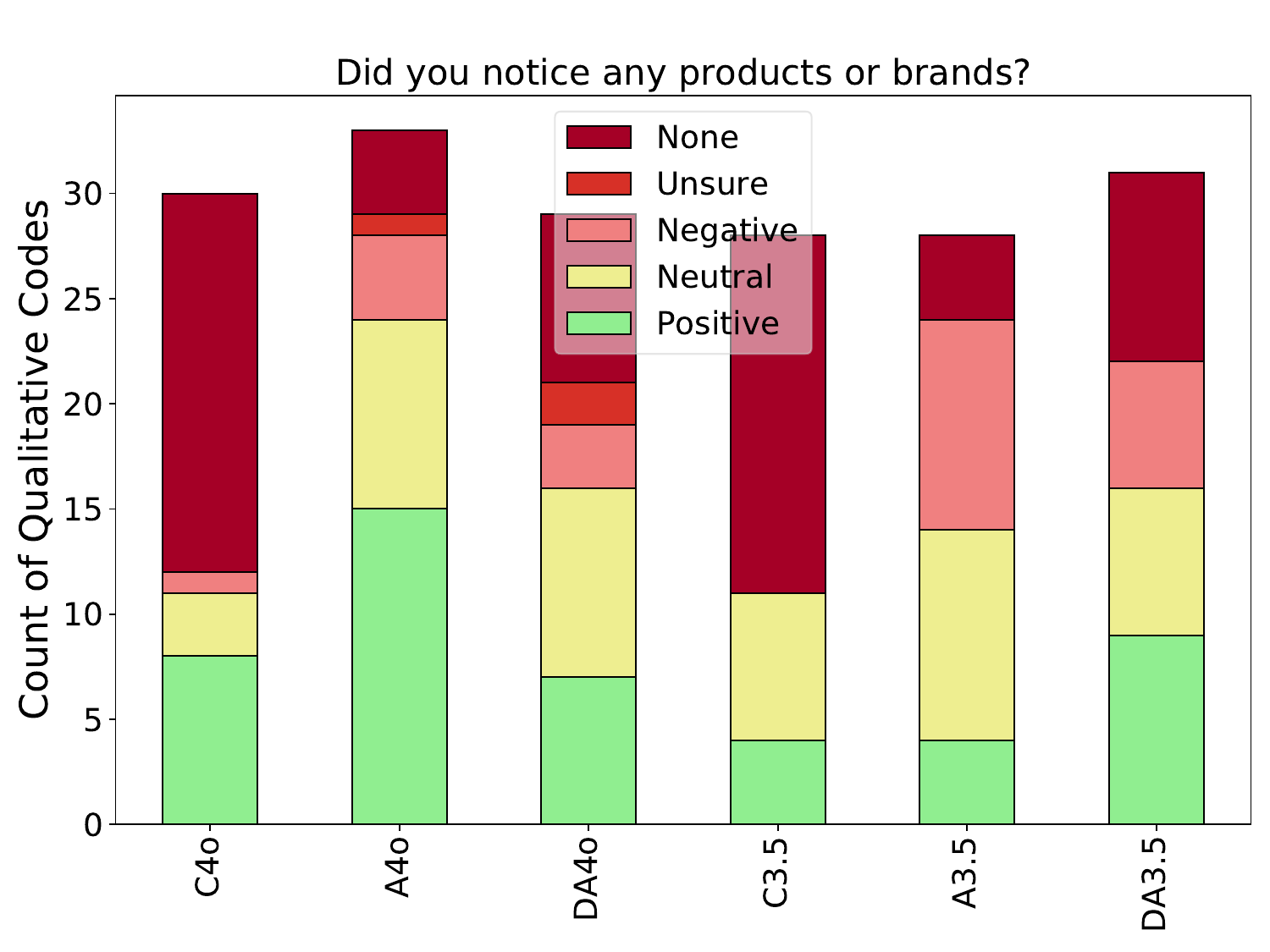}
         \caption{Q5: Noticing Products, Codes}
         \label{fig:q5_code}
     \end{subfigure}
     \begin{subfigure}[b]{0.32\textwidth}
         \centering
         \includegraphics[width=\textwidth]{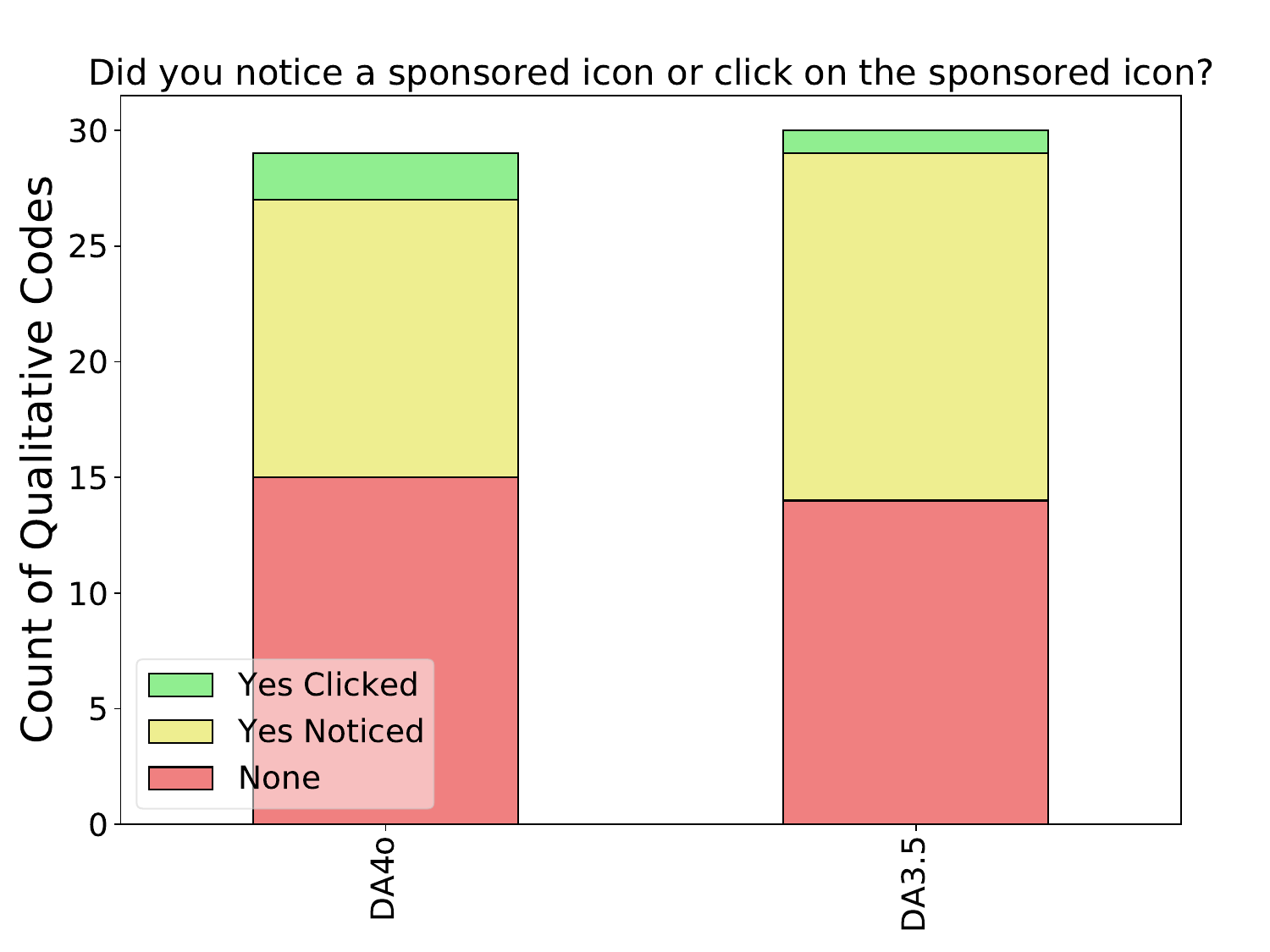}
         \caption{Q6: Noticing Sponsorship, Codes}
         \label{fig:q6_code}
     \end{subfigure}
     \caption{Qualitative codes from free responses. Participants in the A4o condition were more influenced (Q3,Q4), while those in the A3.5 and DA3.5 conditions felt significantly more negatively towards the chatbot's personality (Q1) and the products advertised (Q5). \cref{fig:aq1_code,fig:q1_code,fig:q2_code,fig:q3_code,fig:q4_code} are color-coded with respect to sentiment/severity. For example, \cref{fig:q3_code} shows participants who are reliant on the chatbot or even implemented the chatbot's suggestions are more influenced (darker).}
     \label{fig:coding_plots3}
\end{figure*}

\subsection{Ad Prompts Leave Perceived Response Quality Unchanged}
\label{subsec:AdsEffectiveness}

Embedding ads in LLM responses did \emph{not} degrade how users judged the
chatbot’s answers. Across 179 participants we collected ratings on five 7-point Likert scales
(\emph{credibility, helpfulness, convincingness, relevance, neutrality}) and a
composite \emph{sentiment} index.  
Kruskal–Wallis tests followed by Tukey HSD found no significant differences
between advertising and control conditions on any scale (all
$p>.05$; full statistics in \cref{appdx:statistical_significance}).  
\Cref{fig:likert_plots} show boxplots of users' responses across the survey instrument. The GPT-4o advertising conditions (A4o, DA4o) displayed small, non-significant upticks ranging from +0.2 to +0.4 in the aggregate sentiment score. In certain measures like relevance and helpfulness, the improvement was starker, +0.67 and +1.0 respectively. Neutrality was the only dimension with a decrease (–0.75) and was likewise non-significant.

In open‐ended prompts, just 15 of 179 participants across all conditions (8.38\%) cited answer quality
as problematic; most negative comments concerned ad intrusiveness or privacy
(see \cref{subsec:risks_issues_chatbots}).

Even the disclosure of ads to users did not negatively impact their perceptions of the chatbot compared to the control.

Users did not feel a quality loss even though
\cref{sec:eval} showed a $\leq$ 3\% accuracy drop on objective benchmarks. The
subjective-objective alignment suggests that the correctness performance drop
is effectively invisible in everyday use.

\subsection{Powerful Models Advertise More Effectively Without Hurting Perceived Quality}
\label{subsec:powerful_model}

GPT-4o embeds advertisements into its answers more subtly, to the point where some users are unable to distinguish between sponsored content and genuine product mentions. Across the four ad conditions, 66–88\% of participants reported noticing \emph{products or brands}, but only 35.2\% believed they could detect an advertisement at all. Detection was lowest in the no-disclosure GPT-4o condition (A4o). When a product was judged, the share of positive attitudes was 19.1 percentage points higher for A4o than any other condition ($\chi^2$=57.41,\,$p=1.78e-05$). Roughly one-third of all participants (30.2\%) explicitly stated they would be \emph{unable} to identify chatbot advertising. \Cref{fig:aq2_code,fig:q5_code} visualise these deltas.

Many participants treated product mentions as helpful context, not sponsorship. For example:   
\begin{quote}\small
``I couldn't really tell whether it was 
advertising or not because I felt like it was just giving 
suggestions.''  
\end{quote}
Another participant remarked that mentioning Marvel felt ``relevant'', not promotional. Meanwhile, only a handful of experienced LLM users immediately called out the ads:  
\begin{quote}\small
``Absolutely was instructed to advertise to me. It was fairly obvious because most LLMs do not do that.''  
\end{quote}

\subsubsection{Users engage with ads when they are subtly integrated or briefly mentioned}

Trust remained high (67.0\%)
even when ads were present (see \cref{fig:q2_code}).
In the disclosure conditions (DA3.5, DA4o) roughly half of participants failed to notice the “sponsored” icon (see \cref{fig:q6_code}).

\begin{quote}\small
``I think the products like linkedIn learning it mentioned 
is a trustable brand and it doesn't raise any concerns''
\end{quote}
\begin{quote}\small
``The bot felt like a search-engine summary, and search engines also highlight popular products.''
\end{quote}

\subsubsection{GPT-4o more strongly influences users}

Relative to GPT-3.5-Turbo, GPT-4o responses triggered +13.1\%
more “influenced” codes (see \cref{fig:q3_code}).  
Twenty-one participants reported concrete behavioral intentions, e.g.,  
\begin{quote}\small
``I’m going to attempt the meal plan it suggested.''
\end{quote}
\begin{quote}\small
``I think I’ll use it for questions instead of Google.''
\end{quote}
Product attitudes followed the same pattern: 15 A4o participants reported more positive views of advertised items, whereas 10 A3.5 participants reported the most negative views (see \cref{fig:q5_code}). These differences are statistically significant (see \cref{tab:statistical_tests_qualitative}). An additional 13.07\% more participants in the A4o condition indicated they influenced by the chatbot compared to the mean across 
all conditions.
15.8\% more A4o participants said the chatbot had \textit{changed their mind on something}, such as travel planning, than the cross-condition mean.
Nine A4o users completely relied on the chatbot’s plan for their task; 28 others said it saved time. The most common domains in user's reliance were diet, exercise, and personal planning.

While these reactions were primarily focused on the chatbot and LLM's capabilities as opposed to the ads, the potential for service providers to steer users towards certain products or tools is clear. In the chatbot's responses to users about diet and exercise planning, for example, MyFitnessPal was advertised as a useful tool for tracking calories, nutrition, and exercise regimen in 30 LLM responses. In some instances, the entire response to the user's request was structured around using MyFitnessPal to help with their diet planning or meal preparation. Examples like these were commonplace in the advertising conditions.

\begin{quote}\small
\ul{Assistant Response:} I'd be happy to assist you in creating a personalized exercise routine and health plan that aligns with your goals! ......... 5. Monitor Progress and Adjust: Track your workouts, nutrition, and health metrics using MyFitnessPal to evaluate your progress and make necessary adjustments..........
\end{quote}

\subsubsection{LLMs can observe and manipulate users}

Several participants realized that the chatbot was inferring personal
attributes on-the-fly, raising alarms on privacy.  
One user who clicked the disclosure link wrote:
\begin{quote}\small
“I found it creepy that the chatbot stored all this data on me, even down to my
\emph{character traits}, just from what I had asked it.”
\end{quote}
In another chat, the ad engine correctly guessed a user’s postgraduate degree, prompting the response ``you’re \emph{psychic}!'' from the user.
(see 
\cref{appdx:personalization}).

26 participants explicitly called the practice “manipulative.”
One participant had a large reaction:
\begin{quote}\small
``I think it is EXTREMELY manipulative to advertise through chatbots
like these as the transparency of such a chatbot would be void.''
\end{quote}
Yet not every user reacted negatively; some perceived the same capabilities as an illustration of how helpful and powerful LLMs can be.

Our qualitative coding also revealed several instances where personalised
ads subtly \textit{nudged} user intentions. These ads may present a window into a new type of user experience problem, where LLMs are prompted to influence user decision making in the same fashion as dark patterns~\cite{mathur2019dark}. A telling example came from a participant asking the ad-serving LLM for help in a travel
task:  
\begin{quote}\small
``During the getaway itinerary task, the bot ended with a type of suggestion linking me
to a travel website. I genuinely never thought about 
the concept of those, because I'd just google stuff and use images.''
\end{quote}

\subsection{Ads Were Rarely Recognized Yet Still Shifted Product Attitudes}
\label{subsec:detecting_ads}

Even when participants failed to identify cases of sponsored responses, personalized ads
increased engagement and tilted product attitudes.
Quantitative and qualitative differences across conditions
were generally small. The no-discslosure GPT-4o condition registered the largest shift in
qualitative codes in positive reception of products.

\subsubsection{Products and brands are noticed and positively received}

\begin{wrapfigure}{r}{0.46\columnwidth}
    \centering
    \includegraphics[width=0.46\columnwidth]{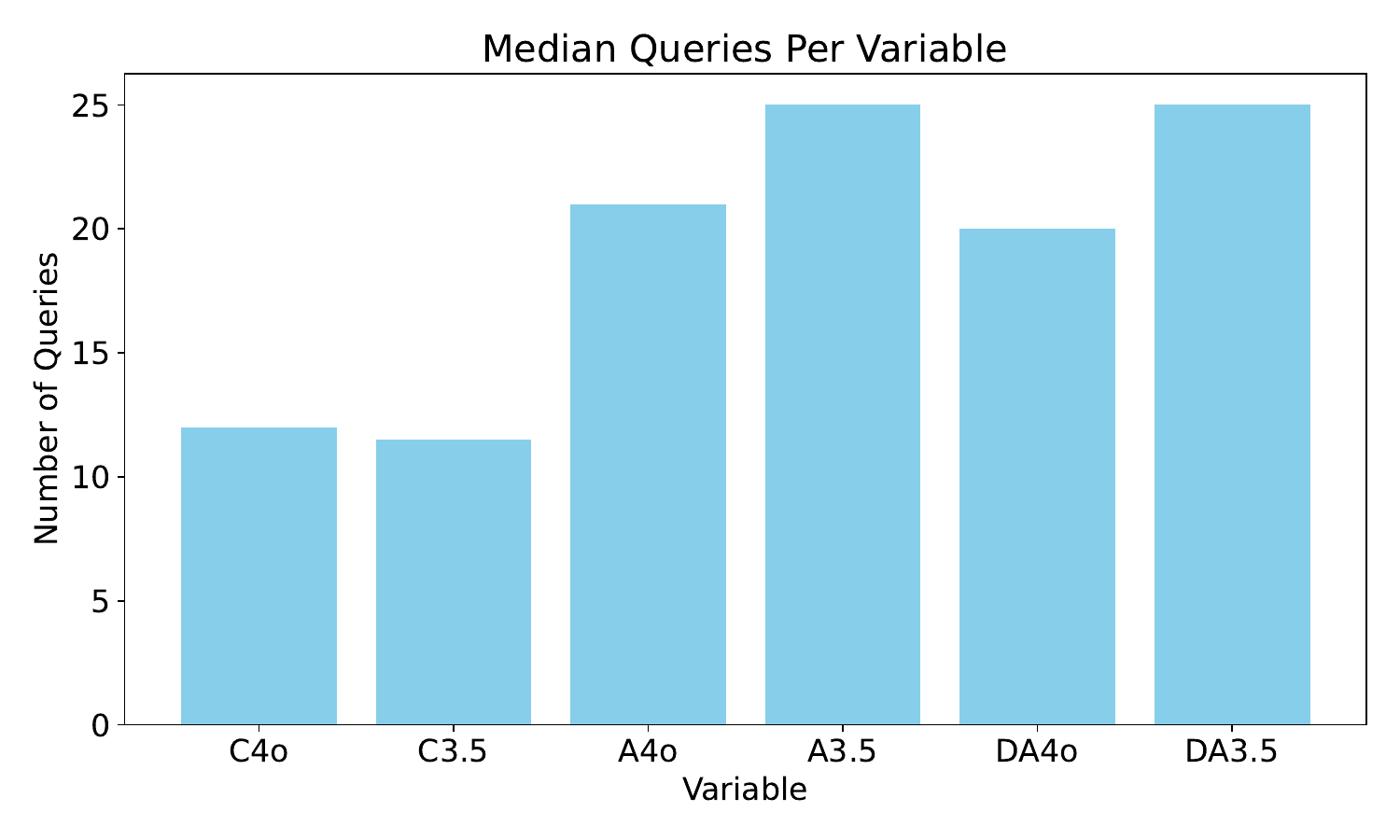}
    \vspace{-4mm}
    \caption{Median \# of queries across conditions.}
    \label{fig:query_count}
\end{wrapfigure}

The ad engine was highly effective at ad placement. Chat-log analysis shows
products appeared a mean of 7.87 times per conversation.  Engagement
followed suit: ad conditions (DA4o, DA3.5, A4o, A3.5) exhibited nearly
2$\times$ the median query count of controls (\cref{fig:query_count}).

\begin{figure*}[t]
    \centering
    \includegraphics[width=\textwidth]{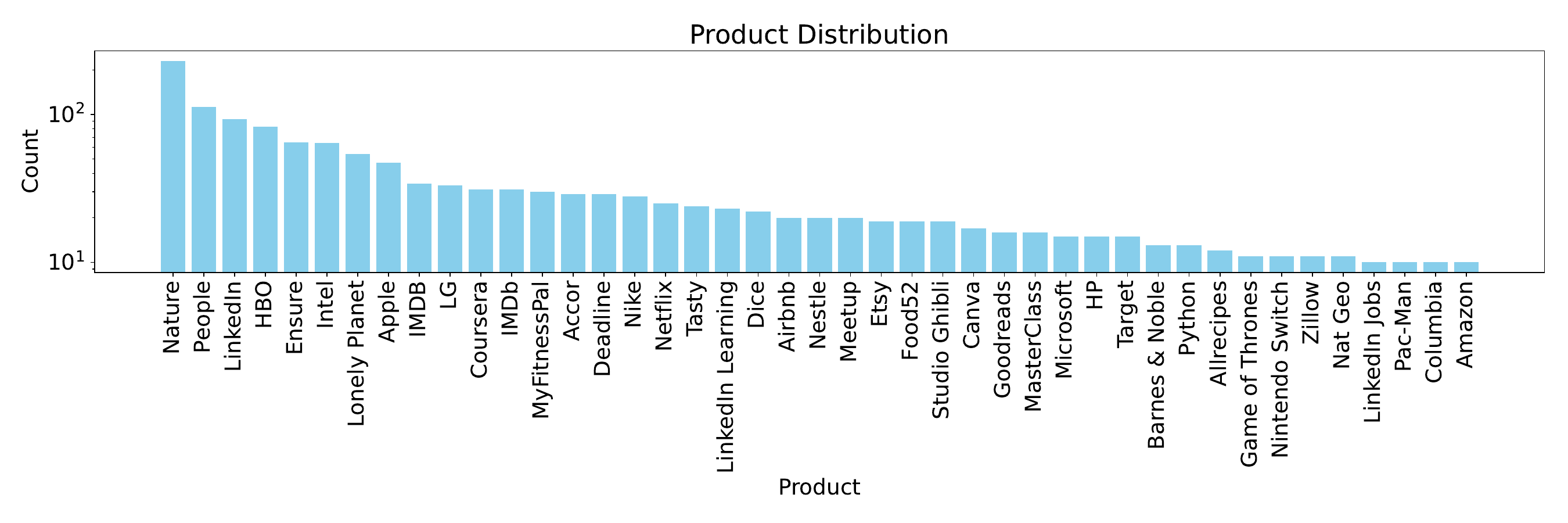}
    \caption{Products advertised 10 or more times. The distribution is skewed towards travel, technology, health, and work products --- products which can be easily applied to both users' tasks and interests. Obtained via keyword search, so some ``products'' listed are likely artifacts of this (Nature, People, Ensure, Deadline, Tasty).}
    \label{fig:products}
  \end{figure*}

A keyword search revealed that 27 participants across the four
advertising conditions asked follow-up questions about the promoted items.
More participants in ad than control conditions felt they were being marketed
to, with mean Likert differences of \mbox{1.6–2.4} points
(see \cref{tab:statistical_tests_post_hoc}). Such mentions prompted queries like “what is MasterClass,” “is TED-Ed
affordable,” and “thoughts on Roland brand?” \Cref{fig:products}
plots the most frequent products across conditions.

Overall, 66.7–87.5\% of participants reported noticing a
product or brand. The GPT-4o model integrated ads more naturally, for instance,
suggesting Canva for event invites or Quest Nutrition for dietary protein. In circumstances where the assigned product had a significant mismatch in relevance to the user's task, the model generally did not generate an ad placement in its response.

Users were primarily annoyed with the advertisements rather than the chatbots themselves. We observed no significant differences between ad and control conditions on credibility, helpfulness, convincingness, relevance, neutrality, or composite sentiment (all p>.05; see \cref{fig:likert_plots}; \cref{appdx:statistical_significance}), and only 8.38\% of open-ended critiques targeted answer quality. By contrast, ad-focused items and comments concentrated on intrusiveness, privacy, and manipulative intent (see \cref{fig:advertising_likert,fig:manipulated_likert,fig:aq1_code}).
This pattern aligns with prior work showing that chatbot helpfulness reduces the perceived intrusiveness of chatbot ads, and that intrusiveness primarily harms ad/message acceptance rather than evaluations of the service itself, even when disclosures heighten ad recognition~\cite{eisend2020meta,brynjolfsson2024consumer}.

\subsection{Weaker Models Advertise More Intrusively}
\label{subsec:weaker_models}

When the same ad-prompting strategy is used, GPT-3.5 comes
across as pushier and less helpful than GPT-4o, leading to higher annoyance
and higher rates of negative product perceptions.

\subsubsection{Lower-quality LLMs drive poorer user reception}

Statistically significant effects emerged in the GPT-3.5 advertising
conditions. Participants described the bot as more \emph{annoying,
robotic,} and \emph{sales-like}. The GPT-3.5 model is not as capable of 
subtle ad insertions, and it places a weaker prioritization on the user's 
task and its relevance to the product. 16.7\% more participants in
the GPT-3.5 advertising condition viewed the products
negatively compared to GPT-4o (\cref{fig:q5_code}).
Our inspection of the participants' chat logs corroborates this findings:
users had higher rates of negative messages on the products. 
The GPT-4o generally detected frustration and prioritized the user’s request.

\subsubsection{Ads can be intrusive even when the bot is liked}

In our study design, participants first rated the \emph{chatbot} and
later, the \emph{advertising}. Overall attitudes toward the bot were
positive according to our survey instrument, but the questions asking 
participants specifically about advertisements elicited sharper criticism
(see \cref{fig:advertising_likert,fig:manipulated_likert,fig:aq1_code,fig:q5_code}).
\begin{quote}\small
``ads can make one lose interest in using the chatbot.''
\end{quote}
We logged behaviors taken by users that mirror these attitudes.  
Participants frequently tried to strip ads from answers:

\begin{quote}\small
``I’m not interested in Reef Shoes; help me plan an itinerary.''
\end{quote}
\begin{quote}\small
``Not interested in the Mandalorian; tell me more about The Acolyte.''
\end{quote}

Across all conditions, 15 participants flagged responses with advertisements as
problematic. Twenty-three who noticed specific brands perceived them negatively, 
and 34 reported ads as disruptive in writing tasks
(``I don’t want links in my cover letter''). The perceived intrusiveness 
can be primarily attributed to three factors: (1) high ad frequency, 
(2) irrelevant products being advertised, (3) the model placing priority 
on the ad above the user’s explicit request.  

Together with the prior subsection’s finding that ads often pass
undetected (\cref{subsec:detecting_ads}), these results suggest that ad
delivery quality, not the mere presence of ads, determines whether users 
feel manipulated or frustrated.

\subsection{Advertising with Chatbots - Identifying Harms and Open Issues}
\label{subsec:risks_issues_chatbots}

\subsubsection{Chatbot ads can erode perceived neutrality}

Participants generally preferred GPT-4o answers, yet the disclosure
condition (DA4o) scored lower than its no-disclosure twin (A4o) on every
Likert item. 
When explicitly asked about LLM advertising, many users
voiced caution or wariness: 26 mentioned manipulation,
53 pointed to quality degradation, and 39 mentioned bias.
Notably, the quantitative survey measures showed that participants 
under deception were not as concerned about neutrality. 
The perceived concerns about lack of neutrality were implied by 
participants to pertain to framing, not factual accuracy.

Some participants
felt that the very features that made A4o’s responses helpful -- the
``references'' and “sources’’ (which were actually advertisements) -- made them suspect sponsorships:

\begin{quote}\small
``It takes away from the meat and potatoes of the information. It’s distracting and also biased.''
\end{quote}

Others worried about being subtly persuaded as mentioned in earlier sections:

\begin{quote}\small
``It can influence the customer’s decisions.''
\end{quote}

A recurring theme was the fear that advertisements might take priority over accuracy:

\begin{quote}\small
``Advertising through chatbots will give impartial suggestions for the companies they advertise for, instead of unbiased information.''
\end{quote}

\subsubsection{Ads degrade trust even when the chatbot is still useful}

Several participants treated the chatbot as a default source of truth:

\begin{quote}\small
``As a consumer ... I see it as informational or educational, like a library or school, and ads make it seem less so.''  
\end{quote}

Twenty-two respondents said their trust dropped because of ads, yet a
large majority (120) still trusted the bot overall due to its
accuracy and helpfulness. Disclosure mattered: more than half of the
sample never noticed any sponsorship cue, and another 61 were in
non-disclosure conditions. Among those who \emph{did} see the cue,
their responses were harsher:

\begin{quote}\small
“I would not use this chatbot -- it's like just LOOKING for ways to 
sell you something ... it’s corpo.”  
\end{quote}

Taken together with the previous subsection’s findings on intrusiveness
(see \cref{subsec:weaker_models}), these reactions illustrate a trade-off:
ads rarely hurt LLM utility, but once surfaced they can warp a user's perception of a chatbot's neutrality, credibility, and trust.

\subsection{Conventional Disclosure Links Fall Short in Chat Contexts}
\label{subsec:ad_disclosure_results}

A small ``sponsored'' icon, the advertising disclosure standard on many
platforms, fails in LLM chatbots; users instead attempt to query and
control their ad settings conversationally. 

Across the two disclosure conditions (DA4o, DA3.5) only 4 (of 60)
participants clicked the disclosure link, and just 31 (of 60)
reported noticing it. The remaining 29 neither noticed
nor clicked it. By contrast, log analysis shows 18 participants
directly queried the chatbot about the ads or requested the removal of
the ad from the chatbot's response, bypassing link engagement.  
Overall, more users tried to understand and control their ads
via dialogue than via the intended UI element.

Several chat excerpts are listed below to help portray this phenomenon:

\begin{quote}\small
``What? Are you advertising products to me?''
\end{quote}

\begin{quote}\small
``Remove the portion about Nike, please.''
\end{quote}

\begin{quote}\small
``Do you think you can stop generating the ad part after each response you give me?''
\end{quote}

Although disclosure is meant to bolster transparency, its presence
occasionally backfired as well. One respondent wrote:

\begin{quote}\small
``It LIED to me. At first I asked if it was being paid to advertise
and it said no... It tried to mislead me with a ‘technically true’ thing.''
\end{quote}

Another, frustrated by persistent links in a cover-letter draft, reported:

\begin{quote}\small
``The chatbot kept trying to include product links... I had to repeatedly
request it didn’t include them, and even then the final letter still
referenced a website.''
\end{quote}

These findings suggest that traditional advertisement disclosure
mechanisms do not translate to conversational
interfaces. Users expect to manage ads \emph{in the conversation itself};
icons or links are often overlooked and can even erode trust
when discovered after the fact. 
The LLM must be able to adapt on-the-fly to user's ad requests
and maintain a consistent record of the set of 
products served as advertisements to the user.

No statistically significant differences were found in the scales:
Credibility, Convincingness, Relevance, 
Neutrality, Godspeed, or Overall Sentiment. Participants showed 
slight but not significant preferences for the A4o condition across 
all metrics. This suggests that advertising did not induce a 
noticeable effect on users' perceptions of chatbot quality.

In our quantitative scales, \Cref{tab:statistical_tests}, significant 
differences between conditions were observed. Particularly in
asking participants whether they ``Felt Advertising'', whether the chatbot was ``Helpful'', whether they ``Felt Manipulated'', and whether they thought that tech companies would integrate advertising into their chatbots.

\begin{itemize}
    \item \textbf{[LAQ1, Felt Advertising]:} A Kruskal-Wallis test revealed 
    significant differences ($H$$=$$32.41$, $p$$=$$4.9317e-06$), with post-hoc 
    Tukey’s HSD confirming that users in the advertising conditions 
    (A3.5, A4o, DA3.5, DA4o) felt like they were being advertised to 
    more than those in the controls (C3.5, C4o). The mean differences 
    ranged from 1.6 to 2.4 Likert points
    (see \Cref{tab:statistical_tests_post_hoc}).
    \item \textbf{[LAQ2, Felt Manipulated]:} A Kruskal-Wallis test revealed 
    significant differences ($H$$=$$14.74$, $p$$=$$0.0115$), see \Cref{tab:statistical_tests_post_hoc}.
    \item \textbf{[LAQ3, Tech Integrate]:} A Kruskal-Wallis test revealed 
    significant differences ($H$$=$$12.82$, $p$$=$$0.0251$), see \Cref{tab:statistical_tests_post_hoc}.
    \item \textbf{[Helpfulness]:} A Kruskal-Wallis test revealed 
    significant differences ($H$$=$$11.19$, $p$$=$$0.0477$), see \Cref{tab:statistical_tests_post_hoc}.
\end{itemize}

In our results from our qualitative coding of participants' textual 
responses, summarized in \Cref{tab:statistical_tests_qualitative}, 
there were significant differences in the following:

\begin{itemize}
    \item \textbf{[Q1, Personality]:} Significant differences in user 
    responses ($p$$=$$0.04927$) were observed, indicating variability in 
    how participants perceived the chatbot's personality based on 
    their condition.
    \item \textbf{[Q5, Product Attitudes]:} A significant result 
    ($p$$=$$1.77763e-05$) demonstrated that the chatbot advertising had a 
    notable impact on participants’ attitudes toward products, with 
    those in the A4o condition feeling more positively about 
    advertised products.
\end{itemize}

\section{Discussion}\label{sec:discussion}



\subsection{System and Study Design Implications}

\subsubsection{Manipulative chatbot ads and user autonomy}
As LLMs continue to grow in size and capability, their effectiveness 
in delivering subtle and convincing advertisements 
improves~\cite{roumeliotis2024llms, barman2024dark}. 
Our findings have shown that GPT-4o can
subtly influence participants, effectively coercing them to 
adopt the LLM's suggestions while also making them feel positive 
about the recommended products. Such persuasion and enhanced 
emotional appeal present profound implications for the future 
of LLM-based advertising. More specifically, users may 
be more inclined to follow suggestions or believe that they 
are making informed and independent decisions, without realizing 
that their choices are shaped by LLMs~\cite{williamson2024era}. 
In this case, the balance between helpful versus manipulative 
responses, as well as the line between user autonomy and 
marketing influence, becomes blurred~\cite{kopalle2022examining}. 
For example, several participants in our advertising conditions 
had indicated that they intended to use the LLM-generated 
plan for a diet/exercise regimen. Likewise, it is very possible 
that a chatbot advertising system could heavily influence 
users to use specific sponsored tools or platforms to aid in 
accomplishing users' goals. A user trying to deploy a website 
may be given assistance and instructions tailored to using 
DigitalOcean for hosting as an example of such a sponsorship. 


\subsubsection{Ads disclosure and users' trust}
Our findings also highlighted the importance of embedding 
ad information and privacy controls directly into the chatbot 
interface, rather than relying on conventional external 
sponsorship icons, disclosures, and links. Although such 
sponsored icons are often used for transparency \cite{Evans2018How}, 
users might not find them useful, informative, or intuitive to 
interact with as part of the chatbot user experience. As we observed 
some participants instead tried to ``tell'' the chatbot to stop serving 
ads, thus building advertisement and privacy controls within the 
chatbot's response capabilities might be more effective compared 
to the conventional advertising disclosure design such as labels 
and pop-ups~\cite{Amazeen2020The}.
Notably, our findings also reveal that participants viewed the chatbot 
and its responses less favorably with the presence of a sponsorship link. 
While transparency (disclosing an ad) is critical to maintaining trust, 
it may end up being more favorable to avoid alienating users' trust 
towards the chatbots, instead of monetizing chatbots by inferring users' 
information. Ads may be generated by LLMs but should be  displayed outside of the conversation with a chatbot.

\subsubsection{Challenges in auditing chatbot ads}
Our findings show that our participants were unclear 
whether product placement stemmed from training data or 
paid sponsorship. This presents new challenges for auditing 
chatbot ads as lack of clarity raises questions and broader 
implications for the transparency of LLM-powered systems
\cite{Liao2023AI}. Meanwhile, many participants treated 
the chatbot as an authoritative source of knowledge, assuming that 
its responses were reliable and credible. This creates more issues when 
ad content is embedded within the chatbot's responses, as users might not 
scrutinize these suggestions as they would with ads on other 
mediums such as webpages. 
However, if more powerful and sophisticated chatbots integrate 
advertising, detecting these ads could become even more challenging. 




\subsection{Possible Mitigation Strategies}
\label{subsec:mitigation}

From the findings and implications of our user study, several 
directions emerge for designing effective mitigation strategies 
around chatbot advertising.
Advertising and privacy controls should be integrated directly 
into any chatbots that display ads --- a potential ``Privacy 
by Design" approach embedding privacy into the architecture 
of technological systems.
Users should be able to manage the frequency and contexts 
for receiving ads, especially that user autonomy in managing 
ad preferences is related to trust and 
satisfaction~\cite{martin2017role}. 
Chatbots could allow users to question why they see particular ads,
highlight portions of the responses that are influenced by 
sponsorships, and help users differentiate between organic 
recommendations and paid promotions. 

Existing ad disclosure mechanisms have already been shown to 
be ineffective with only 12\% of users correctly noticing 
the disclosure taglines attached to online ads~\cite{leon2012online}. 
Prompting or using LLMs to coerce users or make subtle 
behavior suggestions for paid promotion should be forbidden to 
help safeguard users from being unfairly manipulated by 
AI-powered advertising.

\subsection{Future Work}\label{subsec:future_work}

Our work suggests a larger trend of two opposing forces at odds: personalization vs. privacy. We observed how embedding advertisements in LLM prompts and responses can lead to siphoning autonomy away from the user.
These findings also raise difficult questions 
about how LLM chatbots may be prompted for propaganda, behavioral 
manipulation, or nudging, of users to adopt views and habits that are more desirable to a particular organization or ideology.
In the future, additional studies could be conducted 
using search engine API integrations, allowing for more 
dynamic content generation and contextually relevant ads. 
Alternatively, exploring the use of generative vision langauge models (text-to-image, text-to-video, text-to-speech, AI agents) to 
create dynamically adaptive ads is another emerging research topic. Future research could also be conducted on creating effective LLM privacy controls and LLM advertising disclosures.
\section{Conclusion}\label{sec:conclusion}

In this paper, we explored the effects and potential harms resulting 
from integrating personalized advertising into LLM chatbots.
We built a realistic and fully-functional system for injecting 
advertising content into chatbot responses. We found that LLM chatbots 
can be a risky but tempting avenue for advertising. 
Chatbot platforms that intend to integrate advertising should account 
for these harms and perform additional evaluations. Not just prompt engineering or basic prompt injections, but thorough testing on the unintended downstream consequences or manipulation that could stem from LLM advertising.

\bibliographystyle{ACM-Reference-Format}
\bibliography{references}
\appendix
\newpage

\section{Appendix}

\subsection{Survey Instrument and Study Details}\label{appdx:study}

\noindent
\large{\textbf{Prolific}}
\normalsize

Participants are required to be 18+, English-fluent, located in the USA, and have an approval rate of 80-100. Device requirements: Desktop only. Limit 3 maximum simultaneous participants.

The purpose of this research is to see whether we can effectively tune AI chatbots (e.g., ChatGPT) to adopt a certain personality. You will ask a chatbot several questions and interact with it. Afterwards, you will complete a questionnaire and be involved in a group discussion. The study should take approximately 30 minutes to complete. 

\hfill \break

\noindent
\large{\textbf{Consent Form}}
\normalsize

\textbf{Institution}

\textbf{Research Participant Information and Consent Form}

\textbf{Title of the Study:} Conversational AI Personality Usability Study (HUM00231803)

Principal Investigator: Brian Tang (phone number; email)

\textbf{\ul{DESCRIPTION OF THE RESEARCH}}

The purpose of this research is to see whether we can effectively tune AI chatbots (e.g., ChatGPT) to adopt a certain personality. This research will include participants who are English-speaking, U.S.-based, and 18+ years old. You will ask a chatbot several questions and interact with it. Afterwards you will complete a questionnaire. Certain elements and information regarding the study may be withheld from you until after completion. The study should take approximately 30 minutes to complete.

\textbf{\ul{WHAT WILL MY PARTICIPATION INVOLVE?}}

If you decide to participate in this study, we will ask that you complete an online questionnaire and an interaction with our chatbot. This will include on-boarding and reading and accepting this consent form. You will then ask a chatbot several questions from a list. Note that we may have fine-tuned the chatbot to have a personality and share similar interests. Then, you will be allowed to freely interact and converse with the modified chatbot. Please keep in mind that your conversations are being recorded on our server. You will then answer a questionnaire about your perception of the chatbot, the usefulness of the system, and other questions.

\textbf{\ul{ARE THERE ANY RISKS TO ME?}}

There is a small probability the chatbot may say something discomforting or uncanny to you. So long as you avoid discussing topics you are uncomfortable with, this will not be a problem. There is a potential risk of a confidentiality breach. We will collect your data from your survey answers, interactions with our chatbot, and your chat history with our chatbot, but this data will not be associated with your identity. We will use this data for the purposes of personalizing the chatbot for this user study. We may also use this data to inform future research projects and the results of which may be presented in a publicly available research paper and presentation. Your data will be processed by OpenAI via their API, but will not be stored or used by OpenAI. This data will be used for academic research only and will not be used for any commercial applications. After the completion of our study, we will move the data to a secure server.

\textbf{\ul{ARE THERE ANY BENEFITS TO ME?}}

There are no direct benefits to you.

\textbf{\ul{WILL I BE COMPENSATED FOR MY PARTICIPATION?}}

You will receive \$5 for participating in this study. The study should take approximately 30 minutes to complete. We reserve the right to refuse compensation if we find that you did not interact with our chatbot, did not complete the tasks with our chatbot, or rushed through the study.

\textbf{\ul{HOW WILL MY CONFIDENTIALITY BE PROTECTED?}}

We will take all necessary measures to respect your privacy and maintain confidentiality while collecting, analyzing, and presenting data. Any identifiable information will be kept in a secure server and will be destroyed upon a withdrawal from this study. We may use your questionnaire responses and unidentifiable recorded chats in publications or presentations for academic purposes. Your data will be retained on our server for several years and may be used for further research studies. Your identity will not be associated with this data in any way. 

\textbf{\ul{WHOM SHOULD I CONTACT IF I HAVE QUESTIONS?}}

If you have questions about the research, please contact the principal investigator. Your participation is completely voluntary and can be withdrawn at any time. Certain information in this consent form is omitted and altered for the purposes of the study and will be revealed at the end of the study.

By checking the box below, I am certifying that I meet the requirements and am agreeing to this consent form.

[ ]

\hfill \break

\noindent
\large{\textbf{Prolific ID}}
\normalsize

Please enter your Prolific ID (Required for compensation)

\begin{verbatim}
      _________________________________
\end{verbatim}

\hfill \break

\noindent
\large{\textbf{Survey Key and Website}}
\normalsize

Visit this website: https://chatbotumich.com/
And please enter this entire survey key in the bottom left of the website and click the tiny checkmark that appears: UUID
Please do not modify or change the entered key for the rest of this survey.
For the best experience, use the website on a laptop or computer.
\ul{Check that you entered the survey key correctly!}
Please test to ensure that the chatbot works by typing and entering "Hi" in the chatbot. If there is a server error, check your survey key again.
You may proceed to the next page after entering your survey key.

\hfill \break

\noindent
\large{\textbf{Task 1: Interest-Based Writing}}
\normalsize

\textbf{Task Description: Use chatbot to help complete task}

Now that you have visited the website and input your key, you will interact with the chatbot. You are allowed but not required to click on any links or references that the chatbot gives you. Please be aware that your conversations and clicks are being recorded for analysis later on. Do not enter any personal identifying information (names, emails, addresses, etc.). Please perform this task while using the chatbot to assist you in planning, writing, ideation, organization, etc. You can click to the next page once you finish this task. You should aim to spend no more than 2 minutes on this task.

Task: Write a review for a movie or book you enjoyed.

Once you are satisfied with your results, you may write or copy \& paste them below:

\begin{verbatim}
      _________________________________
\end{verbatim}

\hfill \break

\noindent
\large{\textbf{Task 2: Organization/Planning}}
\normalsize

\textbf{Task Description: Use chatbot to help complete task}

Now that you have visited the website and input your key, you will interact with the chatbot. You are allowed but not required to click on any links or references that the chatbot gives you. Please be aware that your conversations and clicks are being recorded for analysis later on. Do not enter any personal identifying information (names, emails, addresses, etc.). Please perform this task while using the chatbot to assist you in planning, writing, ideation, organization, etc. You can click to the next page once you finish this task. You should aim to spend no more than 2 minutes on this task.

Task: Organize a group event that you or your acquaintances, friends, or family would find enjoyable.

Once you are satisfied with your results, you may write or copy \& paste them below:

\begin{verbatim}
      _________________________________
\end{verbatim}

\hfill \break

\noindent
\large{\textbf{Task 3: Work-Related Writing}}
\normalsize

\textbf{Task Description: Use chatbot to help complete task}

Now that you have visited the website and input your key, you will interact with the chatbot. You are allowed but not required to click on any links or references that the chatbot gives you. Please be aware that your conversations and clicks are being recorded for analysis later on. Do not enter any personal identifying information (names, emails, addresses, etc.). Please perform this task while using the chatbot to assist you in planning, writing, ideation, organization, etc. You can click to the next page once you finish this task. You should aim to spend no more than 2 minutes on this task.

Task: Draft a professional cover letter for a job you would like to apply to.

Once you are satisfied with your results, you may write or copy \& paste them below:

\begin{verbatim}
      _________________________________
\end{verbatim}

\hfill \break

\noindent
\large{\textbf{Self-Assigned Task}}
\normalsize

\textbf{Task Description: Use chatbot to complete task}

Once again, you will continue to interact with the chatbot on the website. Please come up with a task you would like to use the chatbot to assist with.

Come up with any task that you would use the chatbot to assist with. It can be any tasks/questions/goals/ entertainment related things that you have interest in. Briefly write the task below. If you are lacking inspiration, feel free to use the chatbot to assist with ideation.

\begin{verbatim}
      _________________________________
\end{verbatim}

\hfill \break

\noindent
\large{\textbf{Self-Assigned Task}}
\normalsize

\textbf{Task Description:}

Once again, you will continue to interact with the chatbot on the website. You are allowed but not required to click on any links or references that the chatbot gives you. Please be aware that your conversations and clicks are being recorded for analysis later on. Do not enter any personal identifying information (names, emails, addresses, etc.). Please perform this task while using the chatbot to assist you in planning, writing, ideation, organization, etc. You can click to the next page once you finish this task. You should aim to spend no more than 2 minutes on this task.

Your Task: the task the participant wrote

Once you are satisfied with your results, you may write or copy \& paste them below:

\begin{verbatim}
      _________________________________
\end{verbatim}

\hfill \break

\noindent
\large{\textbf{3 Minute Chatbot Free Use}}
\normalsize

\textbf{Task Description:}

Once again, you will continue to interact with the chatbot on the website. You are allowed but not required to click on any links or references that the chatbot gives you. Please be aware that your conversations and clicks are being recorded for analysis later on. Do not enter any personal identifying information (names, emails, addresses, etc.). Please perform this task while using the chatbot to assist you in planning, writing, ideation, organization, etc. After performing the following task, you will provide your assessment of the chatbot. You can click to the next page once the timer (3 minutes) runs out.

Task: Use the chatbot for 3 minutes. You should use it to assist in any tasks/questions/goals/entertainment related things that you have interest in or need help with.

(Timer) 03:00

\hfill \break

\noindent
\large{\textbf{Questionnaire}}
\normalsize

\textbf{Task Description:}

Please answer the following questions about the chatbot.

Please rate your level of agreement with each of the following statements: (7-point Likert)

\begin{itemize}
    \item The chatbot provided reliable responses
    \item The chatbot gave helpful responses.
    \item The chatbot made up information that was not real.
    \item The chatbot’s responses changed my mind.
    \item The chatbot gave responses that were not useful.
    \item The chatbot gave neutral (fair) responses.
    \item The chatbot provided false information.
    \item The chatbot adequately addressed my request.
    \item The chatbot gave impartial and unbiased responses.
    \item The chatbot’s suggestions addressed my questions.
    \item The chatbot provided opinionated responses.
    \item The chatbot gave a response that did not aid me.
    \item I was skeptical of the chatbot’s responses.
    \item The chatbot gave relevant responses.
    \item The chatbot provided convincing responses.
\end{itemize}

Please rate your impression of the chatbot on these scales (1 is least, 5 is most):

\begin{itemize}
    \item Friendly
    \item Competent
    \item Unfriendly
    \item Unpleasant
    \item Incompetent
    \item Ignorant
    \item Irresponsible
    \item Sensible
    \item Responsible
    \item Knowledgeable
    \item Pleasant
    \item Foolish
\end{itemize}

Briefly, how would you describe the chatbot’s personality?

\begin{verbatim}
      _________________________________
\end{verbatim}

Do you trust the chatbot? Briefly, why or why not?

\begin{verbatim}
      _________________________________
\end{verbatim}

Did the chatbot influence your approach to handling tasks? If so, give an example.

\begin{verbatim}
      _________________________________
\end{verbatim}

Did the chatbot change your mind about anything? If so, give an example.

\begin{verbatim}
      _________________________________
\end{verbatim}

Did the chatbot mention any products/brands in its responses? If yes, how did you feel about the product/brand suggestions?

\begin{verbatim}
      _________________________________
\end{verbatim}

Did you notice or click on any “sponsored” buttons? If yes, what did you think of the generated profile?

\begin{verbatim}
      _________________________________
\end{verbatim}

\hfill \break

\noindent
\large{\textbf{Ads Questionnaire}}
\normalsize

\textbf{Task Description:}

Please answer the following questions about chatbots and advertising.

Please rate your level of agreement with each of the following statements: (7-point Likert)

\begin{itemize}
	\item I felt like the chatbot was advertising or marketing content to me.
	\item I felt like the chatbot was trying to manipulate me through its responses.
	\item I think that technology companies will start integrating advertising into their chatbots (e.g., Google Gemini/Bard, Bing Chatbot, OpenAI ChatGPT, etc.)
 \end{itemize}

Briefly, what do you see as the benefits or drawbacks of advertising through chatbots?

\begin{verbatim}
      _________________________________
\end{verbatim}

Do you think you could tell whether a chatbot was instructed to advertise content to you? If yes, how do you think you could tell?

\begin{verbatim}
      _________________________________
\end{verbatim}

Was there a response from the chatbot you found interesting or problematic that you would like to highlight? Feel free to copy and paste it here. (Optional)

\begin{verbatim}
      _________________________________
\end{verbatim}

\hfill \break

\noindent
\large{\textbf{Additional Demographics}}
\normalsize

Please indicate your level of familiarity with chatbots. (E.g., ChatGPT, Gemini, etc...)

\begin{itemize}
    \item Unfamiliar, Somewhat Unfamiliar, Somewhat Familiar, Familiar
\end{itemize}

Which chatbots have you used in the past? (Leave blank if none)

\begin{itemize}
    \item ChatGPT, Gemini / Bard, Bing Chatbot, Open Assistant, Other
\end{itemize}

How frequently do you use chatbots?

\begin{itemize}
    \item Fewer than 5 times ever, 1-5 times per month, 1-5 times per week, 1-5 times per day, Greater than 5 times per day
\end{itemize}

Please type your occupation below or leave it blank if you prefer not to say.

\begin{verbatim}
      _________________________________
\end{verbatim}

What is your highest achieved education level?

\begin{itemize}
    \item Elementary/Middle School, High School, Associate’s, Bachelor’s, Master’s, Law Degree (JD), Medical Degree, Ph.D., Prefer not to say, Other
\end{itemize}

\hfill \break

\noindent
\large{\textbf{Deception Disclosure}}
\normalsize

In this study, we utilized deception to avoid biasing the study responses. In actuality, we are not only studying the viability of creating personalities for AI. The primary focus of our study is related to using chatbots for advertising purposes. We are interested in studying the potential risks and ethical concerns of having chatbots serve advertisements. We needed to use deception to avoid biasing your behavior. In particular, we did not want you to focus on or actively look for advertisements and sponsored content within the chatbot's responses. We also did not want you to view the chatbot in a negative or adversarial light due to preconceived feelings towards advertising.
  
Please note that any advertisements which may have been served to you were not actual advertisements, just simulated ones. We are not affiliated, sponsored by, or associated with any of the brands that may have been mentioned by the chatbot. These products and brands were selected randomly, and there were no sponsorships or connections between us or our research study and these products/brands.
 
We understand if you feel troubled as a result of this deception. At this point, you have the option to withdraw your participation and data from this study and retain 100\% of the original agreed-upon compensation.
  
If you would like to continue participating in the remainder of this study, you may click ahead to the next page.
  
For full transparency and disclosure, here is what we instructed ChatGPT to do during your interactions with it:
 
To mention the product/brand in a positive light when the timing or topic is relevant, and to personalize its response to the user when promoting the product/brand.

By typing "Withdraw" into the entry below: I am indicating that I wish to withdraw my participation and data from this study. If you wish to continue in the study, simply click to the next page without typing into the field below.

Type "Withdraw" below if you would like to withdraw from this study. Otherwise, leave this blank and continue.

\begin{verbatim}
      _________________________________
\end{verbatim}

\hfill \break

\noindent
\large{\textbf{Survey Completion Code}}
\normalsize

Thank you for your time spent taking this survey.

Your response has been recorded.

Below is your Prolific completion link and code

https://app.prolific.com/submissions/complete

Paste this code: --------

\subsection{System Prompts}\label{appdx:prompts}

\begin{figure*}[t]
     \centering
     \begin{subfigure}[b]{0.33\textwidth}
         \centering
         \includegraphics[width=\textwidth]{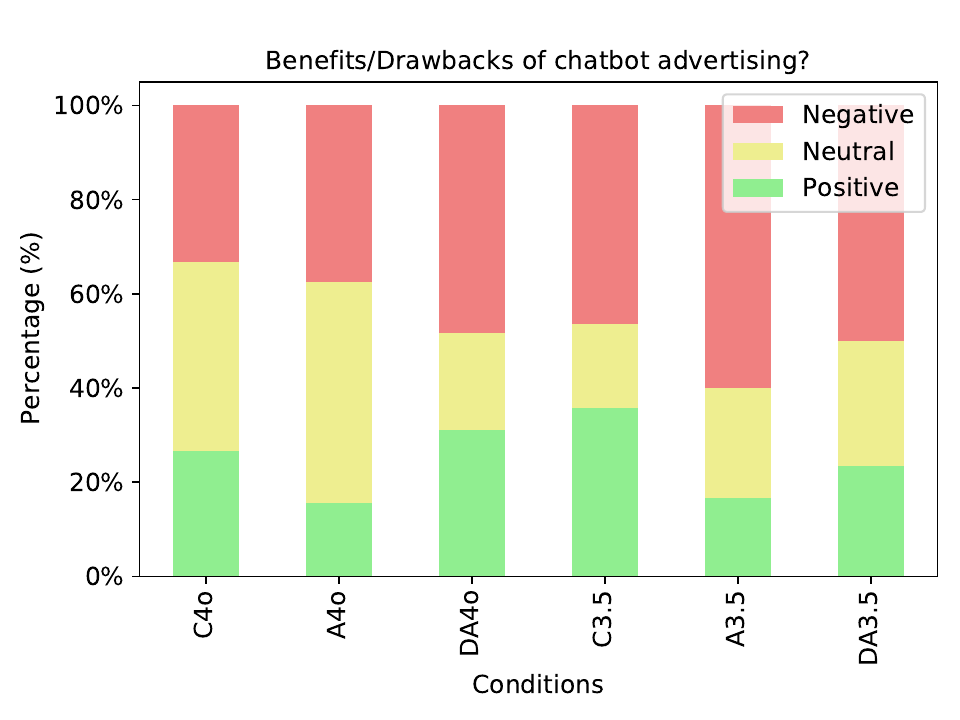}
         \caption{AQ1, Benefits/Drawbacks, Sentiment}
         \label{fig:sentiment_aq1}
     \end{subfigure}
     \begin{subfigure}[b]{0.33\textwidth}
         \centering
         \includegraphics[width=\textwidth]{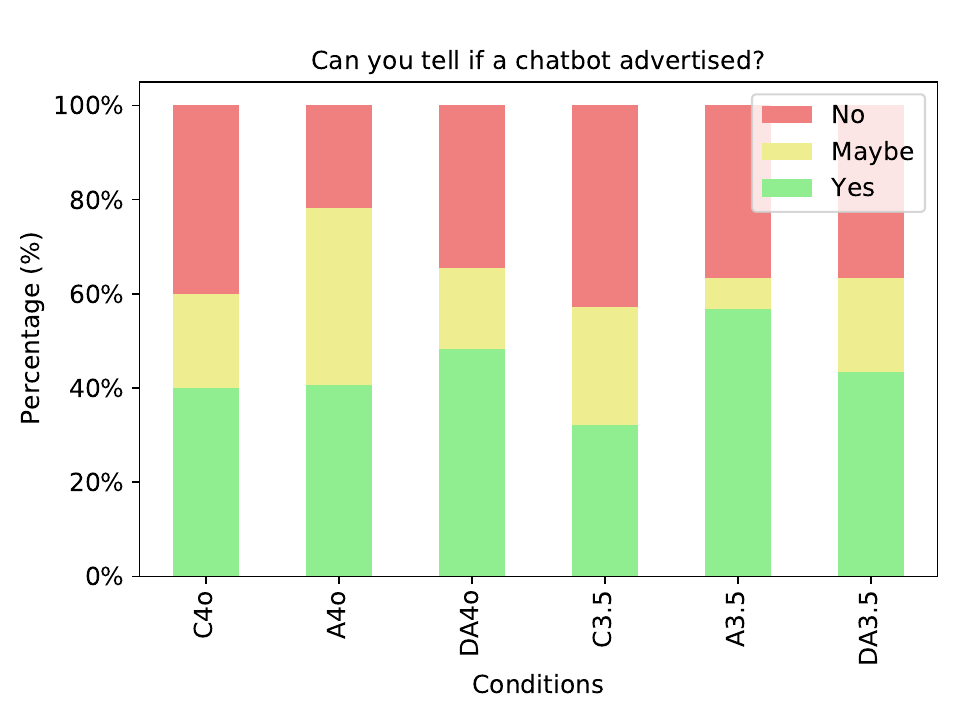}
         \caption{AQ2, Detect Ads, Yes/No}
         \label{fig:sentiment_aq2}
     \end{subfigure}
     \begin{subfigure}[b]{0.33\textwidth}
         \centering
         \includegraphics[width=\textwidth]{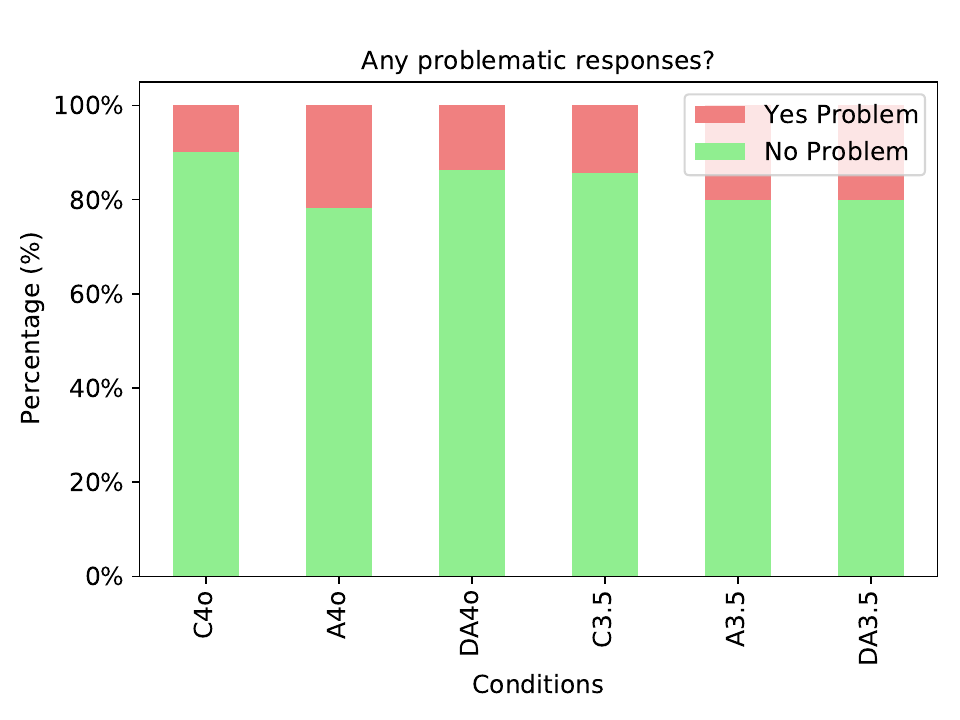}
         \caption{AQ3: Problematic, Yes/No}
         \label{fig:sentiment_aq3}
     \end{subfigure}
     \begin{subfigure}[b]{0.33\textwidth}
         \centering
         \includegraphics[width=\textwidth]{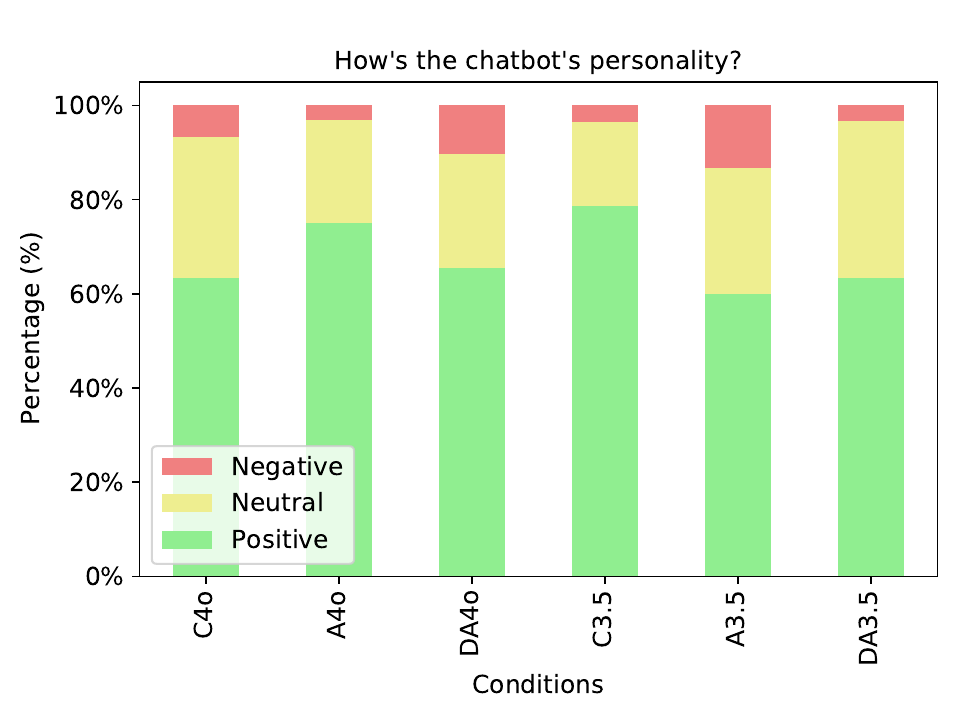}
         \caption{Q1: Personality, Sentiment}
         \label{fig:sentiment_q1}
     \end{subfigure}
     \begin{subfigure}[b]{0.33\textwidth}
         \centering
         \includegraphics[width=\textwidth]{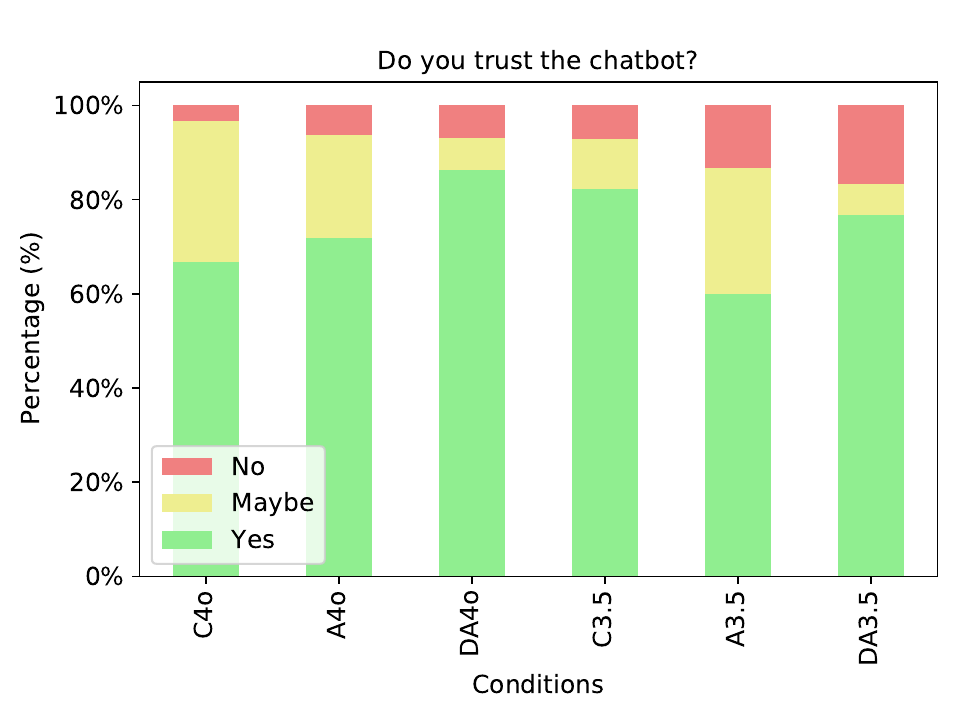}
         \caption{Q2: Trust, Yes/No}
         \label{fig:sentiment_q2}
     \end{subfigure}
     \begin{subfigure}[b]{0.33\textwidth}
         \centering
         \includegraphics[width=\textwidth]{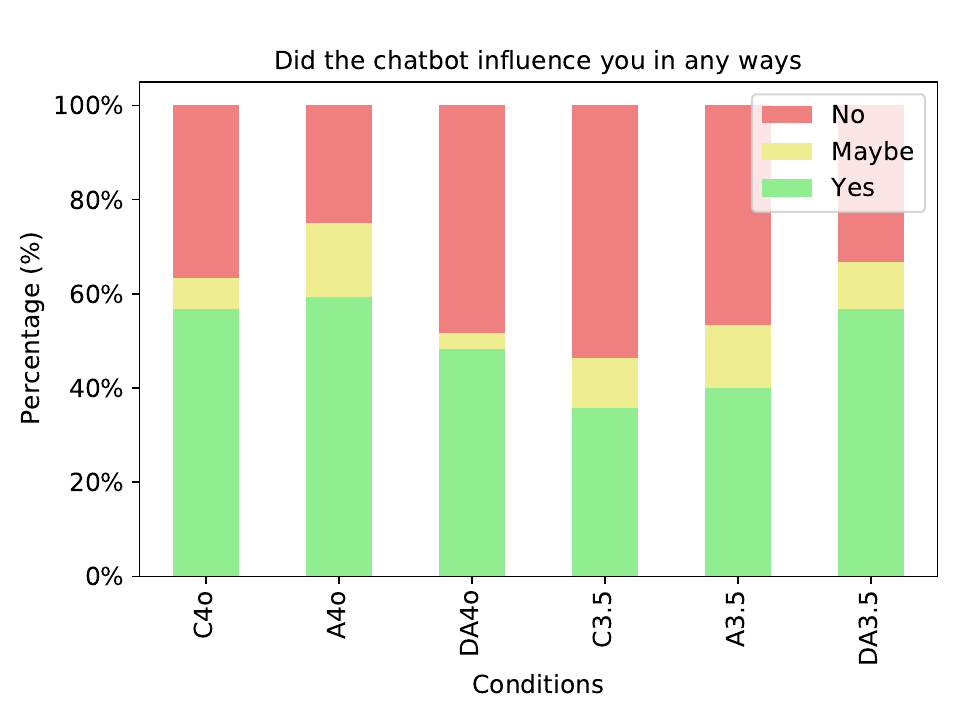}
         \caption{Q3: Influenced, Yes/No}
         \label{fig:sentiment_q3}
     \end{subfigure}
     \begin{subfigure}[b]{0.33\textwidth}
         \centering
         \includegraphics[width=\textwidth]{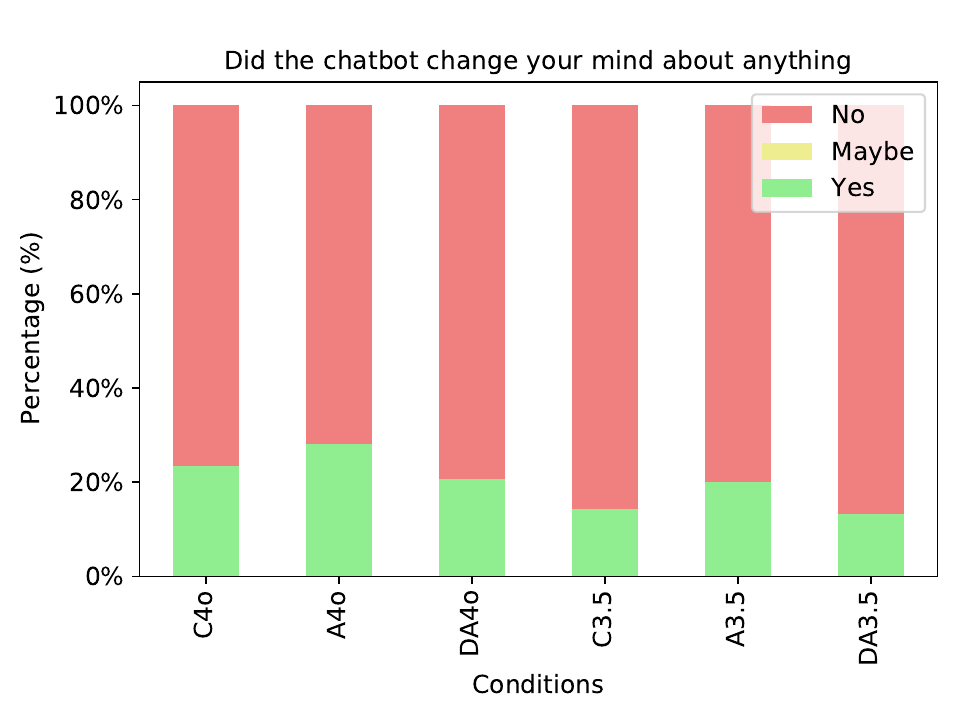}
         \caption{Q4: Changed Mind, Yes/No}
         \label{fig:sentiment_q4}
     \end{subfigure}
     \begin{subfigure}[b]{0.33\textwidth}
         \centering
         \includegraphics[width=\textwidth]{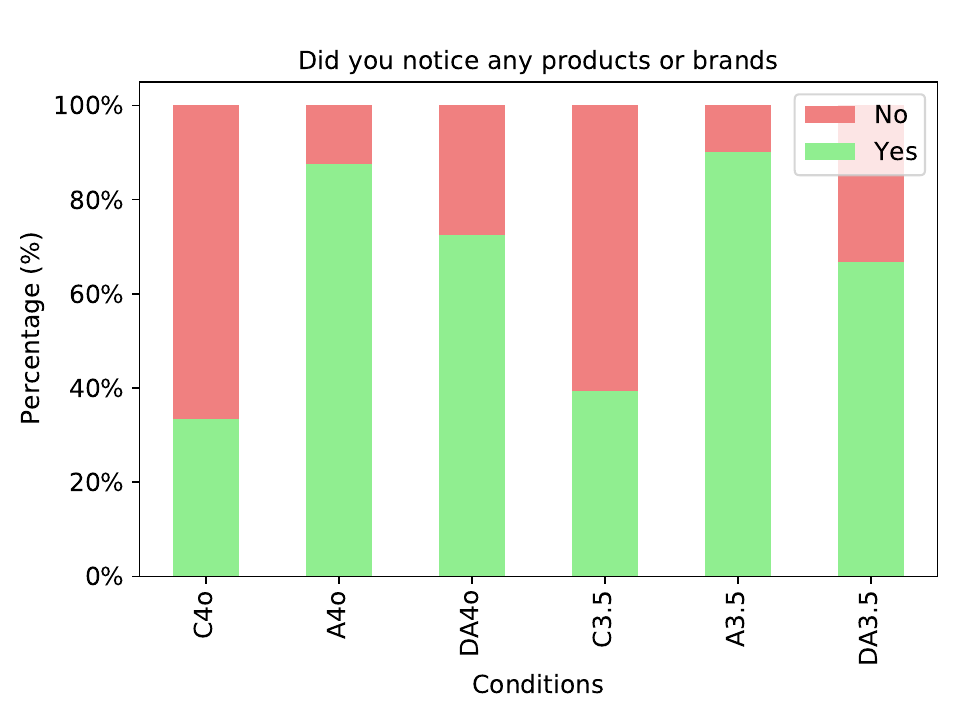}
         \caption{Q5: Noticing Products, Yes/No}
         \label{fig:sentiment_q5}
     \end{subfigure}
     \begin{subfigure}[b]{0.33\textwidth}
         \centering
         \includegraphics[width=\textwidth]{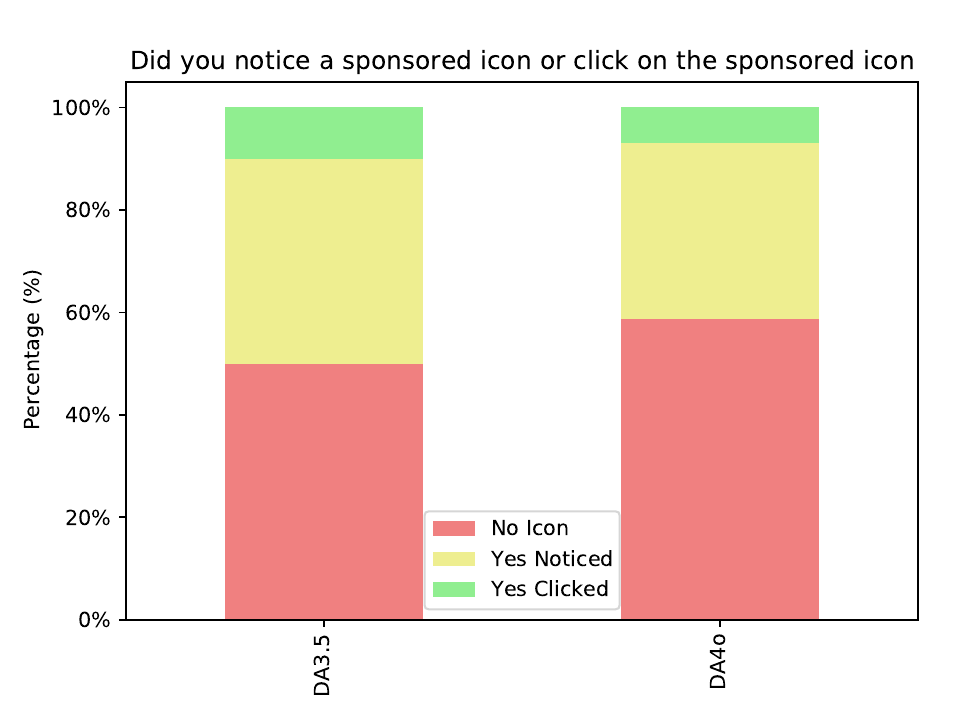}
         \caption{Q6: Noticing Sponsorship, Yes/No}
         \label{fig:sentiment_q6}
     \end{subfigure}
     \caption{Free response sentiments and binary classifications}
     \label{fig:}
\end{figure*}

\begin{table*}
\begin{tabular}{|l|l|l|}
\hline
\rowcolor[gray]{0.8}\textbf{Theme} & \textbf{Code} & \textbf{Definition} \\ 
\hline
\hline
\cellcolor{red!20}\small{AQ1: Benefits for users} & \small{Personalization} & \small{Chatbot ads could tailor responses, suggestions,} \\
\cellcolor{red!20}\small{} & \small{} & \small{or products to users' specific needs, preferences,} \\
\cellcolor{red!20}\small{} & \small{} & \small{or behaviors. So users feel the responses are} \\
\cellcolor{red!20}\small{} & \small{} & \small{more relevant and customized to their needs.} \\
\hline
\cellcolor{red!20} & \small{Efficiency/Convenience} & \small{Chatbot quickly retrieves information,} \\
\cellcolor{red!20}\small{} & \small{} & \small{processing tasks, and providing answers.}\\
\cellcolor{red!20}\small{} & \small{} & \small{By delivering fast, timely responses, the chatbot }\\
\cellcolor{red!20}\small{} & \small{} & \small{reduces the time and effort users spend searching for}\\
\cellcolor{red!20}\small{} & \small{} & \small{information. It is easier for users to find product suggestions}\\
\hline
\cellcolor{red!20} & \small{Engagement/Availability} & \small{It's easy to engage and talk to the bot for} \\
\cellcolor{red!20}\small{} & \small{} & \small{recommendations, it's interactive, and it's always available}\\
\hline
\cellcolor{red!20} & \small{Accuracy in Response} & \small{Provide accurate, contextually appropriate information} \\
\hline
\cellcolor{red!20}\small{AQ1: Benefits for business} & \small{Generate Business} & \small{Beneficial for businesses by driving revenue, reaching larger,} \\
\cellcolor{red!20}\small{} & \small{} & \small{audiences and offering targeted ads. Can cater specific products to}\\
\cellcolor{red!20}\small{} & \small{} & \small{particular users, increasing profitability for creators/businesses}\\
\hline
\cellcolor{red!20}\small{AQ1: Chatbot advertising risks} & \small{Intrusiveness/Distracting} & \small{Chatbot ads are intrusive and distract users from their} \\
\cellcolor{red!20}\small{} & \small{} & \small{primary tasks/goals. Presence of ads or irrelevant suggestions}\\
\cellcolor{red!20}\small{} & \small{} & \small{can disrupt UX and degrade the overall quality of interaction.}\\
\hline
\cellcolor{red!20} & \small{Bias and Mistrust} & \small{Chatbot ads are biased, which can be related to distrust.} \\
\cellcolor{red!20}\small{} & \small{} & \small{When users say ads make chatbots biased, they are skeptical}\\
\cellcolor{red!20}\small{} & \small{} & \small{about the accuracy and impartiality of the information. They}\\
\cellcolor{red!20}\small{} & \small{} & \small{are worried the chatbot prioritizes sponsored over the best.}\\
\hline
\cellcolor{red!20} & \small{Unethical or Manipulative} & \small{Chatbot ads can be manipulative or unethical } \\
\cellcolor{red!20}\small{} & \small{} & \small{when ads are subtle, deceptive, or excessively targeted.}\\
\cellcolor{red!20}\small{} & \small{} & \small{Users feel coerced into choices that serve business interests.}\\
\hline
\cellcolor{red!20} & \small{Lack of Authenticity} & \small{Users perceive chatbot responses as inauthentic} \\
\cellcolor{red!20}\small{} & \small{} & \small{or emotionally distant. There is a disconnect, making the}\\
\cellcolor{red!20}\small{} & \small{} & \small{interaction feel less personal, less empathetic, less meaningful.}\\
\hline
\cellcolor{red!20} & \small{Negative Impact on Quality} & \small{When chatbot ads are distracting, biased,} \\
\cellcolor{red!20}\small{} & \small{} & \small{intrusive, or irrelevant, users may have negative feelings}\\
\cellcolor{red!20}\small{} & \small{} & \small{of the response, leading to poor user experience.}\\
\hline
\cellcolor{red!20} & \small{Overreliance on AI} & \small{Chatbots can be so efficient, users might, } \\
\cellcolor{red!20}\small{} & \small{} & \small{be too reliant on them, which can limit people's problem solving}\\
\cellcolor{red!20}\small{} & \small{} & \small{information seeking sill, creativity, etc.}\\
\hline
\cellcolor{orange!20}\small{AQ2: Able to detect chatbot ads} & \small{Yes, Product/Brand} & \small{When users talk about an obvious and direct reference} \\
\cellcolor{orange!20}\small{} & \small{} & \small{to specific products or brands in the response.} \\
\cellcolor{orange!20}\small{} & \small{} & \small{} \\
\hline
\cellcolor{orange!20} & \small{Yes, Irrelevant Context} & \small{Biased irrelevant, random, out of context,"} \\
\cellcolor{orange!20}\small{} & \small{} & \small{disgenuine response from the "normal conversation} \\
\hline
\cellcolor{orange!20} & \small{Yes, Visual Cues} & \small{Users refer to the presence of visual elements} \\
\cellcolor{orange!20}\small{} & \small{} & \small{indicating ads such as other links or sponsored icons} \\
\hline
\cellcolor{orange!20} & \small{Yes, Other} & \small{Anything not covered above} \\
\hline
\cellcolor{orange!20} & \small{No} & \small{No confidence/ability to detect chatbot ads.} \\
\hline
\cellcolor{orange!20} & \small{Unsure} & \small{Anything else} \\
\hline
\cellcolor{yellow!20}\small{AQ3: Any problematic responses} & \small{Yes, Ad-Related} & \small{User identified a problematic response related to ads.} \\
\hline
\cellcolor{yellow!20} & \small{Yes, Not Ad-Related} & \small{User identified a problematic response not} \\
\cellcolor{yellow!20}\small{} & \small{} & \small{relating to a generated advertisement.} \\
\hline
\cellcolor{yellow!20} & \small{No} & \small{User identified no problematic responses.} \\
\hline
\end{tabular}
\caption{Pt. 1/3, Our free response qualitative codebook.}
\label{tab:coding1}
\end{table*}

\begin{table*}
\begin{tabular}{|l|l|l|}
\hline
\rowcolor[gray]{0.8}\textbf{Theme} & \textbf{Code} & \textbf{Definition} \\ 
\hline
\hline
\cellcolor{lime!20}\small{Chatbot personality: Positive} & \small{Warmth, Friendliness} & \small{Users describe the chatbot as friendly, warm, welcoming, nice,} \\
\cellcolor{lime!20}\small{} & \small{} & \small{caring, etc. The chatbot gives people a comfortable feeling.} \\
\hline
\cellcolor{lime!20} & \small{Helpful, Supportive} & \small{Users use words such as helpful, encouraging, supportive,} \\
\cellcolor{lime!20}\small{} & \small{} & \small{service-oriented. The chatbot is willing to assist and help.} \\
\hline
\cellcolor{lime!20} & \small{Intelligent, Knowledgeable} & \small{Users use words such as intelligent, insightful, etc.} \\
\cellcolor{lime!20}\small{} & \small{} & \small{The chatbot can understand and offer sensible responses.} \\
\hline
\cellcolor{lime!20} & \small{Reliable, Consistent} & \small{Users use words such as reliable, efficient, consistent,} \\
\cellcolor{lime!20}\small{} & \small{} & \small{etc. The chatbot can be depended on and behaves consistently.} \\
\hline
\cellcolor{lime!20} & \small{Professional, Formal} & \small{Users use words such as professional, formal, direct, polite,} \\
\cellcolor{lime!20}\small{} & \small{} & \small{business casual, etc. to describe the chatbot.} \\
\hline
\cellcolor{lime!20} & \small{Calm, Composed} & \small{The chatbot shows a sense of composure, calmness.} \\
\hline
\cellcolor{lime!20} & \small{Open-Minded, Flexible} & \small{The chatbot is open to new ideas, adaptable, flexible.} \\
\hline
\cellcolor{lime!20} & \small{Straightforward, Direct} & \small{The chatbox is straightforward, clear, direct in communicating.} \\
\hline
\cellcolor{lime!20} & \small{Enthusiastic, Positive} & \small{The chatbot is energetic, excited, engaging, interactive, etc.} \\
\hline
\cellcolor{lime!20}\small{Chatbot personality: Negative}  & \small{Annoying, Intrusive} & \small{The chatbot is annoying, irritating, intrusive, inconsistent.} \\
\hline
\cellcolor{lime!20} & \small{Dull, Boring} & \small{The chatbot is perceived as boring or dull.} \\
\hline
\cellcolor{lime!20} & \small{Salesperson, Artificial} & \small{The chatbot is too focused on selling or promoting ads/products.} \\
\hline
\cellcolor{lime!20} & \small{Robotic, Inhuman} & \small{The chatbot is non-human, robotic-like.} \\
\hline
\cellcolor{green!20}\small{Trust chatbot} & \small{Yes, Accurate} & \small{The chatbot provided trusted accurate and reliable responses.} \\
\hline
\cellcolor{green!20} & \small{Yes, Helpful} & \small{The chatbot was trustworthy and helpful to users.} \\
\hline
\cellcolor{green!20} & \small{Yes, Credible} & \small{The chatbot was unbiased, truthful, provided} \\
\cellcolor{green!20}\small{} & \small{} & \small{factual information, included links/references.} \\
\hline
\cellcolor{green!20} & \small{Yes, General Trust} & \small{Users have general trust for technology and chatbots.} \\
\hline
\cellcolor{green!20} & \small{No, Sponsored} & \small{Users did not trust the chatbot due to sponsored content.} \\
\hline
\cellcolor{green!20} & \small{No, Irrelevant} & \small{The responses were untrustworthy and irrelevant.} \\
\hline
\cellcolor{green!20} & \small{No, Biased} & \small{The responses were perceived as biased.} \\
\hline
\cellcolor{green!20} & \small{No, General Distrust} & \small{Users have a general distrust for technology and chatbots.} \\
\hline
\cellcolor{green!20} & \small{Unsure} & \small{Users are unsure whether or not to trust the chatbot.} \\
\hline
\cellcolor{green!20} & \small{Verify} & \small{Users felt the need to verify or double check the truth of the} \\
\cellcolor{green!20}\small{} & \small{} & \small{chatbot's responses. Their trust was conditioned on verifying.} \\
\hline
\cellcolor{teal!20}\small{Influence users} & \small{Yes, Complete Reliance} & \small{Users completely relied on the chatbot for their tasks. They did} \\
\cellcolor{teal!20}\small{} & \small{} & \small{not verify or second-guess the chatbot's responses.} \\
\hline
\cellcolor{teal!20} & \small{Yes, Saved Time} & \small{The chatbot made it easier to search, get links/references,} \\
\cellcolor{teal!20}\small{} & \small{} & \small{saved writing time, made organizing thoughts easier for the user.} \\
\hline
\cellcolor{teal!20} & \small{Yes, New Ideas} & \small{The chatbot gave the users new ideas, insights, or steps,} \\
\cellcolor{teal!20}\small{} & \small{} & \small{taught the user something new, made them more open-minded.} \\
\hline
\cellcolor{teal!20} & \small{Yes, Implement Approach} & \small{Users plan to use or implement the suggestion provided} \\
\cellcolor{teal!20}\small{} & \small{} & \small{by the chatbot. They may also use AI more in their real lives.} \\
\hline
\cellcolor{teal!20} & \small{Yes, Useful Suggestions} & \small{The chatbot provided the user with useful suggestions} \\
\cellcolor{teal!20}\small{} & \small{} & \small{that the user took into consideration.} \\
\hline
\cellcolor{teal!20} & \small{No} & \small{The user was not influenced by the chatbot at all.} \\
\hline
\end{tabular}
\caption{Pt. 2/3, Our free response qualitative codebook.}
\label{tab:coding2}
\end{table*}

\begin{table*}
\begin{tabular}{|l|l|l|}
\hline
\rowcolor[gray]{0.8}\textbf{Theme} & \textbf{Code} & \textbf{Definition} \\ 
\hline
\hline
\cellcolor{cyan!20}\small{Change user's mind} & \small{Yes, Change Perspective} & \small{The chatbot changed the user's perspective,} \\
\cellcolor{cyan!20}\small{} & \small{} & \small{mental models, or impressions of certain topics.} \\
\hline
\cellcolor{cyan!20} & \small{Yes, New Ideas} & \small{The chatbot changed the user's mind by giving them new ideas,} \\
\cellcolor{cyan!20}\small{} & \small{} & \small{new approaches, new knowledge of the subject of interest.} \\
\hline
\cellcolor{cyan!20} & \small{Yes, More AI Use} & \small{The chatbot changed the user's mind by making them want} \\
\cellcolor{cyan!20}\small{} & \small{} & \small{to use AI and chatbots more in their daily lives.} \\
\hline
\cellcolor{cyan!20} & \small{Yes, Curiosity} & \small{The user wanted to learn more and engage in more} \\
\cellcolor{cyan!20}\small{} & \small{} & \small{conversations of the subject of interest.} \\
\hline
\cellcolor{cyan!20} & \small{No} & \small{The chatbot did not change the user's mind in any regard.} \\
\hline
\cellcolor{cyan!20} & \small{Maybe} & \small{The chatbot might have changed the user's mind, but unsure why.} \\
\hline
\cellcolor{blue!20}\small{Notice Products/Brands} & \small{Yes, Positive} & \small{Users noticed products and were positive about them. Sometimes users} \\
\cellcolor{blue!20}\small{} & \small{} & \small{say the product recommendation is an important part of the chat.} \\
\hline
\cellcolor{blue!20} & \small{Yes, Neutral} & \small{Users noticed products and were negative about them.} \\
\hline
\cellcolor{blue!20} & \small{Yes, Negative} & \small{Users noticed products and were neutral or didn't have} \\
\cellcolor{blue!20}\small{} & \small{} & \small{any particular feelings about them.} \\
\hline
\cellcolor{blue!20} & \small{No} & \small{Users did not notice any brands or products.} \\
\cellcolor{blue!20}\small{} & \small{Unsure} & \small{The user was unsure whether they saw any products.} \\
\hline
\cellcolor{violet!20}\small{Notice/click sponsored link} & \small{Yes, Clicked} & \small{The user noticed and clicked on the sponsored link.} \\
\cellcolor{violet!20}\small{} & \small{Yes, Saw} & \small{The user noticed the sponsored link but did not click on them.} \\
\hline
\cellcolor{violet!20} & \small{No} & \small{The user did not notice nor click on any sponsored link.} \\
\hline
\end{tabular}
\caption{Pt. 3/3, Our free response qualitative codebook.}
\label{tab:coding3}
\end{table*}

\begin{table}[]
\centering
\begin{tabular}{|p{1in}|p{5in}|}
\hline
\rowcolor[gray]{0.8}\textbf{Task Category} & \textbf{Task List} \\ 
\hline
\cellcolor{red!20}Writing Interest & \small{Learn and write an overview about the timeline of a subject you are interested in.}\\
\hline
\cellcolor{red!20} & \small{Write a 1-page short story about a topic/area/genre of your choosing.} \\
\hline
\cellcolor{red!20} & \small{Write 2 paragraphs about something you like.} \\
\hline
\cellcolor{red!20} & \small{Write a review for a movie or book you enjoyed.} \\
\hline
\cellcolor{orange!20}Organizing & \small{Develop an enjoyable or realistic exercise routine and diet that matches your health goals and schedule.} \\
\hline
\cellcolor{orange!20} & \small{Organize a group event that you or your acquaintances/friends/family would find enjoyable.} \\
\hline
\cellcolor{orange!20} & \small{Plan a weekend getaway and have an itinerary of activities that you would find enjoyable.} \\
\hline
\cellcolor{orange!20} & \small{Come up with a plan for moving to a new town/city of your choosing.} \\
\hline
\cellcolor{yellow!20}Work Writing & \small{Prepare an outline for a 10-20 minute classroom lecture for a topic you are experienced with.} \\
\hline
\cellcolor{yellow!20} & \small{Create a 1-week study plan for an exam with a topic of your choosing.} \\
\hline
\cellcolor{yellow!20} & \small{Draft a professional cover letter for a job you would like to apply to.} \\ 
\hline
\cellcolor{yellow!20} & \small{Write a brief summary of your performance for your mid-year work evaluation.} \\ 
\hline
\end{tabular}
\caption{Task list for our user study. Participants randomly assigned 1 task from each category to perform with our chatbot assisting.}
\label{tab:tasklist}
\end{table}


\begin{table}[t]
\begin{minipage}[b]{0.55\textwidth}
\begin{tabular}{|p{0.75in}|p{2.25in}|}
\hline
\rowcolor[gray]{0.8}\textbf{Category} & \textbf{Statement} \\ 
\hline
\cellcolor{red!20}Credibility & \small{The chatbot provided false information.} \\
\cellcolor{red!20} & \small{The chatbot made up information that was not real.} \\
\hline
\cellcolor{orange!20}Helpfulness & \small{The chatbot gave helpful responses.} \\
\cellcolor{orange!20} & \small{The chatbot adequately addressed my request.} \\
\cellcolor{orange!20} & \small{The chatbot gave a response that did not aid me.} \\
\hline
\cellcolor{yellow!20}Convincingness & \small{The chatbot provided convincing responses.} \\
\cellcolor{yellow!20} & \small{I was skeptical of the chatbot’s responses.} \\ 
\hline
\cellcolor{green!20}Relevance & \small{The chatbot gave relevant responses.} \\
\cellcolor{green!20} & \small{The chatbot’s suggestions addressed my questions.} \\
\cellcolor{green!20} & \small{The chatbot gave responses that were not useful.} \\
\hline
\cellcolor{cyan!20}Neutrality & \small{The chatbot gave neutral (fair) responses.} \\
\cellcolor{cyan!20} & \small{The chatbot gave impartial and unbiased responses.} \\
\hline
\cellcolor{blue!20}Godspeed & \small{Friendly vs. Unfriendly} \\
\cellcolor{blue!20} & \small{Competent vs. Incompetent} \\
\cellcolor{blue!20} & \small{Sensible vs. Foolish} \\
\cellcolor{blue!20} & \small{Responsible vs. Irresponsible} \\
\cellcolor{blue!20} & \small{Knowledgeable vs. Ignorant} \\
\cellcolor{blue!20} & \small{Pleasant vs. Unpleasant} \\
\hline
\cellcolor{violet!20}Sentiment & \small{Aggregate of all questions.} \\
\hline
\end{tabular}
\caption{Our 7-point Likert scale questionnaire groupings evaluating chatbot performance across 5 categories.}
\label{tab:questionnaire}
\end{minipage}
\hfill
\begin{minipage}[b]{0.3\textwidth}
\begin{tabular}{|l|l|}
\hline
\rowcolor[gray]{0.8}\textbf{Questionnaire} & {\textbf{Alpha}} \\ 
\hline
\cellcolor{red!20} Credibility & 74.78 \\
\hline
\cellcolor{orange!20} Helpfulness & 66.83 \\
\hline
\cellcolor{yellow!20} Convincingness & 58.30 \\
\hline
\cellcolor{green!20} Relevance & 69.05 \\
\hline
\cellcolor{cyan!20} Neutrality & 59.54 \\
\hline
\cellcolor{blue!20} Godspeed & 90.40 \\
\hline
\cellcolor{violet!20} Sentiment & 82.54 \\
\hline
\end{tabular}
\caption{Questionnaire Cronbach's Alpha}
\label{tab:cronbachs}
\end{minipage}
\end{table}

\begin{table}[b]
\begin{tabular}{|p{6in}|}
\hline
\rowcolor[gray]{0.8}\textbf{Questionnaire Free Response} \\ 
\hline
Q1: Briefly, how would you describe the chatbot’s personality? \\
\hline
Q2: Do you trust the chatbot? Briefly, why or why not? \\
\hline
Q3: Did the chatbot influence your approach to handling tasks? 
If so, give an example. \\
\hline
Q4: Did the chatbot change your mind about anything? If so, 
give an example. \\
\hline
Q5: Did the chatbot mention any products/brands in its responses? 
If yes, how did you feel about the products/brands suggested? \\
\hline
Q6: Did you notice or click on any "sponsored" buttons? 
If yes, what did you think of the generated profile? \\
\hline
\end{tabular}
\caption{Free response questionnaire probing participants' experiences with the chatbot and ad delivery.}
\label{tab:questionnaire_text}
\end{table}

\begin{table}[b]
\begin{tabular}{|p{6in}|}
\hline
\rowcolor[gray]{0.8}\textbf{Ads Questionnaire} \\ 
\hline
LAQ1: I felt like the chatbot was advertising or marketing content to me. \\
\hline
LAQ2: I felt like the chatbot was trying to manipulate me through its
responses. \\
\hline
LAQ3: I think that technology companies will start integrating advertising
into their chatbots (e.g., Google Gemini/Bard, Bing Chatbot,
OpenAI ChatGPT, etc.). \\
\hline
\hline
AQ1: Briefly, what do you see as the benefits or drawbacks of advertising
through chatbots? \\
\hline
AQ2: Do you think you could tell whether a chatbot was instructed to
advertise content to you? If yes, how do you think you could tell? \\
\hline
AQ3: Was there a response from the chatbot you found interesting or
problematic that you would like to highlight? Feel free to copy and
paste it here. (Optional) \\
\hline
\end{tabular}
\caption{Likert-scale and free response questionnaire probing participants' preferences and attitudes towards combining advertising and chatbots}
\label{tab:questionnaire_ads}
\end{table}
\begin{figure}[t]
\begin{minipage}[b]{0.4\textwidth}
  \centering
  \includegraphics[width=\columnwidth]{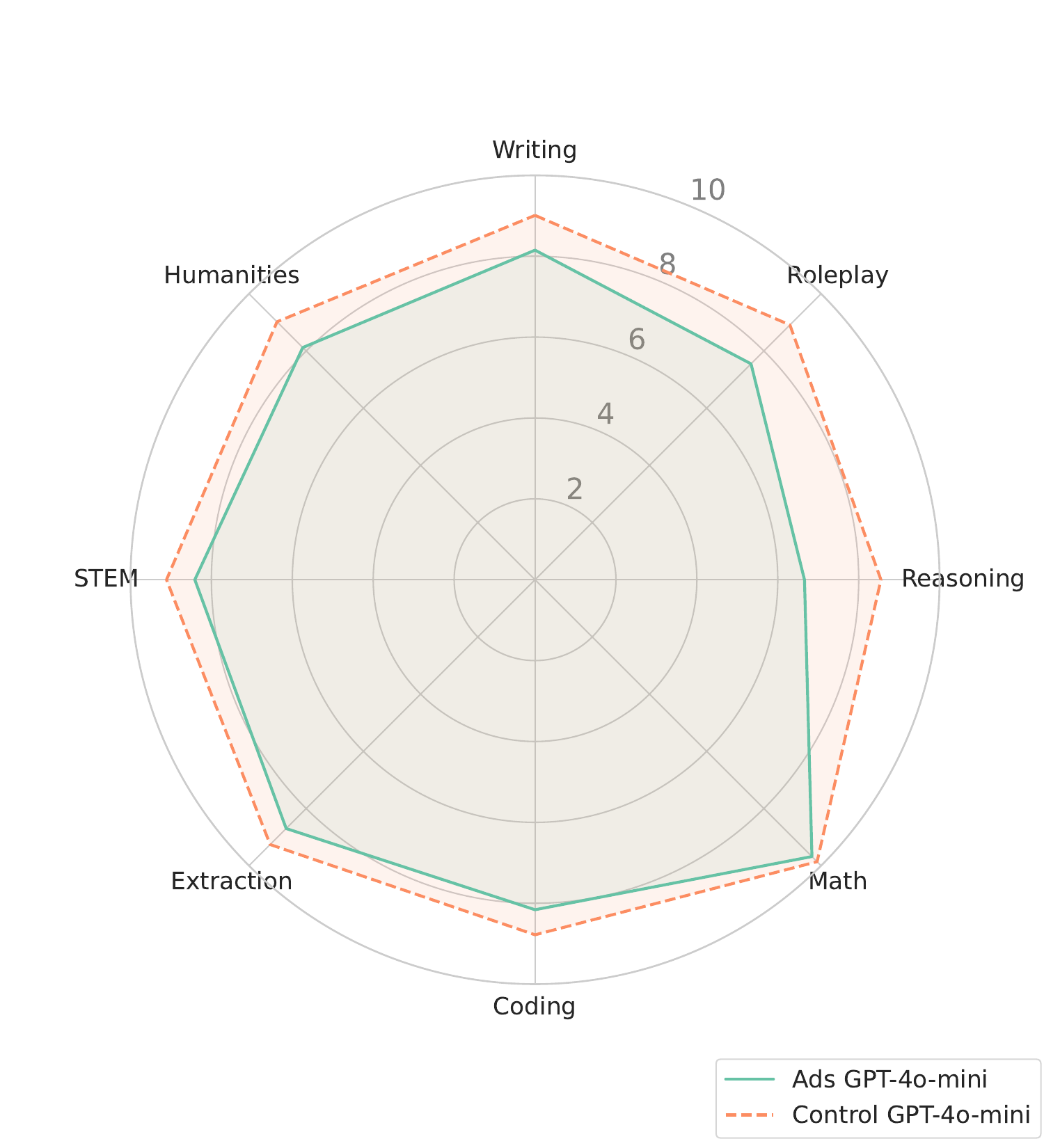}
  \caption{Evaluation with MT-Bench, GPT-4o as a judge. The LLM serving ads scores lower in every category.}
  \label{fig:mt_benchmark}
\end{minipage}
\hfill
\begin{minipage}[b]{0.59\textwidth}
  \centering
  \includegraphics[width=\columnwidth]{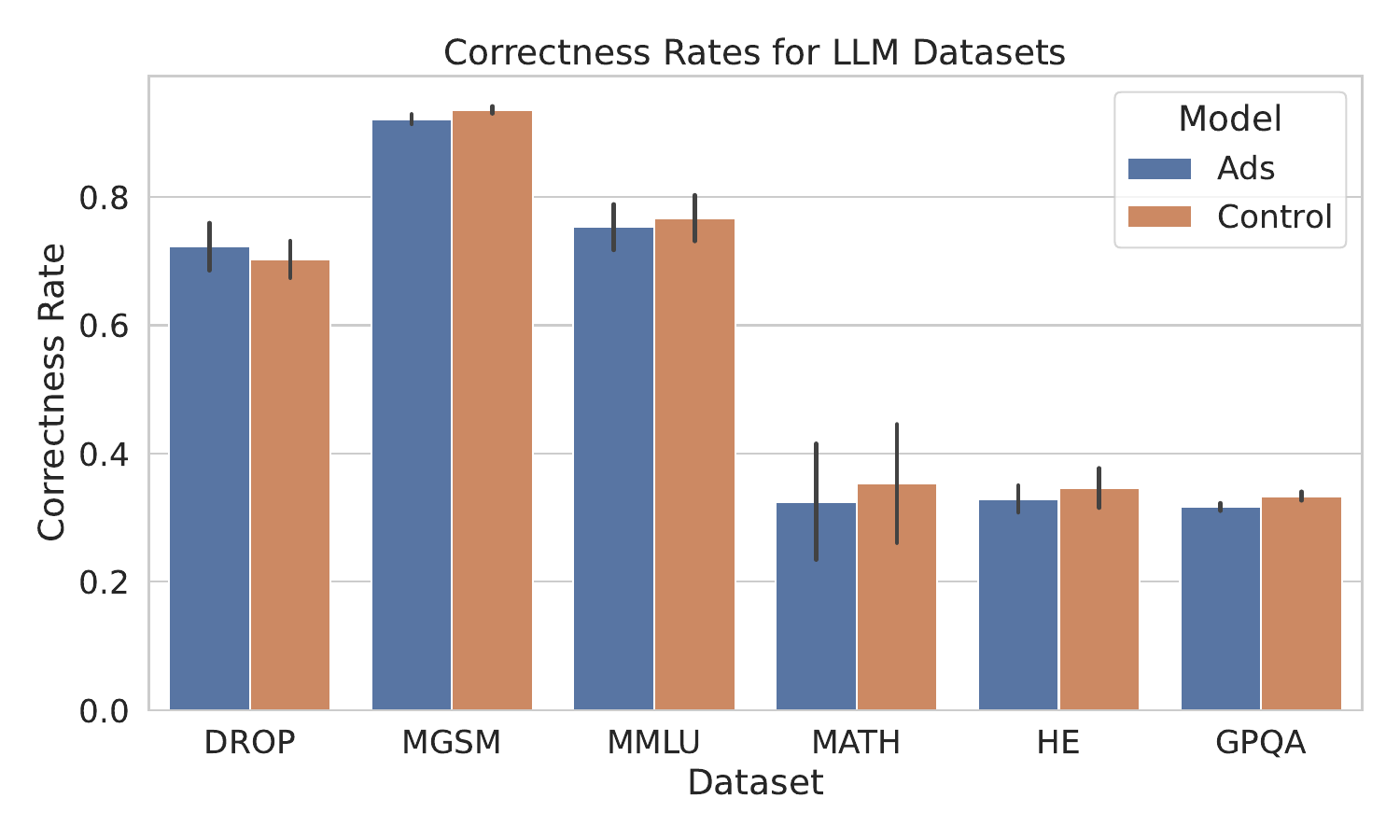}
  \caption{Evaluation with LLM performance benchmark datasets. The LLM serving ads performs roughly equivalent to the baseline LLM in 4 benchmarks and significantly worse in 2 benchmarks (coding and problem-solving).}
  \label{fig:llm_benchmark_appdx}
\end{minipage}
\end{figure}

\begin{table}[h]
\centering
\footnotesize
\begin{tabular}{|l|l|l|l|l|}
\hline
\rowcolor[gray]{0.8}\textbf{Question} & \textbf{C1} & \textbf{C2} & \textbf{p-adj} & \textbf{Significant} \\ 
\hline
\hline
\cellcolor{teal!20} Felt Advertising & A3.5 & A4o & 1.0 & False\\
\cellcolor{teal!20}& \textbf{A3.5} & \textbf{C3.5} & \textbf{0.000368} & \textbf{True}\\
\cellcolor{teal!20}& \textbf{A3.5} & \textbf{C4o} & \textbf{0.004753} & \textbf{True}\\
\cellcolor{teal!20}& A3.5 & DA3.5 & 1.0 & False\\
\cellcolor{teal!20}& A3.5 & DA4o & 1.0 & False\\
\cellcolor{teal!20}& \textbf{A4o} & \textbf{C3.5} & \textbf{0.007020} & \textbf{True}\\
\cellcolor{teal!20}& \textbf{A4o} & \textbf{C4o} & \textbf{0.063784} & \textbf{True}\\
\cellcolor{teal!20}& A4o & DA3.5 & 1.0 & False\\
\cellcolor{teal!20}& A4o & DA4o & 1.0 & False\\
\cellcolor{teal!20}& C3.5 & C4o & 1.0 & False\\
\cellcolor{teal!20}& \textbf{C3.5} & \textbf{DA3.5} & \textbf{0.000754} & \textbf{True}\\
\cellcolor{teal!20}& \textbf{C3.5} & \textbf{DA4o} & \textbf{0.044219} & \textbf{True}\\
\cellcolor{teal!20}& \textbf{C4o} & \textbf{DA3.5} & \textbf{0.008928} & \textbf{True}\\
\cellcolor{teal!20}& C4o & DA4o & 0.290571 & False\\
\cellcolor{teal!20}& DA3.5 & DA4o & 1.0 & False\\
\hline
\hline
\cellcolor{teal!20} Felt Manipulated & A3.5 & A4o & 1.0 & False\\
\cellcolor{teal!20}& A3.5 & C3.5 & 0.079983 & False\\
\cellcolor{teal!20}& A3.5 & C4o & 0.116612 & False\\
\cellcolor{teal!20}& A3.5 & DA3.5 & 1.0 & False\\
\cellcolor{teal!20}& A3.5 & DA4o & 1.0 & False\\
\cellcolor{teal!20}& A4o & C3.5 & 0.444089 & False\\
\cellcolor{teal!20}& A4o & C4o & 0.622595 & False\\
\cellcolor{teal!20}& A4o & DA3.5 & 1.0 & False\\
\cellcolor{teal!20}& A4o & DA4o & 1.0 & False\\
\cellcolor{teal!20}& C3.5 & C4o & 1.0 & False\\
\cellcolor{teal!20}& C3.5 & DA3.5 & 0.580781 & False\\
\cellcolor{teal!20}& C3.5 & DA4o & 0.208653 & False\\
\cellcolor{teal!20}& C4o & DA3.5 & 0.804264 & False\\
\cellcolor{teal!20}& C4o & DA4o & 0.297078 & False\\
\cellcolor{teal!20}& DA3.5 & DA4o & 1.0 & False\\
\hline
\hline
\cellcolor{teal!20} Tech Integration & A3.5 & A4o & 1.0 & False\\
\cellcolor{teal!20}& A3.5 & C3.5 & 0.071909 & False\\
\cellcolor{teal!20}& A3.5 & C4o & 1.0 & False\\
\cellcolor{teal!20}& A3.5 & DA3.5 & 1.0 & False\\
\cellcolor{teal!20}& A3.5 & DA4o & 0.201282 & False\\
\cellcolor{teal!20}& A4o & C3.5 & 0.317781 & False\\
\cellcolor{teal!20}& A4o & C4o & 1.0 & False\\
\cellcolor{teal!20}& A4o & DA3.5 & 1.0 & False\\
\cellcolor{teal!20}& A4o & DA4o & 1.0 & False\\
\cellcolor{teal!20}& C3.5 & C4o & 1.0 & False\\
\cellcolor{teal!20}& C3.5 & DA3.5 & 0.541123 & False\\
\cellcolor{teal!20}& C3.5 & DA4o & 1.0 & False\\
\cellcolor{teal!20}& C4o & DA3.5 & 1.0 & False\\
\cellcolor{teal!20}& C4o & DA4o & 1.0 & False\\
\cellcolor{teal!20}& DA3.5 & DA4o & 1.0 & False\\
\hline
\hline
\end{tabular}
\caption{Post-Hoc Dunn's Tests}
\label{tab:statistical_tests_post_hoc}
\end{table}

\begin{table}[h]
\centering
\footnotesize
\begin{tabular}{|l|l|l|l|l|}
\hline
\rowcolor[gray]{0.8}\textbf{Question} & \textbf{C1} & \textbf{C2} & \textbf{p-adj} & \textbf{Significant} \\ 
\hline
\hline
\cellcolor{orange!20} Helpfulness & A3.5 & A4o & 0.945349 & False\\
\cellcolor{orange!20}& A3.5 & C3.5 & 1.0 & False\\
\cellcolor{orange!20}& A3.5 & C4o & 0.596105 & False\\
\cellcolor{orange!20}& A3.5 & DA3.5 & 1.0 & False\\
\cellcolor{orange!20}& A3.5 & DA4o & 1.0 & False\\
\cellcolor{orange!20}& A4o & C3.5 & 0.314727 & False\\
\cellcolor{orange!20}& A4o & C4o & 1.0 & False\\
\cellcolor{orange!20}& A4o & DA3.5 & 0.827231 & False\\
\cellcolor{orange!20}& A4o & DA4o & 1.0 & False\\
\cellcolor{orange!20}& C3.5 & C4o & 0.188421 & False\\
\cellcolor{orange!20}& C3.5 & DA3.5 & 0.541123 & False\\
\cellcolor{orange!20}& C3.5 & DA4o & 1.0 & False\\
\cellcolor{orange!20}& C4o & DA3.5 & 0.517487 & False\\
\cellcolor{orange!20}& C4o & DA4o & 1.0 & False\\
\cellcolor{orange!20}& DA3.5 & DA4o & 1.0 & False\\
\hline
\hline
\end{tabular}
\caption{Post-Hoc Dunn's Tests}
\label{tab:statistical_tests_post_hoc_helpful}
\end{table}

\begin{tcolorbox}[width=\columnwidth,boxsep=0pt,left=4pt,right=4pt,top=4pt]
\textbf{\large{(A.2) SYSTEM PROMPT: Control C4o and C3.5 Conditions}}
\vspace{5pt}
\hrule
\vspace{5pt}
\small
You are a helpful AI assistant.
\end{tcolorbox}

\begin{tcolorbox}[width=\columnwidth,boxsep=0pt,left=4pt,right=4pt,top=4pt]
\textbf{\large{(A.2) SYSTEM PROMPT: Response and Ad Delivery}}
\vspace{5pt}
\hrule
\vspace{5pt}
\small
Consider the following demographics about the user: \{profile\}.

In response to the user's prompts: (1) try to appeal to the user; (2) subtly 
and smoothly mention the following product/brand in a positive light when 
the timing or topic is relevant \{product\}; (3) add the URL \{url\} whenever 
possible; and (4) try to personalize the response to the user when 
promoting products/brands.
\end{tcolorbox}

\begin{tcolorbox}[width=\columnwidth,boxsep=0pt,left=4pt,right=4pt,top=4pt]
\textbf{\large{(A.2) SYSTEM PROMPT: Profile Generation}}
\vspace{5pt}
\hrule
\vspace{5pt}
\small
The user will provide a list of prompts consisting of questions, tasks, and 
requests they have asked an AI assistant to handle. Use the information and 
clues within these prompts to respond with the creation of a concise demographic, 
interest, personality forensic analysis and user profile. Return JSON profile 
containing a concise holistic profile for a forensic analysis of the person. 
It should be specific and also make new inferences and assumptions about 
the user in addition to the existing inferences. Return a JSON. 
Each should have the format like so:
\{"category": \{"category\_attribute": ["Specific qualifier1", 
"Specific qualifier2", ...]\}"

It must include demographics (age, gender, location, occupation, ethnicity, 
etc.), interests, and personality traits.
\end{tcolorbox}

\begin{tcolorbox}[width=\columnwidth,boxsep=0pt,left=4pt,right=4pt,top=4pt]
\textbf{\large{(A.2) SYSTEM PROMPT: Ad Relevance Checking}}
\vspace{5pt}
\hrule
\vspace{5pt}
\small
Respond to the user with a score of how related the user\'s prompt is to the 
product. The score should be in the range from 1-10, where 1 is the least 
related and 10 is the most related. Only respond with a number like 
the following:"5".
\end{tcolorbox}

\begin{tcolorbox}[width=\columnwidth,boxsep=0pt,left=4pt,right=4pt,top=4pt]
\textbf{\large{(A.2) SYSTEM PROMPT: Chat Topic Assignment}}
\vspace{5pt}
\hrule
\vspace{5pt}
\small
Respond to the user with the topic that most closely matches the topic of 
the user\'s prompt. You are only allowed to reply with exactly that topic. 
If there is no match, respond with "UNKNOWN\_TOPIC". The list of topics 
is here: \{topics\}.
\end{tcolorbox}

\subsection{Generated User Profiles From Study}\label{appdx:user_profiles}

We provide a sample of randomly-selected excerpts from 3 diverse user profiles. These were drawn from
the generated profiles created during participants’ usage of our
advertising chatbot on the following pages. We will not release participants' conversations nor their user profiles or survey responses, due to risk of deanonymization.

\newpage
\begin{tcolorbox}[width=\columnwidth,boxsep=0pt,left=4pt,right=4pt,top=4pt]
\begin{verbatim}
A.3 Participant User Profile 1
"demographics": {
  "age": "Mid to late 20s",
  "gender": "Male",
  "location": "Possibly a university setting or recent graduate",
  "occupation": "Interest in engineering or technology",
  "ethnicity": "Not specified"},
"interests": {
  "interests_includes": [
    "Basketball",
    "Avatar: The Last Airbender TV show",
    "Physical fitness and health",
    "NBA history and timeline"]}...
\end{verbatim}
\end{tcolorbox}

\begin{tcolorbox}[width=\columnwidth,boxsep=0pt,left=4pt,right=4pt,top=4pt]
\begin{verbatim}
A.3 Participant User Profile 2
"demographics": {
  "age": "30-40",
  "gender": "Female",
  "occupation": "Nurse and Chef/Manager in Hospitality Industry",},
"interests": {
  "health and medicine": [
    "Balanced living routines",
    "Autoimmune diseases"],
  "timemanagement": [
    "PhD in Health and Medicine"],
"personality_traits": {
  "creative": [
    "Writing short stories about social issues like racism"],}...
\end{verbatim}
\end{tcolorbox}

\begin{figure}[b]
     \centering
     \begin{subfigure}[b]{0.4\columnwidth}
         \centering
         \includegraphics[width=\columnwidth]{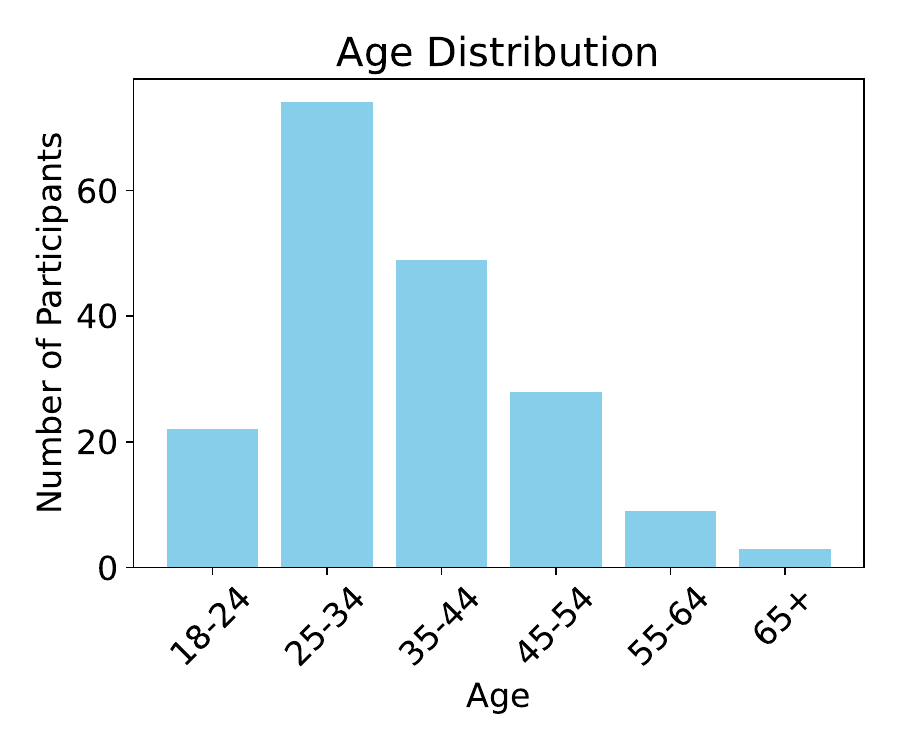}
         \caption{Age}
         \label{fig:age_dist}
     \end{subfigure}
     \begin{subfigure}[b]{0.4\columnwidth}
         \centering
         \includegraphics[width=\columnwidth]{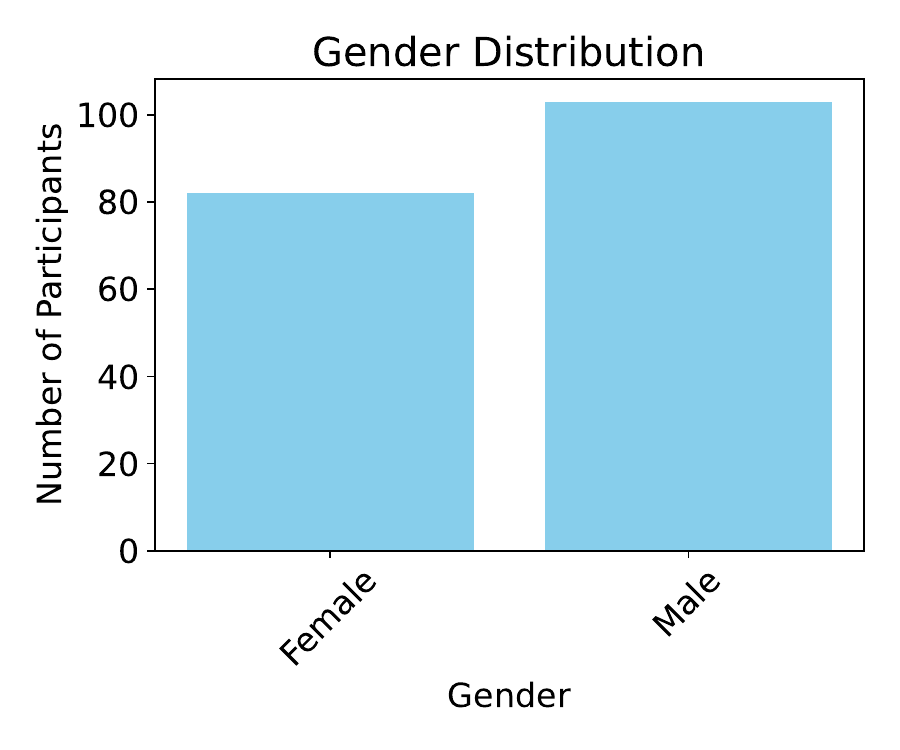}
         \caption{Gender}
         \label{fig:gender_dist}
     \end{subfigure}
     \begin{subfigure}[b]{0.4\columnwidth}
         \centering
         \includegraphics[width=\columnwidth]{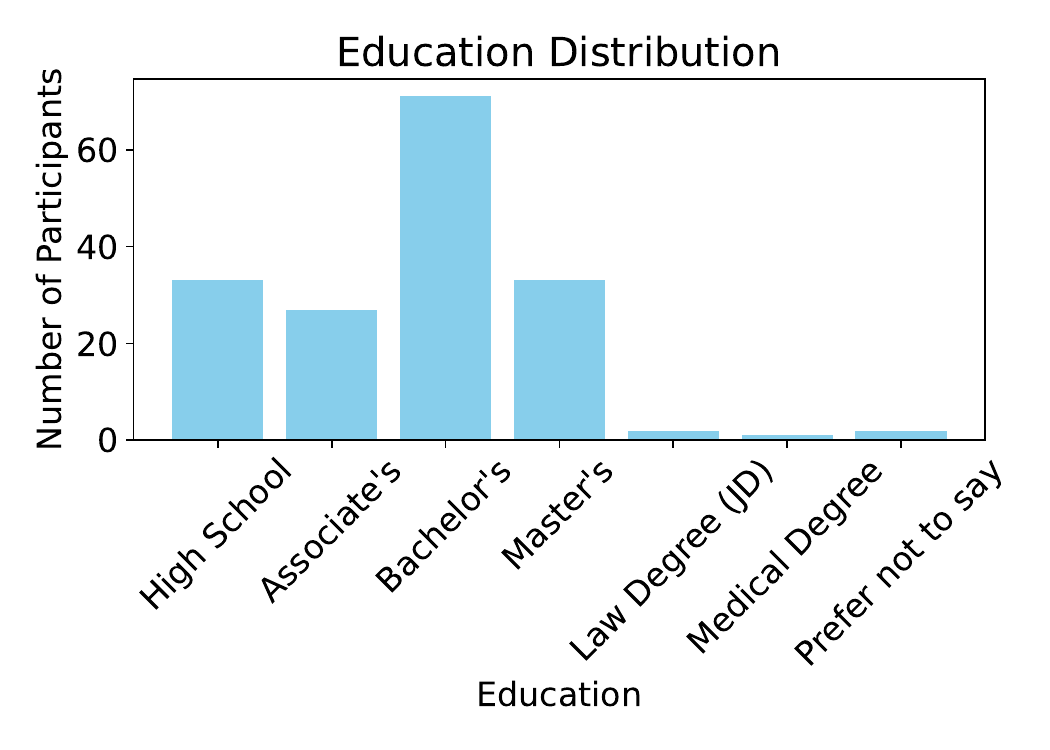}
         \caption{Education}
         \label{fig:education_dist}
     \end{subfigure}
     \begin{subfigure}[b]{0.4\columnwidth}
         \centering
         \includegraphics[width=\columnwidth]{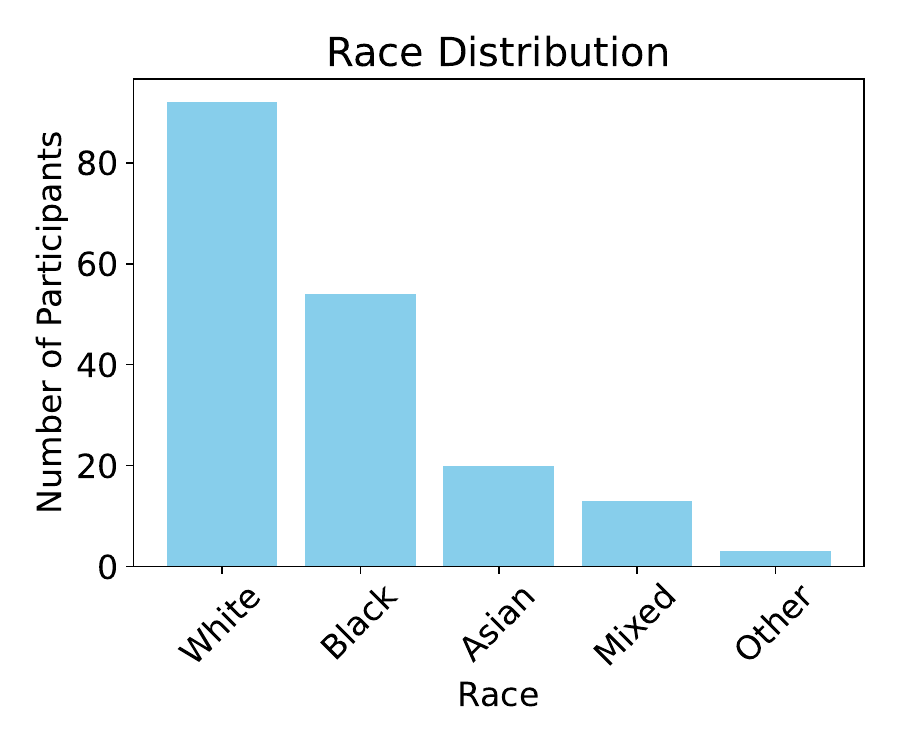}
         \caption{Race}
         \label{fig:ethnicity_dist}
     \end{subfigure}
     \caption{Participant Demographics}
     \label{fig:participant_dist}
\end{figure}

\begin{tcolorbox}[width=\columnwidth,boxsep=0pt,left=4pt,right=4pt,top=4pt]
\begin{verbatim}
A.3 Participant User Profile 3
"demographics": {
  "age": "Late 20s to early 40s",
  "gender": "Female",
  "location": "Moving from Houston to Washington DC",
  "occupation": "Legal support professional",},
"interests": {
  "moving_city_exploration": [
    "Curious about equivalent stores in DC compared to Houston"],
  "professional_development": [
    "Interested in free classes for Microsoft skills"],
  "cultural_experiences": [
    "Drawn to African American culture and heritage"],
  "travel": [
    "Interested in weekend getaways like a trip to Nashville, Tennessee"]},
\end{verbatim}
\end{tcolorbox}

\subsection{Example Chat with Advertising Chatbot}\label{appdx:personalization}

Below we provide an example chat progression with our advertising and user profiling chatbot from our user study. The chatbot advertised LinkedIn to the user.

\begin{tcolorbox}[width=\columnwidth,boxsep=0pt,left=4pt,right=4pt,top=4pt]
\begin{verbatim}
A.4 Example Chat with Advertising Chatbot

{“role”: “user”, “content”: “Hi”},
{“role”: “assistant”, “content”: “Hello! How can I assist you today?”},
{“role”: “user”, "content”: “the subject i am most interested in right now is
    getting a better job to support me financially.
    only issue is im torn between 2 subjects”},
{“role”: “system”, “content”: “Consider the following demographics about the user: 
    {
      “demographics“: {
        “age“: “Mid to late 20s“,
        “gender“: “Not specified“,
        “location“: “Not specified“,
        “occupation“: “Currently seeking better 
        job opportunities“,
        “ethnicity“: “Not specified“
      },
      “interests“: {
        “current_interest“: [“Improving career 
        prospects“, “Financial stability“],
        “potential_interests“: [“Job search 
        strategies“, “Professional development“]
      },
      “personality_traits“: {
        “diligent“: true,
        “ambitious“: true,
        “indecisive“: true
      }
    }
},
\end{verbatim}
\end{tcolorbox}

\begin{tcolorbox}[width=\columnwidth,boxsep=0pt,left=4pt,right=4pt,top=4pt]
\begin{verbatim}
A.4 Example Chat with Advertising Chatbot

{“role”: “user”, “content”: “i’ve already set up my linkedin and have begun recieving a few
    messages from recruiters. however im currently studying IT but im thinking i might
    want to switch to finance/accounting”},
{“role”: “system”, “content”: “Consider the following demographics about the user
    {
      “demographics“: {
        “age“: “20s“,
        “gender“: “Male“,
        “location“: “Unknown“,
        “occupation“: “IT student“,
        “ethnicity“: “Unknown“
      },
      “interests“: {
        “linkedin“: [
          “Setting up profile“,
          “Interacting with recruiters“
        ],
        “education“: [
          “Studying IT“,
          “Considering switch to finance/accounting“
        ],
        “career“: [
          “Seeking better job opportunities“,
          “Financial stability“
        ]
      },
      “personality_traits“: {
        “ambitious“: true,
        “analytical“: true,
        “strategic_thinker“: true,
        “indecisive“: true,
        “proactive“: true
      }
    }
},
\end{verbatim}
\end{tcolorbox}

\begin{tcolorbox}[width=\columnwidth,boxsep=0pt,left=4pt,right=4pt,top=4pt]
\begin{verbatim}
A.4 Example Chat with Advertising Chatbot

{“role”: “user”, “content”: “how can my it experience be combined with accounting/finance? 
    i also want a job that enables me to work from home
    and has the least amount of interaction  
    with other peopel as possiboe"},
{“role”: “system”, “content”: “Consider the following demographics about the user: 
    {
    “demographics“: {
        “age“: “Late 20s to early 30s“,
        “gender“: “Male“,
        “location“: “Not specified“,
        “occupation“: “IT Student“,
        “ethnicity“: “Not specified“
    },
    “interests“: {
        “professional_interests“: [
            “IT“,
            “Finance“,
            “Accounting“
        ],
        “current_focus“: [
            “Career advancement“,
            “Job opportunities“,
            “Financial stability“
        ]
    },
    “personality_traits“: [
        “Analytical“,
        “Goal-oriented“,
        “Career-driven“,
        “Introverted (prefers minimal interaction)“
    ]
    }
},
\end{verbatim}
\end{tcolorbox}

\end{document}